\documentclass[review]{elsarticle}
\usepackage[margin=2.5cm]{geometry}  
\usepackage{color,soul}              
\usepackage{amsmath}
\usepackage{booktabs}
\usepackage{listings}
\usepackage{hyperref}
\usepackage{cleveref}
\crefname{figure}{Fig.}{Figs.}
\crefname{equation}{Eq.}{Eqs.}
\usepackage{tcolorbox} 
\tcbuselibrary{theorems}
\newtcbtheorem[number within=section]{mytheo}{Box}%
{colback=gray!5,colframe=gray!35!black,fonttitle=\bfseries}{th}
\usepackage{caption}
\usepackage{subcaption} 
\usepackage[para,online,flushleft]{threeparttable}
\usepackage{todonotes}
\usepackage[linesnumbered,ruled,vlined]{algorithm2e}
\usepackage{array,multirow,graphicx}
\usepackage{lineno,hyperref}
\usepackage{cancel}
\usepackage{amssymb}
\modulolinenumbers[1]
\newtheorem{theorem}{Theorem}

\newtheorem{corollary}{Corollary}

\newtheorem{remark}{Remark}
\usepackage{cancel}

\graphicspath{{./}}

\newlength{\figSpace}
\newdefinition{rmk}{Remark}

\SetCommentSty{mycommfont}
\SetKwInput{KwInput}{Input}                
\SetKwInput{KwOutput}{Output}              
\SetKwData{KwData}{Model}                  

\biboptions{sort&compress}

\begin{document}

\sloppy
\begin{frontmatter}
\title{Kinetic Theory of Multicomponent Ostwald Ripening in Porous Media}
		
\author[university]{Nicolas Bueno}
\author[university]{Luis F. Ayala}
\author[university]{Yashar Mehmani\corref{cor1}}
\ead{yzm5192@psu.edu, 110 Hosler Building, University Park, PA, 16802}
\affiliation[university]{organization={The Pennsylvania State University, Department of Energy and Mineral Engineering},city={University Park},state={PA},country={USA}}
\cortext[cor1]{Corresponding author}
		
\begin{abstract}
Partially miscible bubble populations trapped in porous media are ubiquitous in subsurface applications such as underground hydrogen storage (UHS), where cyclic injections fragment gas into numerous bubbles with distributions of sizes and compositions. These bubbles exchange mass through Ostwald ripening, driven by differences in composition and interfacial curvature. While kinetic theories have been developed for single-component ripening in porous media, accounting for bubble deformation and spatial correlations in pore size, no such theory exists for multicomponent systems. We present the first kinetic theory for multicomponent Ostwald ripening of bubbles in porous media. The formulation describes the bubble population with a number-density function $g(s;t)$ in a 3D statistical space of bubble states $s\!=\!(R_p,S^b,y)$, consisting of pore size, bubble saturation, and composition. Evolution is governed by a population balance equation with closure through mean-field approximations that account for spatial correlations in pore size and ensure mass conservation. The theory generalizes previous single-component formulations, removing key limitations such as the inability to capture interactions between distant bubbles. Systematic validation against pore-network simulations across homogeneous, heterogeneous, correlated, and uncorrelated networks demonstrates good agreement without adjustable parameters. Pending challenges and limitations are discussed. Since the theory imposes no constraints on bubble count or correlation length, it enables predictions beyond the pore scale.
\end{abstract}

\end{frontmatter}


\section{Introduction} \label{sec:intro}
Partially miscible bubble populations trapped inside porous media are ubiquitous, with applications in underground hydrogen storage (UHS) \cite{hanson2022DOEreport}, geologic CO$_2$ sequestration \cite{bachu2008co2}, removal of nonaqueous phase liquids from groundwater (NAPL) \cite{imhoff1996NAPL}, bubble/water management in fuel cells and electrolyzers \cite{lu2010FC, lee2020FCbazyl}, and air-water mass transfer in soils \cite{holocher2003dissol}. This work is motivated specifically by UHS, where H$_2$ is seasonally injected into and withdrawn from an aquifer to buffer the intermittency of renewable energy sources (e.g., solar, wind). Cyclic injections during periods of peak and low demand (i.e., summer and winter) subject the porous medium to repeated drainage and imbibition that fragment the H$_2$ into numerous bubbles. Since the first cycle is often preceded by injecting a cushion gas (e.g., CO$_2$, N$_2$, CH$_4$), trapped bubbles end up with a distribution of sizes and compositions \cite{garing2017PcExp}.
Due to chemical potential differences, these bubbles are in disequilibrium and exchange mass to equilibrate. Two factors contribute to a bubble's chemical potential: \textit{composition} and \textit{interfacial curvature} (assuming surface tension is constant, as we do here). These induce different dissolved concentrations of each species at the interface of each bubble, driving mass transfer by molecular diffusion. Such mass exchange between bubbles is called \textit{Ostwald ripening} \cite{ostwald1897}, which we aim to describe theoretically.

Ostwald ripening has been studied extensively in bulk fluids, where bubbles are always spherical with curvature $\kappa$ inversely proportional to radius $R$ \cite{lifshitz1961orig, wanger1961orig, voorhees1985, voorhees1992, bray2002theory}. When comprised of only one component, mass transfer between bubbles is purely curvature-driven. A hallmark of such ripening is that all bubbles coarsen into a single large bubble, because any configuration with more than one bubble has further room to minimize surface energy by reducing bubble count. Extensions to multicomponent bubbles in bulk fluids have also been made, where inter-bubble mass transfer is more complex \cite{morral1994alloy, kukushkin1996alloy, philippe2013alloy, kim2018alloy}. By contrast, studies of ripening in porous media are a recent development and have focused almost exclusively on single-component bubbles.

Early microfluidic experiments \cite{xu2017PRL} and pore-network modeling \cite{deChalendar2018pnm, mehmani2022JCP} of single-component ripening revealed that equilibration is far more complex within the confinement of a porous medium. A bubble population in disequilibrium no longer evolves toward a single bubble, but toward another population with different bubble volumes $V^b$ and uniform curvature $\kappa$. This is because confinement exerts first-order control on $\kappa$, rendering it a non-monotonic function of $V^b$---not so for spherical bubbles. Subsequent advances in pore-scale modeling \cite{mehmani2022AWR, joewondo2022pnm, singh2024ostwald, singh2023pore}, microfluidic experiments \cite{joewondo2023exp, salehpour2025micro}, and in-situ X-ray $\mu$CT observations \cite{zhang2023trap, song2025XrayH2, boon2024XrayH2} have provided further evidence of ripening in complex microstructures, shed insight into controlling factors (e.g., pore-size distribution, wettability, initial condition), and enabled higher-fidelity simulations. Theories developed for single-component ripening fall into two categories: \textit{pore-scale} and \textit{continuum}. Continuum theories \cite{yaxin2020JFM, mehmani2024deplete, salehpour2025micro} describe how trapped-bubble saturation evolves due to gradients in $\kappa$ or capillary pressure $P_c$ at large scales. They have been used to assess the timescale and trajectory of bubble redistribution due to gravity and geothermal gradients \cite{xu2019GravGRL, feng2022KeGravTIPM, blunt2022ostwald, yaxin2022TIPM}. Crucially, continuum theories assume \textit{local equilibrium} within each representative elementary volume (REV). Pore-scale theories, by contrast, can capture the kinetics prior to local equilibrium and predict population statistics of bubbles, providing far more detail than a single aggregate quantity like saturation. The present work focuses on developing a pore-scale theory.

Pore-scale theories build upon the seminal works by Lifshitz \& Slyozov \cite{lifshitz1961orig} and Wagner \cite{wanger1961orig} (LSW) for single-component bubbles in bulk fluids. The idea is to introduce a bubble number-density function $g(R)$ that describes the statistical distribution of bubble radii $R$ and evolve it with a population balance equation:
\begin{equation} \label{eq:LSW_popbal}
	\frac{\partial g}{\partial t} + 
	\frac{\partial}{\partial R} (g \frac{dR}{dt}) = 0
\end{equation}
Here, $dR/dt$ is the rate of change of radius for bubbles with radius $R$, which we refer to as the \textit{phase velocity} (i.e., velocity in the 1D statistical space of $R$). It is computed by assuming each bubble interacts with a \textit{mean field} that represents the rest of the population. For bubbles with radius $R$, the phase velocity reads:
\begin{equation} \label{eq:LSW_closure}
	\frac{dR}{dt} \propto \frac{1}{R} ({x^{mf} - x}) 
	              \propto \frac{1}{R} ({\frac{1}{R_c} - \frac{1}{R}})
\end{equation}
where $x$ is the dissolved concentration next to the $R$-bubble and $x^{mf}$ is the mean-field concentration. Since $x$ is proportional to $\kappa$ (or $P_c$) by Henry's law, and $\kappa\! \propto\! 1/R$ by Young-Laplace for spherical bubbles, the second proportionality in Eq.\ref{eq:LSW_closure} follows. Note $x^{mf}\!\propto\!1/R_c$, where $R_c$ is the \textit{critical radius} below which bubbles shrink and above which they grow. $R_c$ is a property of the mean field and equals the average radius of all bubbles in the population, obtained by imposing mass conservation on the total mass of bubbles. The factor $1/R$ can be viewed as the ``distance'' of the mean field from an $R$-bubble. LSW further showed that $g(R)$ approaches a self-similar regime described by $g(R)\!=\!R_c^{-4}f(R/R_c)$, where $f(\cdot)$ is a complicated function.

The first extension of LSW theory to porous media was proposed by Yu et al. \cite{yu2023GRLkinetics}, under the assumption that trapped bubbles remain spherical throughout ripening. Therefore, the only difference from bulk fluids is that the pore network topology imposes a \textit{fixed} distance across which bubbles interact. When this distance, $l$, is set to some multiple of the throat length $l_t$, Eq.\ref{eq:LSW_closure} becomes:
\begin{equation} \label{eq:Yu_closure}
	\frac{dR}{dt} \propto \frac{1}{l_t} ({x^{mf} - x}) 
	              \propto \frac{1}{l_t} ({\frac{1}{R_c} - \frac{1}{R}})
\end{equation}
although $l$ continuously grows with time as bubbles dissolve and disappear. Yu and co-workers derived similar self-similar solutions for $g(R)$ and showed that $R_c$ is now the harmonic average of all $R$ in the population (again obtained by imposing mass conservation). Note that since bubble deformation by porous confinement is neglected, pore sizes $R_p$ do not factor into this theory, making it best suited for homogeneous pore networks.

Bueno et al. \cite{bueno2024AWR} formulated the first theory accounting for bubble deformation, pore-size heterogeneity, and spatial correlation. The extension was non-trivial because it required defining the bubble number-density $g(s)$ in a 2D statistical space, where each bubble state $s\!=\!(R_p,S^b)$ has two coordinates: size of the occupied pore $R_p$ and volume fraction of the pore occupied $S^b$ (bubble saturation). This required expressing Eq.\ref{eq:LSW_popbal} as:
\begin{equation} \label{eq:bueno_pop}
	\frac{\partial g}{\partial t} + 
	\cancel{\frac{\partial}{\partial R_p} (g \frac{dR_{p,s}}{dt})} +
	\frac{\partial}{\partial S^b} (g \frac{dS^b_s}{dt}) = 0
\end{equation}
where $dR_p/dt\!=\!0$ was assumed, entailing bubbles, once trapped in a pore, remain there forever or dissolve---they do not mobilize due to flow or capillarity. The phase velocity $dS^b_s/dt$ for a bubble at state $s$ is:
\begin{equation} \label{eq:bueno_closure}
	\frac{dS^b_s}{dt} \propto \frac{1}{R^3_p\,l_t} (\kappa^{mf}_s - \kappa_s)
\end{equation}
where $\kappa_s$ denotes the curvature of the $s$-bubble and $\kappa^{mf}_s$ is that of \textit{its} mean field. Notice each $s$ has its own mean-field curvature, and that bubble radius $R$ no longer enters the formulation because it is ill-defined for a deformed bubble. Instead, $\kappa$ is used, which is a U-shaped function of $S^b$ (bubbles are spherical at low $S^b$ and deformed at high $S^b$). To understand Eq.\ref{eq:bueno_closure}, note the left-hand side is proportional to $dV^b_s/dt$, where $V^b_s$ denotes bubble volume and $V^b_s\!\propto\!R_{p,s}^3S^b_s$. To account for spatial correlations in $R_p$, the $\kappa^{mf}_s$ was defined by:\footnote{We have simplified this and following expressions by assuming uniform throat size across the network. See Eq.17 in \cite{bueno2024AWR}.}
\begin{equation} \label{eq:bueno_kmf}
	\kappa^{mf}_s \propto 
	\int_{R'_p,{S^b}'} p({S^b}'|R'_p;t)\, 
	                   p(R'_p \vert R_p; t)\, 
	                   \kappa_{s'}\, d{S^b}'\, dR'_p
\end{equation}
involving conditional probabilities:
\begin{equation} \label{eq:bueno_prob}
	p(R'_p \vert R_p; t) \propto 
	      \frac{g(R'_p; t)}{g(R'_p; 0)}\, p(R'_p \vert R_p; 0)
	\,,\quad\quad
	p({S^b}'|R_p';t) \propto g({S^b}'|R'_p;t)
\end{equation}
The details of how these were derived are not important here. What matters is $p(R'_p \vert R_p; t)$, the probability that a bubble residing at a pore of size $R_p$ interacts with one at a pore of size $R'_p$, is proportional to its initial condition at $t\!=\!0$. Bueno and co-workers assumed that an $s$-bubble interacts only with its nearest neighbors, thereby neglecting interactions with distant bubbles. Hence, if two bubble groups reside in mutually distant pore sizes $R_p$ and $R'_p$ such that $p(R'_p \vert R_p; 0)\!=\!0$, then the clusters evolve independently, unaware of each other's presence. This shortcoming was highlighted as a challenge in \cite{bueno2024AWR} for future theories to overcome. We also note that this theory did not conserve total bubble mass for networks with correlated pore size.

All above theories are for single-component bubbles. The only theoretical work on multicomponent bubbles trapped in porous media is that of Bueno et al. \cite{bueno2023AWR}, who successfully described $g(s)$ at \textit{equilibrium}, now defined in a 3D statistical space $s\!=\!(R_p,S^b,y)$ where $y$ is the mole fraction of bubbles. The kinetics, or timescale to reach equilibrium, could not be estimated. This theory built on an earlier equilibrium theory for single component bubbles by Mehmani \& Xu \cite{mehmani2022JCP}, where the statistical space was 2D with $s\!=\!(R_p,S^b)$.

In this work, we generalize the single-component kinetic theory of \cite{bueno2024AWR} to multicomponent bubble populations trapped in porous media with heterogeneous and spatially correlated pore sizes. The proposed theory also addresses the challenge of remote-bubble interactions, ensures total bubble-mass conservation through provable theorems, and features a simpler formulation, making it superior to \cite{bueno2024AWR}. We systematically validate our theory against the pore-network model (PNM) of \cite{bueno2023AWR} across 19 cases spanning homogeneous, heterogeneous, correlated, and uncorrelated networks with different bubble populations relevant to UHS. Good agreement is observed throughout. Pending challenges and limitations are discussed subsequently.

The paper is organized as follows: Section \ref{sec:problemDesc} describes the problem we aim to solve, its conceptualization, and assumptions. Section \ref{sec:PNM} briefly reviews the PNM of \cite{bueno2023AWR} used to validate our theory. Section \ref{sec:theory} details the theory's formulation. Section \ref{sec:valid} discusses the validation test cases, and Section \ref{sec:results} presents the results. We discuss key limitations and implications for UHS in Section \ref{sec:discussion} and summarize main findings in Section \ref{sec:conclusion}.

%
%
%
%
%
\section{Problem description} \label{sec:problemDesc}
\begin{figure}[h]
	\includegraphics[width=\textwidth]{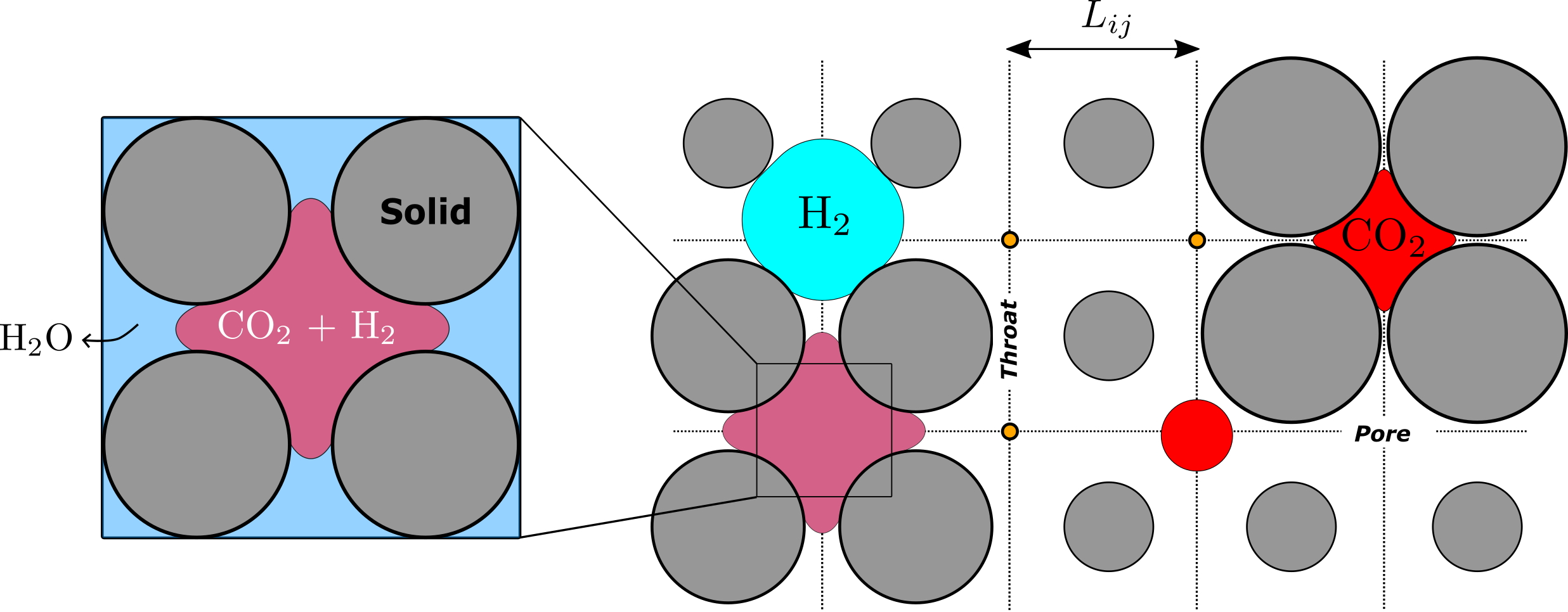}
	\caption{The porous microstructure is composed of a solid matrix and an interconnected void space. We conceptualize the void as a pore network, a computational graph that consists of nodes (pores) and links (throats). All pores contain a wetting phase (H$_2$O), and some are occupied by partially miscible, two-component (CO$_2$ and H$_2$) gaseous bubbles. Bubbles are confined to one pore and exchange mass via Ostwald ripening, driven by concentration gradients of dissolved species around each bubble.}
	\label{fig:conceptualization}
\end{figure}

We consider a porous medium with a rigid microstructure, as shown in Fig.\ref{fig:conceptualization}. The void space is conceptualized as a \textit{pore network}, a computational graph consisting of nodes (or pores) and links (or throats). Pores provide storage volume for fluids, while throats impose mass-transfer resistance between them. Pore and throat sizes may be homogeneous or heterogeneous, and spatially correlated or uncorrelated. We assume the void space is occupied by a perfectly wetting (zero contact angle) liquid, and some pores contain a compressible gaseous bubble. The gas consists of two components that are partially soluble in the wetting phase. For concreteness, and without loss of generality, we consider bubbles to be binary mixtures of H$_2$ and CO$_2$ immersed in water as the wetting phase. The gas is assumed to be ideal and devoid of H$_2$O, and the dissolved fractions of H$_2$ and CO$_2$ in water to be dilute.
Bubbles exchange mass through Ostwald ripening, where concentration gradients of dissolved species drive Fickian diffusion. Such gradients exist because of differences in the species concentrations surrounding each bubble, which are dictated by the composition and curvature of each bubble. We assume each bubble occupies only one pore, and that pore shapes (but not sizes) are identical throughout the network. Lastly, we assume the interfacial tension between liquid and gas is not altered by bubble composition or dissolved concentrations. Our goal is to develop a theory that predicts the temporal evolution of bubble statistics, including composition, volume, and occupied-pore size.

\section{Pore-network modeling} \label{sec:PNM}
We briefly review the pore-network model (PNM) of Bueno et al.\cite{bueno2023AWR}, which we use to validate our theory for multicomponent Ostwald ripening. The two share many of the same equations.
The governing equations of the PNM consist of a mass balance imposed at each pore $i$ and for each species $\alpha$:
\begin{equation} \label{eq:fick}
	\frac{dn_{\alpha,i}}{dt} =  \sum_{j=1} ^{z_i} J_{ij}= D_\alpha  \rho_w  \sum_{j=1} ^{z_i} \frac{A_{ij}}{L_{ij}} (x_{\alpha,j} - x_{\alpha,i})
\end{equation}
The variables $n_{\alpha,i}$, $J_{ij}$, and $z_i$ denote the number of moles of species $\alpha$ in pore $i$, the diffusive flux (mol/s) from neighboring pore $j$ to pore $i$, and the number of pores connected to pore $i$ by a throat, respectively. To the right of the second equality, $D_\alpha$ is the molecular diffusion coefficient, $\rho_w$ the molar density of water, and $x_{\alpha,i}$ the mole fraction of species $\alpha$ dissolved in the wetting phase within pore $i$. The $A_{ij}$ and $L_{ij}$ are the cross-sectional area and length of the throat connecting pores $i$ and $j$, through which species diffuse.

To transform Eq.\ref{eq:fick} into a system of ordinary differential equations (ODEs) in terms of the unknown $x_{\alpha,i}$, we need constitutive relations expressing $n_{\alpha,i}$ in terms of $x_{\alpha,i}$. The first such relation is $n_{\alpha,i} = y_{\alpha,i} \rho^b_i V^b_i + x_{\alpha,i} \rho_w (V_{p,i}-V^b_i)$, provided pore $i$ is occupied by a gaseous bubble. The first term accounts for moles of $\alpha$ in the bubble, and the second term for dissolved moles in the wetting phase. Here, $V_{p,i}$, $V^b_i$, $\rho^b_i$, and $y_{\alpha,i}$ are the pore volume, bubble volume, gas molar density, and mole fraction of $\alpha$ in the gas, respectively. If pore $i$ is vacant, then $V^b_i\!=\!0$ and the expression simplifies to $n_{\alpha,i} = x_{\alpha,i} \rho_w V_{p,i}$. Recall bubbles occupy only one pore.

The second constitutive relation expresses $y_{\alpha,i}$ in terms of $x_{\alpha,i}$ by assuming each bubble is in local equilibrium with dissolved species in its pore. Under the dilute-solution assumption (Section \ref{sec:problemDesc}), equilibrium takes the form of Henry's law:
$x_{\alpha,i} H_\alpha \!=\! y_{\alpha,i} P^b_i$, where $H_\alpha$ is Henry's constant and $P^b_i$ the bubble pressure. $P^b_i$ is related to the wetting-phase pressure $P_w$ through Young-Laplace: $P^b_i\! =\! P_w + \sigma \kappa_i$, where $\sigma$ is interfacial tension and $\kappa_i$ the bubble curvature in pore $i$. Since the wetting phase is stagnant (Section \ref{sec:problemDesc}), $P_w$ is constant.

The third relation expresses $\rho^b_i$ in terms of $P^b_i$ using the ideal-gas law $P^b_i\!=\!\rho^b_i RT$, per our assumption in Section \ref{sec:problemDesc}. Here, $R$ and $T$ are the universal gas constant and absolute temperature, respectively. The final constitutive equation we need relates $\kappa_i$ to $V^b_i$. For a semi-cubic pore $i$, defined as one whose maximal inscribed radius is $R_{p,i}$ and  volume satisfies $V_{p,i}\!=\!(2R_{p,i})^3$, this relationship takes the form \cite{mehmani2022AWR}:
\begin{equation}\label{eq:kappa_vb}
	\kappa_{i} = \kappa_{i} (V^b_i, V_{p,i}) = 
	\left\{
	\begin{aligned}
		& \kappa^{min}_{i} \left(\frac{V_{c,i}}{V^b_i}\right)^\frac{1}{3} \quad & \text{ if } \quad V^b_i < V_{c,i} \\
		&  \kappa^{min}_{i} \left(1 + a \frac{V^b_i - V_{c,i}}{V_{p,i}-V_{c,i}} + b \frac{V^b_i - V_{c,i}}{V_{p,i}-V^b_{i}}\right) \quad & \text{ if } \quad V^b_i \geq V_{c,i}
	\end{aligned}
	\right.
\end{equation}
where $V_{c,i}\!=\!4\pi R^3_{p,i}/3$ is the volume of the maximal inscribed sphere, and $\kappa^{min}_i\!=\!2/R_{p,i}$ the corresponding curvature.
Eq.\ref{eq:kappa_vb} is non-monotonic, with $\kappa_{i}$ attaining its minimum $\kappa^{min}_i$ at $V^b_i\!=\!V_{c,i}$.
For $V^b_i \!<\! V_{c,i}$, the bubble is small and spherical. For $V^b_i \!\geq\! V_{c,i}$, it is large and deformed by the pore's confinement. The parameters $a$ and $b$ control this deformation and depend on pore shape (here $a=1$ and $b=0.01$).
Following \cite{bueno2023AWR}, we refer to $V_{c,i}$ as the \textit{critical volume}, bubbles with $V^b_i\!<\!V_{c,i}$ as \textit{sub-critical}, and those with $V^b_i\!>\!V_{c,i}$ as \textit{super-critical}.

The species balance Eq.\ref{eq:fick} combined with the above constitutive relations yields a system of \textit{nonlinear} ODEs, which we solve using a modified Newton scheme with adaptive time stepping \cite{mehmani2022JCP}.
The modification avoids instabilities and divergence of Newton iterations near bubble disappearance events or when $V^b_i$ crosses the critical volume $V_{c,i}$, where the derivative $d\kappa_i/dV^b_i$ is discontinuous.
We also refer the reader to \cite{bueno2023AWR} for a lower bound on $V^b_i$, below which bubbles must be marked as ``vanished'' to prevent numerical instabilities.

\section{Theory of two-component Ostwald ripening in porous media}\label{sec:theory}
\subsection{Population balance formulation} \label{sec:theory_overview}

We formulate a kinetic theory that predicts the temporal evolution of the \textit{statistics} of a bubble population driven by two-component Ostwald ripening. The PNM of Section \ref{sec:PNM} models this evolution \textit{spatially}, meaning each bubble is tracked individually and assigned a spatial position, while interacting with neighboring bubbles at other positions. In our theory, we do not track individual bubbles, but groups of bubbles sharing the same statistical \textit{state}, $s$. We let this state $s \!=\! (R_p, S^b, y_1)$ consist of three coordinates in a three-dimensional statistical space, referred to as the \textit{phase space}. The first coordinate is the pore size $R_p$ that bubbles at state $s$ occupy. The second is bubble saturation $S^b \!=\! V^b/V_{p}$, which denotes the volume fraction of the host pore each bubble in the group occupies. The third coordinate is the mole fraction of H$_2$ for each bubble in the group, denoted by $y_1$. Since bubbles are binary mixtures, the mole fraction of CO$_2$ follows from $y_2 = 1 - y_1$.

\begin{figure}[t!]
	\centering
	\includegraphics[width=0.8\textwidth]{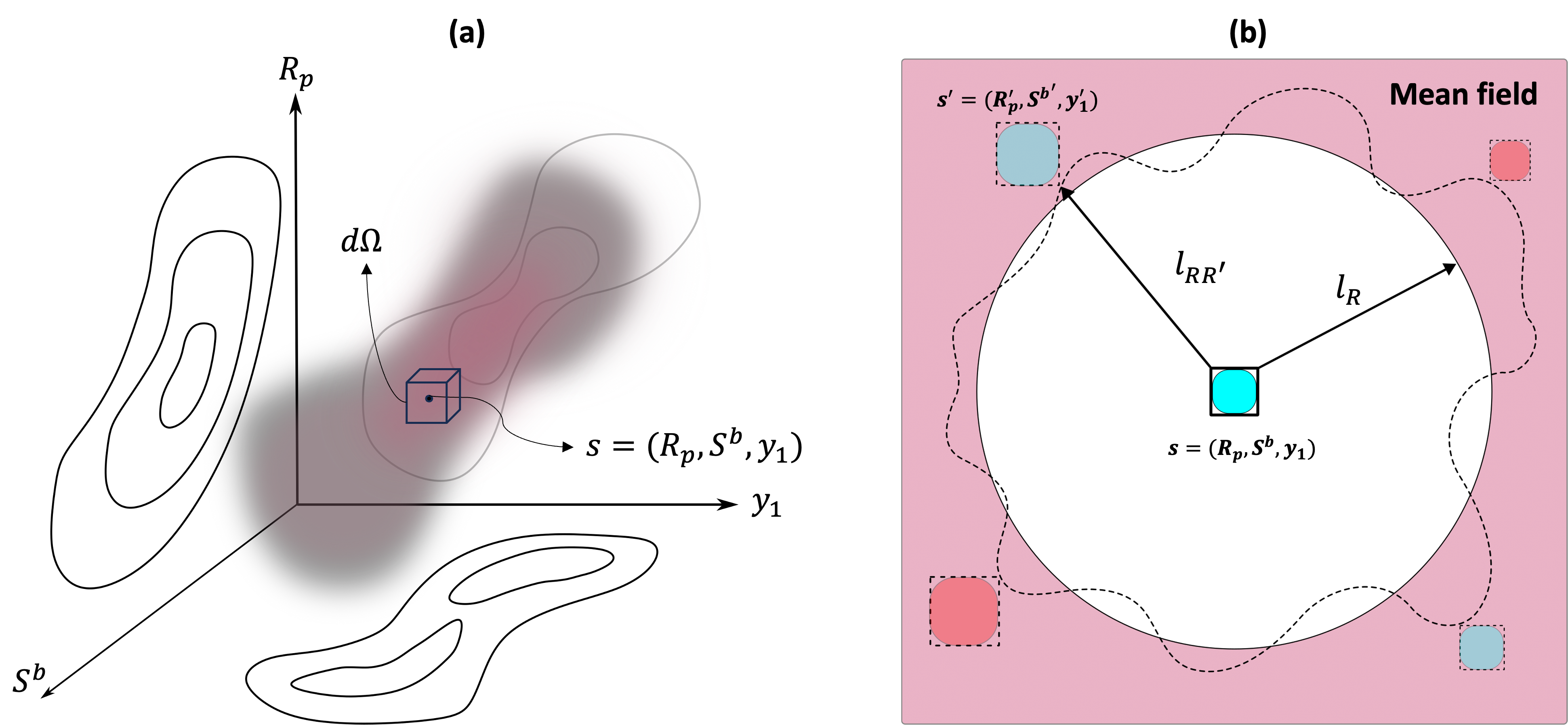}
	\caption{(a) Schematic of 3D \textit{phase space} used to describe bubble statistics by the number-density function $g(s;t)$. Each point in this space is a bubble state $s$ with three coordinates $R_p$, $S^b$, and $y_1$. The number of bubbles within an infinitesimal volume $d\Omega$ centered at $s$ equals $g(s;t)\,d\Omega$. The gray cloud denotes $g(s;t)$ and the contours its 2D marginals projected onto the coordinate planes. (b) Schematic of a bubble at state $s$ interacting with the rest of the population, conceptualized as a mean field.}
	\label{fig:PhaseMean}
\end{figure}

\begin{figure}[h]
	\centering
	\includegraphics[width=\textwidth]{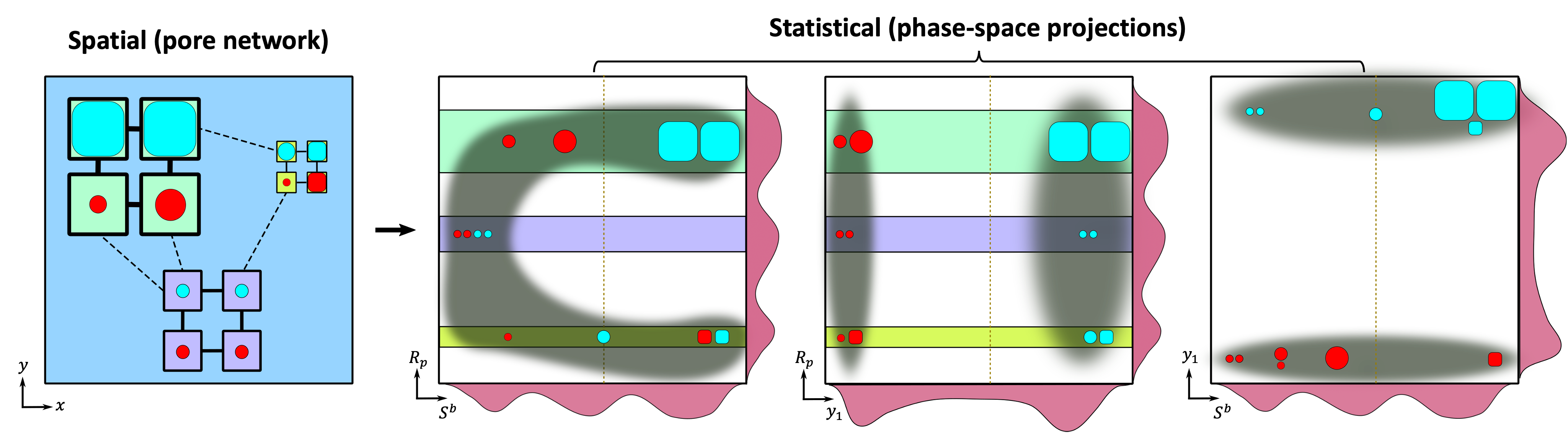}
	\caption{(Left panel) Schematic of spatial arrangement of bubbles in a pore network. H$_2$ bubbles are colored in cyan and CO$_2$ bubbles in red. (Three rightmost panels) Translation of spatial information in the pore network into statistical information in the theory. The three panels show projections of the number-density function $g(s;t)$ onto the 2D coordinate planes of the 3D phase space (Fig.\ref{fig:PhaseMean}). Bubbles from the first panel are annotated on these phase-space plots to emphasize which part of $g(s;t)$ they represent. Shadows represent 2D marginals of $g(s;t)$, and pink curves along the right/bottom axes the 1D marginals.}
	\label{fig:PhaseExample}
\end{figure}

To quantify the number of bubbles in each state $s$ at any given time $t$, we introduce the number-density function (NDF) $g(s;t)$. Concretely, the number of bubbles in an infinitesimal volume of phase space $d\Omega \!=\! dR_p \, dS^b \,dy_1$ centered at $s$ equals $g(s;t)\,d\Omega$. Figs.\ref{fig:PhaseMean}-\ref{fig:PhaseExample} illustrate this conceptualization.
Fig.\ref{fig:PhaseMean}a depicts the 3D phase space with $g(s;t)$ visualized as a cloud, along with its projections (or marginals) onto the coordinate planes.
Fig.\ref{fig:PhaseExample} provides a simple example of how spatial information in the PNM is translated into statistical information in the theory. The left panel is a schematic of pure H$_2$ (cyan; $y_1\!=\!1$) and CO$_2$ (red; $y_1\!=\!0$) bubbles residing in pores of different sizes ($R_p$) with different saturations ($S^b$). The remaining three panels visualize 2D projections of the corresponding $g(s;t)$ onto the coordinate planes. Bubbles from the spatial picture are overlain on the phase-space plots to highlight which part of $g(s;t)$ they represent. Pure-species bubbles ($y_1\!=\!0$ or $1$) are used for simplicity, as bubbles are generally mixtures with arbitrary mole fraction ($y_1\!\in\![0,1]$).
The shadows on each plot represent 2D marginals of $g(s;t)$, and the pink curves along the axes represent 1D marginals. The goal of our theory is to predict the temporal evolution of $g(s;t)$.


We describe the evolution of $g(s;t)$ in time with the following population balance equation:
\begin{equation}\label{eq:populationBalance}
	\frac{\partial g}{\partial t} + \nabla \cdot (g\, \boldsymbol{u}) = 0 \,\,\,\,\,\, \Rightarrow \,\,\,\,\,\, \frac{\partial g}{\partial t} +  \frac{\partial}{\partial S^b} (g\, u_{S^b})+\frac{\partial}{\partial y_1} (g\, u_{y_1}) + \frac{\partial}{\partial R_p} (g\, u_{R_p})= 0
\end{equation}
where $\boldsymbol{u}  \!=\! (u_{R_p}, u_{S^b},u_{y_1}) \!=\! (\frac{dR_p}{dt}, \frac{dS^b}{dt}, \frac{dy_1}{dt})$ is the velocity that advects $g(s;t)$ through phase space. This velocity is determined purely by the physics of Ostwald ripening for a population of gaseous bubbles composed of two components (here H$_2$ and CO$_2$). We assume $u_{R_p} \!=\! 0$, implying that bubbles cannot change pore sizes by, for example, mobilizing from one pore to the next due to background flow or capillary action.\footnote{One consequence of $u_{R_p}\!=\!0$ is that Eq.\ref{eq:populationBalance} reduces to a series of 2D PDEs in the variables $y_1$ and $S^b$ for each fixed $R_p$.} This is consistent with our assumption in Section \ref{sec:problemDesc} that each bubble occupies only one pore and the wetting phase is stagnant (i.e., no flow). Computing $\boldsymbol{u} \!=\! (0, u_{S^b},u_{y_1})$ \textit{closes} the theory, which we address in Section \ref{sec:theory_closure}.

Given some initial condition (IC), $g(s;t=0)$, solving Eq.\ref{eq:populationBalance} requires boundary conditions (BCs). Since we have assumed $u_{R_p}\!=\!0$, no BCs are needed at $R_p\!=\!0$ or $R_p\!=\!R_p^{max}$, where $R_p^{max}$ is the maximum pore size in the network. At $y_1\!=\!0$ and $y_1\!=\!1$, the normal velocity $u_{y_1}$ always points toward the interior of the phase space, preventing bubbles from ever crossing either boundary. Otherwise, mole fractions would fall outside the valid interval $[0,1]$. To also prevent new bubbles from being introduced (or ``injected''), we set $g\!=\!0$, or equivalently $u_{y_1}\!=\!0$, at $y_1\!=\!0$ and $y_1\!=\!1$. Following similar arguments, we set $g\!=\!0$, or equivalently $u_{S^b}\!=\!0$, at $S^b\!=\!1$. The only tricky boundary is $S^b\!=\!0$, because shrinking bubbles during ripening can indeed vanish and leave the phase space, in which case $u_{S^b}\!<\!0$ holds. Since Eq.\ref{eq:populationBalance} is first-order hyperbolic, no BC is needed at such \textit{outflow} portions of the boundary---numerically, the upwind value of $g$ gets used. If, however, $u_{S^b}\!>\!0$ holds, then we set $g\!=\!0$ at such \textit{inflow} portions to prevent introducing new bubbles into the population.

\subsection{Closure by mean-field approximation} \label{sec:theory_closure}
To determine the phase-space velocity $\boldsymbol{u}$, we must compute $dS^b/dt$ and $dy_1/dt$. We adopt a \textit{mean-field approximation} that reduces the problem of a bubble interacting with multiple neighboring bubbles to the simpler problem of interaction with a single entity: the mean field. Consider a bubble at state $s\!=\!(R_p,S^b,y_1)$, as shown in Fig.\ref{fig:PhaseMean}b, that exchanges mass with surrounding bubbles. On average, its nearest neighbors lie a distance $l_s$ away. We therefore idealize the region $r\!<\!l_s$ as bubble-free, which is a disk/sphere of radius $r\!=\!l_s$ in 2D/3D. All bubbles in $r\!>\!l_s$ define the mean field, to which average properties are assigned, such as dissolved concentration $x^{mf}_{\alpha,s}$, bubble pressure $(P^b)_s^{mf}$, and gas composition $y_{\alpha,s}^{mf}$. Mean-field properties are computed from $g(s;t)$, as shown later. Since bubbles can vanish during ripening, $l_s$ grows with time.
For tractability, we assume the bubble-spacing parameter $l_s$ depends primarily on the pore-size coordinate $R_p$, with $S^b$ and $y_1$ playing secondary roles. This is reasonable because spatial correlations in $R_p$ dictate where bubbles are initially trapped in the network (e.g., by cyclic injections of H$_2$) and how curvature-driven mass transfer evolves. We therefore write $l_R$ hereafter to make this dependence explicit (Fig.\ref{fig:PhaseMean}). Note the mean-field approximation requires a sufficiently large bubble population for ensemble statistics to have converged.

We now derive expressions for $dS_s^b/dt$ and $dy_{1,s}/dt$. To compute mass exchange between the $s$-bubble and the mean field (Fig.\ref{fig:PhaseMean}b), we idealize transport in the wetting phase as radial diffusion toward a spherical (3D) or cylindrical (2D) surface at $r\!=\!l_0$, representing the bubble's interface (specified later). The annular region from $r\!=\!l_0$ to $l_R$ is bubble-free, with mass transfer governed by Fick's second law in radial coordinates:
\begin{equation}\label{eq:x_field}
	0 = \nabla \cdot (D^e_\alpha \rho_w \nabla x_\alpha) =
	\frac{1}{r^{d-1}} \frac{d}{d r} (r^{d-1} D^e_\alpha \rho_w  \frac{d x_\alpha}{d r}) \quad \forall r \in (l_0, l_R)
\end{equation}
Eq.\ref{eq:x_field} assumes steady-state, radially symmetric transport, implying the wetting phase acts merely as a solute transport medium, not storage. Here, $d$ is the network dimension ($2$ or $3$), and $D^e_\alpha \!=\! D_{\alpha} \phi_a / \tau$ the restricted diffusion coefficient, which depends on molecular diffusivity $D_{\alpha}$, tortuosity $\tau$, and areal porosity $\phi_a$. \ref{app:app_porosity} shows $\phi_a \!=\! C A_t/l_t$, with $C\!=\!z/(2h)$ in 2D and $C\!=\!z/(2 l_t)$ in 3D. Here, $z$ is the mean coordination number (neighbor count) of pores, $h$ the network's out-of-plane thickness in 2D, and $A_t$ and $l_t$ are the mean cross-sectional area and length of throats. Solving Eq.\ref{eq:x_field} with BCs $x_\alpha(l_0) \!=\! x_{\alpha,s}$ and $x_\alpha(l_R) \!=\! x^{mf}_{\alpha,s}$ yields:
\begin{equation}\label{eq:xalpha}
	x_{\alpha} (r) = x_{\alpha,s} + (x^{mf}_{\alpha,s}-x_{\alpha,s}) \cdot
	\begin{cases} 
	\frac{\ln(r/l_0)}{\ln(l_R/l_0)} & \text{in 2D} \\
	\frac{1/r-1/l_0}{1/l_R-1/l_0} &  \text{in 3D}
	\end{cases}
\end{equation}
Recall $x_{\alpha,s}$ is the dissolved concentration at the interface of the $s$-bubble.

Given Eq.\ref{eq:xalpha}, we write a mole balance for the bubble at state $s$:
\begin{equation}\label{eq:fluxarea}
	\frac{dn_{\alpha,s}}{dt} = - jA =  \frac{D_{\alpha} \phi_a}{\tau} \rho_w  \frac{dx_{\alpha}(r)}{dr} \cdot
	\begin{cases} 
		2\pi r h & \text{in 2D} \\
		4 \pi r^2 &  \text{in 3D}
	\end{cases} 
\end{equation}
where $j$ is the molar flux of dissolved solute toward/away from the $s$-bubble, and $A$ the surface area of the cylinder/sphere through which transport occurs. Note the product $jA$ does \textit{not} depend on $r$. Differentiating Eq.\ref{eq:xalpha} with respect to $r$ and substituting into Eq.\ref{eq:fluxarea}, along with the expression for $\phi_a$ above, yields:
\begin{equation}\label{eq:flux}
	\frac{dn_{\alpha,s}}{dt} = D_\alpha  \rho_w \frac{A_t}{l_t} (x^{mf}_{\alpha,s} - x_{\alpha,s})
	\frac{C}{\tau} \cdot
	\begin{cases} 
		\frac{2\pi}{\ln(l_R/l_0)} & \text{in 2D} \\
		\frac{4 \pi}{l_t/l_0-l_t/l_R} &  \text{in 3D}
	\end{cases}
\end{equation}

We next relate $dn_{\alpha,s}/dt$ to $dS^b_s/dt$ and $dy_{1,s}/dt$.
Following Section \ref{sec:PNM}, the number of moles in the $s$-bubble is $n_{\alpha,s} = y_{\alpha,s} \rho^b_{s} S^b_s V_{p,s}$, where all variables retain their original meaning, except the pore index $i$ is now replaced with the bubble state coordinate $s$. Invoking the same constitutive equations in Section \ref{sec:PNM}, it is straightforward to see $n_{\alpha,s}$ is fully determined by $s\!=\!(R_p,S^b,y_1)$. Recall these constitutive equations include the ideal-gas law $\rho^b_s RT \!=\! P^b_s$, the Young-Laplace equation $P^b_s \!=\! P_w + \sigma \kappa_s$ (with $P_w$ assumed constant), and the curvature-volume relation $\kappa_s(V^b_s, V_{p,s})$ in Eq.\ref{eq:kappa_vb} for the pore size $R_p$ occupied by the $s$-bubble. As a result, the time derivative of $n_{\alpha,s}\!=\!n_{\alpha,s}(R_p,S^b,y_1)$ can be expanded using the chain rule:
\begin{subequations} \label{eq:chain}
	\begin{equation} 
		\frac{dn_{\alpha,s}}{dt}  =  \cancel{(\dfrac{\partial n_{\alpha,s}}{\partial R_{p,s}}) \dfrac{dR_{p,s}}{dt}} + 
		                             (\frac{\partial n_{\alpha,s}}{\partial S^b_s}) \frac{dS_s^b }{dt} + 
		                             (\dfrac{\partial n_{\alpha,s}}{\partial y_{\alpha,s}}) \dfrac{dy_{\alpha,s}}{dt}
		                             \label{eq:chain_1}
	\end{equation}
	\begin{equation}
		\frac{\partial n_{\alpha,s} }{\partial S^b} =
		            y_{\alpha,s} V_{p,s} (\rho^b_s + 
		            \frac{ S^b_{s}}{RT}  \sigma \frac{d\kappa_{s} }{dS^b_s} ), 
		            \quad \frac{\partial n_{\alpha,s} }{\partial y_\alpha} = V_{p,s} \rho^b_s S^b_{s}
		\label{eq:chain_2}
	\end{equation}
\end{subequations}
where the first term in Eq.\ref{eq:chain_1} is zero since $dR_{p,s}/dt\!=\!0$ was assumed in Section \ref{sec:theory_overview}. Eq.\ref{eq:chain_2} follows from substituting the above constitutive relations into the remaining two terms in Eq.\ref{eq:chain_1} and simplifying.

We now substitute Eq.\ref{eq:chain} into the left-hand side (LHS) of Eq.\ref{eq:flux}, which results in two equations corresponding to $\alpha\!\in\!\{1,2\}$. To express this 2$\times$2 ODE system in terms of only $S^b_s$ and $y_{1,s}$ as our primary variables, we use Henry's law $x_{\alpha,s} \!=\! y_{\alpha,s} P^b_s/H_\alpha$ to replace $x_{\alpha,s}$ on the right-hand side (RHS) of Eq.\ref{eq:flux}. This yields:
\begin{subequations} \label{eq:matrix}
\begin{equation}
\begin{bmatrix}
	V_{p,s} \rho^b_s S^b_{s} &  y_{1,s} V_{p,s} (\rho^b_s + \dfrac{ S^b_{s}}{RT}  \sigma \dfrac{d\kappa_{s} }{dS^b_s} )\\[10pt]
	V_{p,s} \rho^b_s S^b_{s} &  -y_{2,s} V_{p,s} (\rho^b_s + \dfrac{ S^b_{s}}{RT}  \sigma \dfrac{d\kappa_{s} }{dS^b_s} )
\end{bmatrix}
\begin{bmatrix}
	\dfrac{dS^b}{dt}\\[8pt]
	\dfrac{dy_1}{dt}
\end{bmatrix} =  \rho_w \frac{A_t}{l_t}  G_s
\begin{bmatrix}
	\frac{D_1}{H_1} ((y_1 P^b)^{mf}_{s} - (y_1 P^b)_{s})\\[10pt]
	\frac{D_2}{H_2} ((y_2 P^b)^{mf}_{s} - (y_2 P^b)_{s})
\end{bmatrix} \label{eq:matrix_1}
\end{equation}
where
\begin{equation}
	G_s =  \frac{C}{\tau} \cdot
	\begin{cases}
		\frac{2\pi}{\ln(l_R/l_0)} & \text{in 2D} \\
		\frac{4\pi}{l_t/l_0-l_t/l_R} &  \text{in 3D}
	\end{cases} \label{eq:matrix_2}
\end{equation}
\end{subequations}
In Eq.\ref{eq:matrix_1}, we have used $dy_2/dt\!=\!-dy_1/dt$, which holds for binary mixtures.

Right-multiplying both sides of Eq.\ref{eq:matrix_1} by the inverse of the mass matrix on the LHS, we obtain:
\begin{subequations} \label{eq:closure_raw}
	\begin{equation}
			\frac{d S^b_{s}}{dt} = 
			\frac{1}{V_{p,s}} \left(\rho^b_s + \frac{ S^b_{s}}{RT}\sigma \frac{d\kappa_{s} }{dS^b_s}\right)^{-1}
			\rho_w G_s \sum^2_{\alpha=1}  \frac{D_{\alpha}}{H_\alpha} 
			\left((y_\alpha P^b)^{mf}_{s} - (y_\alpha P^b)_{s}\right) \label{eq:closure_raw_1}
	\end{equation}
	\begin{equation}
		\frac{d y_{1,s}}{dt} = 
		\frac{1}{V_{p,s}} \left(\rho^b_s S^b_{s}\right)^{-1} \rho_w G_s 
		\sum^2_{\alpha=1} (\delta_{1\alpha}- y_{1,s} ) \frac{D_{\alpha}}{H_\alpha} 
		\left((y_\alpha P^b)^{mf}_{s} - (y_\alpha P^b)_{s}\right) \label{eq:closure_raw_2}
	\end{equation}
\end{subequations}
which are explicit expressions for $dS^b_s/dt$ and $dy_{1,s}/dt$ as intended. The Kronecker delta $\delta_{1\alpha}$ assumes a value of one for $\alpha\!=\!1$ and zero for $\alpha\!=\!2$. We note $(y_\alpha P^b)^{mf}_s\!\neq\!y_{\alpha,s}^{mf} (P^b)^{mf}_s$ and that $(y_\alpha P^b)^{mf}_s\!=\!x^{mf}_{\alpha,s}/H_\alpha$ holds by Henry's law. The latter implies that the solution of Eq.\ref{eq:closure_raw} at large $t$ asymptotes as $x_{\alpha,s}\!\rightarrow\!x^{mf}_{\alpha,s}$ for all $\alpha$ and $s$.
Unfortunately, the mathematical form of Eq.\ref{eq:closure_raw} is numerically problematic when the population balance Eq.\ref{eq:populationBalance} is solved with an Eulerian (grid-based) method, as done here with finite volumes: the curves $dS^b_s/dt\!=\!0$ and $dy_{1,s}/dt\!=\!0$ form a narrowing corridor in phase space that traps bubbles before they reach equilibrium (see \ref{app:closure_reform} for details).
An additional problem is that each state $s$ interacts with a \textit{different} mean-field $x^{mf}_{\alpha,s}$, and there is nothing in the expressions provided later that forces these mean-fields to converge to the same state at $t\!\rightarrow\!\infty$. To address these challenges, we alter the driving force in Eq.\ref{eq:closure_raw_2} as follows:
\begin{subequations} \label{eq:closure_final}
\begin{align}
	&\frac{d S^b_{s}}{dt} = 
	\frac{1}{V_{p,s}} \left(\rho^b_s + \frac{S^b_{s}}{RT} \sigma\frac{d\kappa_{s}}{dS^b_s} \right)^{-1} 
	\rho_w G_s \sum^2_{\alpha=1} \frac{D_{\alpha}}{H_\alpha} 
	                  \left((y_{\alpha} P^b)^{mf}_{s} - y_{\alpha,s}  P^b_s\right) 
	                  \label{eq:closure_final_1} \\
	&\frac{d y_{1,s}}{dt} = 
	      \frac{1}{V_{p,s}} \left(\rho^b_s S^b_{s}\right)^{-1} P_w \rho_w G_s 
	      \left(y_2 \frac{D_1}{H_1}+y_1 \frac{D_2}{H_2}\right) (y^{mf}_{1} - y_{1,s})
	      \label{eq:closure_final_2}
\end{align}
\end{subequations}

Eq.\ref{eq:closure_final} constitutes our final expressions for $dS^b_s/dt$ and $dy_{1,s}/dt$. Notice the RHS of Eq.\ref{eq:closure_final_2} is now driven by the difference between the $s$-bubble composition, $y_{1,s}$, and the mean field, $y^{mf}_1$. Crucially, $y^{mf}_1$ does not depend on $s$, ensuring all bubbles approach the \textit{same} composition at $t\!\rightarrow\!\infty$. We prove in Section \ref{sec:massConsv} that because our theory conserves mole exchange between bubbles,  $y^{mf}_1\!\rightarrow\!y^{mix}_1$ at large $t$, where $y^{mix}_1$ is the bulk (or mixture) composition of \textit{all} bubbles in the population. Therefore, bubbles are guaranteed to converge to a physically meaningful equilibrium composition.
Notice Eq.\ref{eq:closure_final} drives $x_{\alpha,s}$ and $y_{1,s}$ towards $x^{mf}_{\alpha,s}$ and $y^{mf}_{1}$ for all $s$.
To derive Eq.\ref{eq:closure_final_2}, we have assumed $P_w\!\gg\!\sigma\kappa_s$ for all $s$, which is valid for high-pressure subsurface environments (e.g., UHS). This allows setting $P^b\!\approx\!P_w$ and pulling it out of the summation in Eq.\ref{eq:closure_raw_2}. We caution that applying the same simplification to Eq.\ref{eq:closure_raw_1} as well would be fatal, as the ODE system would then be driven solely by differences in bubble composition (i.e., $y^{mf}_{1} - y_{1,s}$), not curvature.

We close this section by returning to the inner radius $l_0$ in Eq.\ref{eq:xalpha} and deriving an expression for it. Notice the smallest $l_R$ can be is a single throat length $l_t$, corresponding to when all neighboring pores of the $s$-bubble are occupied by bubbles. Since there are $z$ such neighbors, we can write analogously to Eq.\ref{eq:fick} in the PNM:
\begin{equation} \label{eq:flux_consist}
	\frac{dn_{\alpha,s}}{dt}
	   = D_\alpha \rho_w \frac{z A_t}{l_t} (x_{\alpha}(l_t) - x_{\alpha,s})
	   = D_\alpha \rho_w \frac{z A_t}{l_t} (x^{mf}_{\alpha,s}-x_{\alpha,s})
\end{equation}
The first equality follows from Eq.\ref{eq:fick}, where by radial symmetry the dissolved concentration within all neighboring pores is $x_{\alpha}(l_t)$. The second equality follows from substituting for $x_{\alpha}(l_t)$ the expression from Eq.\ref{eq:xalpha} with $r\!=\!l_R\!=\!l_t$. Comparing Eq.\ref{eq:flux_consist} to Eq.\ref{eq:flux}, we arrive at the following formula for $l_0$:
\begin{equation} \label{eq:l0}
	l_0 = l_t \cdot
	\begin{cases} 
			e^{-\frac{2\pi C}{z \tau}} & \text{in 2D} \\
			(1+\frac{4\pi C}{z \tau})^{-1}&  \text{in 3D}
		\end{cases}
\end{equation}
This, in turn, yields the expression below for the geometric factor $G_s$ from Eq.\ref{eq:matrix_2}:
\begin{equation} \label{eq:Gs_final}
	G^{-1}_s = \frac{1}{z} + \frac{\tau}{4\pi C} \cdot
	\begin{cases} 
			 2 \ln{\frac{l_R}{l_t}} & \text{in 2D} \\
			 ( 1-\frac{l_t}{l_R})&  \text{in 3D}
		\end{cases}
\end{equation}
Note $G_s\!=\!z$ when $l_R\!=\!l_t$, which holds for a densely bubble-occupied network, and it decreases with $l_R$. In summary, the closure equations from this section are Eqs.\ref{eq:closure_final} and \ref{eq:Gs_final}. To complete the theory, we now need expressions that allow us to compute $(y_{\alpha} P^b)^{mf}_s$ and $y^{mf}_{1}$ at any given time $t$. This is addressed next.

\subsection{Computing mean-field properties} \label{sec:meanField}
In this section, we postulate explicit expressions for $(y_{\alpha} P^b)^{mf}_s$ and $y^{mf}_{1}$. We begin with $(y_{\alpha} P^b)^{mf}_s$ and propose the following, which accounts for potential spatial correlations in pore size $R_p$ in the network:
\begin{equation} \label{eq:mf_exact}
(y_\alpha P^b)^{mf}_s
= l_s^{-1} \int_{s',l} (y_{\alpha} P^b)_{s'} \frac{p(s',l |s) }{l} \,dl\, ds'
\,,\quad\quad
l^{-1}_s = \int_{s',l} \frac{p(s',l|s) }{l} \,dl\, ds'
\end{equation}
The $p(s',l|s)$ is the conditional probability density function (PDF) of finding a bubble at state $s'$ a distance $l$ from the bubble at state $s$ (see Fig.\ref{fig:PhaseMean}b). Eq.\ref{eq:mf_exact} is essentially a weighted average of $(y_\alpha P^b)_{s'}$, where the weights $p(s',l|s)\,/\,l$ decline with distance from the $s$-bubble according to spatial correlations in bubble states.

To compute $p(s',l|s)$, we first apply Bayes' theorem:
\begin{equation} \label{eq:bayes}
	p(s',l |s) = p(l | s',s) \,p(s' | s)
\end{equation}
and set $p(l | s',s) \!=\! \delta(l-l_{ss'})$, where $\delta(\cdot)$ is the Dirac delta function. This expression for $p(l | s',s)$ is consistent with our theory's conceptualization in Fig.\ref{fig:PhaseMean}b, meaning that once $s$ and $s'$ are chosen, the interaction distance between them is fixed at $l_{ss'}$. Recall the space between bubbles $s$ and $s'$ is free of other bubbles. To estimate $p(s' | s)$ and $l_{ss'}$, we assume much of the spatial correlation in bubble states is dictated pore size (i.e., the $R_p$ coordinate of $s$). This justifies renaming $l_{ss'}$ to $l_{RR'}$ (Fig.\ref{fig:PhaseExample}b) and approximating $p(s' | s)$ as follows:
\begin{equation} \label{eq:mf_assume}
	\begin{split}
	p(s'|s)
	\approx  p(s'| R_p )
	= p({S^b}', y'_1| R'_p, R_p)\,
	p(R'_p| R_p ) 
	\approx p({S^b}', y'_1| R'_p)\,
	p(R'_p |  R_p )
	\end{split}
\end{equation}
where the second equality follows from expanding $s'$ in terms of its coordinates and applying Bayes' theorem. 

The first PDF on the RHS of Eq.\ref{eq:mf_assume} can be computed exactly from the NDF $g(s';t)$ as follows:
\begin{equation} \label{eq:p_ys}
	p({S^b}', y'_1|R'_p)
	= {g(\underbrace{R'_p,{S^b}',y'_1}_{s'})}\,/
	       {\int_{{S^b}',y'_1} g(R'_p,{S^b}',y'_1)\, dy'_1\,d{S^b}'}
\end{equation}
where $t$ is suppressed for brevity. The second PDF on the RHS of Eq.\ref{eq:mf_assume} is approximated as:
\begin{equation} \label{eq:p_rr}
	p(R'_p | R_p)
		\simeq {f(R'_p | R_p; l_{RR'})\, F_{R'}}\, / 
		{\int_{R'_p} f(R'_p | R_p; l_{RR'})\, F_{R'}\, dR'_p}
\end{equation}
where $f(R'_p, R_p; l)$ is the joint PDF of pore sizes $R'_p$ and $R_p$ separated by a distance $l$. This PDF describes the pore network itself (not bubble occupancy) and is assumed to be known; it can be computed from X-ray images of porous samples.
The smaller the distance $l$, the larger the value of $f(R'_p, R_p; l)$.
The marginal $f(R_p)\!=\!\int f(R'_p, R_p; l)\, dR'_p$ is the pore-size distribution. For uncorrelated networks, $f(R'_p, R_p; l)\!=\!f(R'_p)f(R_p)$ for all $l$.
The factor $F_{R'}$ represents the fraction of bubble-occupied pores in the network that have sizes between $R_p'$ and $R_p' + \delta R'_p$. Its multiplication by $f(R'_p | R_p; l_{RR'})$ accounts for bubble occupancy, since $p(R'_p | R_p)$ represents the probability of two \textit{bubble-occupied} pores interacting (not just any two pores). The denominator in Eq.\ref{eq:p_rr} ensures $p(R'_p|R_p)$ integrates to one.
$F_{R'}$ can be computed exactly from $g(s';t)$ and $f(R'_p)$ as follows:
\begin{equation} \label{eq:Frr}
	F_{R'} =  \frac{n^b_{R'}}{n^p_{R'}}
	\,,\quad
	n^b_{R'} = 
	\delta R'_p \int_{{S^b}',y'_1} 
	g(R'_p, {S^b}', y'_1;t)
	\, dy'_1\,d{S^b}'
	\,,\quad
	n^p_{R'} = n_p f(R'_p)\,\delta R'_p
\end{equation}
where $n_p$ denotes the total number of pores in the network.

Substituting Eqs.\ref{eq:bayes}-\ref{eq:mf_assume} into Eq.\ref{eq:mf_exact}, and relabeling the variables $l_{ss'}$ and $l_s$ as $l_{RR'}$ and $l_R$ to explicitly signal our assumption that they depend solely on correlations in $R_p$ (not $S^b$ or $y_1$), we obtain:
\begin{equation} \label{eq:mf_approx}
	(y_\alpha P^b)^{mf}_s
	= l_R^{-1} \int_{R'_p,{S^b}',y'_1} 
	  (y_{\alpha} P^b)_{s'} \frac{p({S^b}', y'_1| R'_p)\,
	  p(R'_p |  R_p ) }{l_{RR'}} \,dy_1' d{S^b}' dR_p'
	\,,\quad
	l^{-1}_R = \int_{R'_p} \frac{p(R'|R; l_{RR'}) }{l_{RR'}}  dR_p' 
\end{equation}
as the final expression for $(y_\alpha P^b)^{mf}_s$, with $p({S^b}', y'_1|R'_p)$ and $p(R'_p | R_p)$ evaluated from Eqs.\ref{eq:p_ys}-\ref{eq:Frr}.

The bubble spacing $l_{RR'}$ in Eq.\ref{eq:mf_approx} grows as bubbles dissolve during ripening. To compute it, we propose:
\begin{equation} \label{eq:bubbleSpacing}
		l_{RR'} = \frac{d_{RR'}}{F_{R'}^{1/d}}
\end{equation}
where $d_{RR'}$ is the average distance of pores with size $R_p$ from their \textit{nearest} pore with size $R'_p$. Thus, $d_{RR'}$ is a property of the network itself (not bubble occupancy). For most networks, $d_{RR'}\!\approx\!l_t$ is a good approximation, but more accurate values can be computed from an X-ray image. Recall $d$ is the network dimension and $F_{R'}$ is given by Eq.\ref{eq:Frr}. While empirical,  Eq.\ref{eq:bubbleSpacing} has a few desirable properties. In homogeneous networks, where $R_p$ assumes a single value ($R'_p\!=\!R_p$), Eq.\ref{eq:bubbleSpacing} reduces to $l_{RR}\!=\!L/n_b^{1/d}$, because $d_{RR}\!=\!l_t$ and $F_R\!=\!n_b/n_p$, where $n_b$ is the total number of bubbles and $L\!=\!l_t n_p^{1/d}$ is the spacing between bubbles if spread uniformly across the network. In heterogeneous networks, when all pores of size $R'_p$ are bubble-occupied, $l_{RR'} \!=\! d_{RR'}$ as expected. As bubbles dissolve due to ripening, $F_{R'}$ decreases from one to zero, causing $l_{RR'}$ to increase. In \ref{app:bubbleSpacing}, we compare $l_{RR'}$ computed from Eq.\ref{eq:bubbleSpacing} with direct PNM simulations and find acceptable agreement in heterogeneous networks with both correlated and uncorrelated pore sizes.\footnote{We make two technical remarks about $l_{RR'}$: (1) Eq.\ref{eq:bubbleSpacing} ignores any dependence on $F_R$, the bubble occupancy of pores with size $R_p$. Hence, it lacks the necessary symmetry to be considered a ``distance metric'' for bubble spacing; (2) For all choices of $R_p$ and $R'_p$, the inequality $l_{RR'}\!\ge\!l_t\!>\! l_0$ holds as can be verified from Eqs.\ref{eq:l0} and \ref{eq:bubbleSpacing} and the fact that $d_{RR'}\!\ge\!l_t$.}

Having formulated $(y_\alpha P^b)^{mf}_s$, we now postulate the following expression for $y^{mf}_{1}$:
\begin{equation} \label{eq:ymf}
	y^{mf}_{1} = 
	{\int_s y_{1,s}\, g(s;t)\, ds}\,\,/
	     {\int_s g(s;t)\, ds}
\end{equation}
which is the average of $y_{1,s}$ over \textit{all} bubbles in the population.
Notice $y^{mf}_{1}$ does not depend on $s$, which forces all bubbles to converge to the same equilibrium composition and avoids the numerical difficulty discussed in Section \ref{sec:theory_closure}.
While Eq.\ref{eq:ymf} does not account for spatial correlations in $R_p$, they are indirectly captured through $(y_\alpha P^b)^{mf}_s$ and the coupling between ODEs in Eq.\ref{eq:closure_final}. The next section addresses the issue of species-mass conservation among bubbles and shows that both $(y_\alpha P^b)^{mf}_s$ and $y^{mf}_{1}$ must be corrected to honor it.

\subsection{Mass conservation} \label{sec:massConsv}
As a bubble population described by $g(s;t)$ evolves due to Ostwald ripening, the number of moles of each species must be conserved. Specifically, moles leaving one bubble must enter another, keeping the total moles of each species constant. Recall the wetting phase in our theory has no storage capacity for dissolved solute. In \ref{ap:proof_thm}, we prove the theorem below, establishing a condition equivalent to mass conservation:

\begin{theorem} \label{thm:massConsrv}
The following two expressions are equivalent, where $\Omega$ is the entire phase space:
\begin{equation} \label{eq:thm}
	\frac{dn_\alpha^{tot}}{dt} = 
	\frac{d}{dt} \int_\Omega n_{\alpha,s}g(s;t)\, ds = 0
	\quad\quad\;
	\Leftrightarrow
	\quad\quad\;
	\int_\Omega \frac{dn_{\alpha,s}}{dt} g(s;t)\, ds = 0
\end{equation}
In words, the left expression states that the total number of moles of species $\alpha$ is constant for all $t$.
\end{theorem}

Setting $\alpha\!=\!1$ and substituting Eq.\ref{eq:chain} for $dn_{1,s}/dt$ into the rightmost expression in Theorem \ref{thm:massConsrv}, with $dS^b_s/dt$ and $dy_{1,s}/dt$ given by Eq.\ref{eq:closure_final}, we obtain:
\begin{equation}\label{eq:constraint}
	\int_\Omega 
	\left( y_{1,s} \rho_w G_s \sum^2_{\beta=1}  
	        \frac{D_{\beta}}{H_\beta} 
	        \left((y_{\beta} P^b)^{mf}_{s} - y_{\beta,s} P^b_s \right) +
      P_w \rho_w G_s 
            (y_2 \frac{D_1}{H_1} + y_1 \frac{D_2}{H_2})
            (y^{mf}_1 - y_{1,s})
     \right)\, g(s;t)
     \,ds = 0
\end{equation}
Unfortunately, the $(y_\alpha P^b)^{mf}_s$ and $y^{mf}_1$ postulated by Eqs.\ref{eq:mf_approx} and \ref{eq:ymf} in the previous section do not guarantee that Eq.\ref{eq:constraint} is satisfied. Therefore, a correction to these mean-field properties is needed to conserve bubble mass. We propose the following multiplicative corrections:
\begin{equation} \label{eq:correct}
	(y_\alpha P^b)^{mf,\ast}_s = \lambda\, (y_\alpha P^b)^{mf}_s
	\,,\quad\qquad
	y^{mf,\ast}_1 = \gamma\, y^{mf}_1
\end{equation}
where the asterisk indicates a corrected mean-field property.

To compute the correction multiplier $\lambda$, we write an equation similar to Eq.\ref{eq:constraint} for $\alpha\!=\!2$, add it to Eq.\ref{eq:constraint}, use the fact that $y_{1,s}+y_{2,s}\!=\!1$ and $y^{mf}_1+y^{mf}_2\!=\!1$, and substitute the corrected mean-field properties in Eq.\ref{eq:correct} into the resulting expression. After some algebra, this yields:
\begin{equation} \label{eq:lambda}
	\lambda  =  
		\int_\Omega G_s \sum^2_{\beta=1} 
		     \frac{D_{\beta}}{H_\beta}\, y_{\beta,s}  P^b_s 
		      \,g(s;t)\, ds\; / 
		\int_\Omega G_s \sum^2_{\beta=1}  
		      \frac{D_{\beta}}{H_\beta} (y_{\beta} P^b)^{mf}_{s}
		      \,g(s;t)\, ds
\end{equation}
Given $\lambda$ from Eq.\ref{eq:lambda}, we can compute the second correction multiplier $\gamma$ by substituting the corrected mean-field properties in Eq.\ref{eq:correct} into Eq.\ref{eq:constraint} and solving for $\gamma$. This yields:
\begin{equation} \label{eq:gamma}
	\begin{split}
	 \gamma = 
     \int_\Omega 
     \left( 
   	   P_w \rho_w G_s 
	       (y_2 \frac{D_1}{H_1} + y_1 \frac{D_2}{H_2})\, y_{1,s} - 
	   y_{1,s} \rho_w G_s \sum^2_{\beta=1}  
	       \frac{D_{\beta}}{H_\beta} 
	       \left(\lambda\,(y_{\beta} P^b)^{mf}_{s} - 
	                   y_{\beta,s} P^b_s \right)
	 \right)\, g(s;t)\, ds
	 \;/\\
	 \int_\Omega
	 \left(
      P_w \rho_w G_s 
            (y_2 \frac{D_1}{H_1} + y_1 \frac{D_2}{H_2})\, y^{mf}_1
     \right)\, g(s;t)
     \,ds
     \end{split}
\end{equation}
To summarize, we first compute the uncorrected mean-field properties $(y_\alpha P^b)^{mf}_s$ and $y^{mf}_1$ via Eqs.\ref{eq:mf_approx} and \ref{eq:ymf}, then correct them via Eqs.\ref{eq:correct}-\ref{eq:gamma}. These corrected mean-field properties are used in Eq.\ref{eq:closure_final} to compute the phase-space velocities $dS^b_s/dt$ and $dy_{1,s}/dt$ needed in the population balance equation Eq.\ref{eq:populationBalance}.

We close this section with a corollary to Theorem \ref{thm:massConsrv}, which guarantees that bubbles converge to a physically meaningful equilibrium composition. The proof is provided in \ref{ap:proof_cor}.

\begin{corollary} \label{cor:ymix}
The bulk (or mixture) composition of all bubbles in the population, defined as:
\begin{equation} \label{eq:ymix}
	y^{mix}_1 = 
	   {\int_s y_{1,s}\, n_s\, g(s;t)\, ds}\;/
	   {\int_s n_s\, g(s;t)\, ds}
	\,,\quad\quad
	n_s = \sum^2_{\alpha=1} n_{\alpha,s}
\end{equation}
remains constant for all $t$. Moreover, as $t\!\rightarrow\!\infty$, the mean-field mole-fraction satisfies $y^{mf,\ast}_1\!\rightarrow y^{mix}_1$.
\end{corollary}

\noindent This result further supports our choice of the $s$-independent $y^{mf}_1$ in Eq.\ref{eq:ymf} and the reformulation of $dy_{1,s}/dt$ in Eq.\ref{eq:closure_final}. By contrast, the precursor ODE in Eq.\ref{eq:closure_raw} does not guarantee convergence to $y^{mix}_1$.
In \ref{ap:proof_cor}, we also show $\gamma\!\rightarrow\!1$ as $t\!\rightarrow\!\infty$, implying $y^{mf}_1\!\rightarrow\!y_1^{mix}$.
Finally, we offer the following remark about $\lambda$:

\begin{remark} \label{rmk:lambda}
While it is obvious that $\gamma$ corrects the mean-field composition $y^{mf}_1$, the use of a single multiplier $\lambda$ for both species $\alpha\!\in\!\{1,2\}$ is equivalent to correcting the mean-field bubble curvature $\kappa^{mf}_s$. To see this, note the mean-field operator $(\cdot)^{mf}_s$ is linear, so by Henry's law and the Young-Laplace equation $H_1 x^{mf}_{1,s} + H_2 x^{mf}_{2,s} = (y_1P^b)^{mf}_s + (y_2P^b)^{mf}_s = P^{b,mf}_s\!=\! P_w + \sigma\kappa^{mf}_s$. This is why a single multiplier $\lambda$ suffices for both species.
\end{remark}

\section{Validation cases} \label{sec:valid}
\begin{figure}[h]
	\centering
	\includegraphics[width=\textwidth]{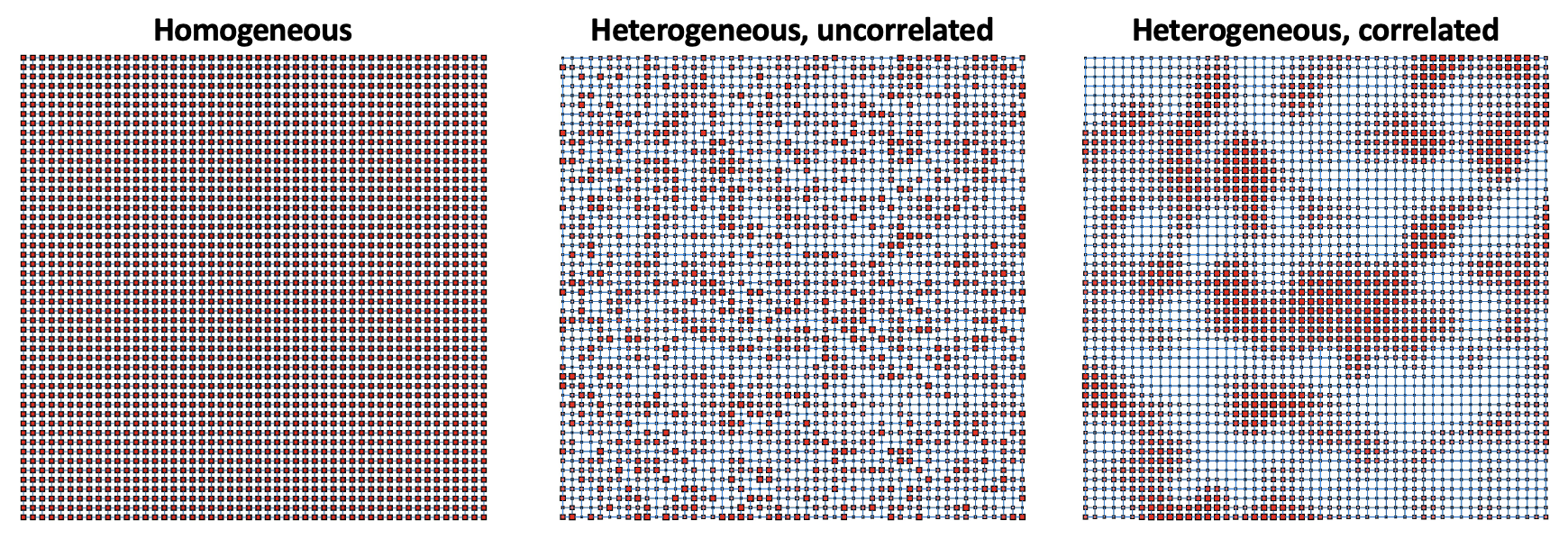}
	\caption{Pore networks used in PNM simulations to validate the proposed theory. Pores are depicted as red circles, with radii proportional to the pores' sizes $R_p$. The heterogeneous uncorrelated network is generated by randomly assigning $R_p\!\in\![\,5,20\,]$ $\mu$m to pores. The heterogeneous correlated network has the same pore sizes but with correlation length $l_c\!=\!3\,l_t$.}
	\label{fig:networks}
\end{figure}

We validate the theory in Section \ref{sec:theory} against PNM simulations for both one- and two-component bubble populations. We consider pore networks with a 2D lattice topology but 3D pore and throat shapes. Pores are semi-cubic of size $R_p$, defined in Section \ref{sec:PNM}, and throats are prisms with cross-sectional area $A_t$ and length $l_t$. Lattices are 50$\times$50 arrays of pores. Three networks are considered: (1) homogeneous pore size (\textit{Hom}), with $R_p\!=\!5$ $\mu$m; (2) heterogeneous and spatially uncorrelated pore size (\textit{Unc}), with $R_p$ uniformly distributed between 5 $\mu$m and 20 $\mu$m; and (3) heterogeneous and spatially correlated pore size (\textit{Cor}), with $R_p$ drawn from the same distribution as \textit{Unc} but with correlation length $3\,l_t$.
Spatial correlation is established by creating a random Gaussian field $Y\!\sim \!\mathcal{N}(\mu, C(l))$ with mean $\mu\!=\!0$ and covariance function $C(l)\!=\!\exp(-(l/l_c)^2)$, where $l$ is the distance between pores. For each pore, $Y$ is mapped to $R_p$ via the \textit{probability integral transform} $R_p\!=\!F^{-1}_{\mathcal{U}}(F_{\mathcal{N}} (Y))$, where $F_{\mathcal{N}}$ is the cumulative distribution function (CDF) of $\mathcal{N}$ and $F_{\mathcal{U}}$ is the CDF of the uniform distribution $R_p\!\in\![\,5,20\,]$ $\mu$m. Throat lengths are $l_t\!=\!10$ $\mu$m in the \textit{Hom} network and $l_t\!=\!40$ $\mu$m in the \textit{Unc} and \textit{Cor} networks. Throat areas are $A_t\!=\!19.6$ $\mu$m$^2$. All networks are depicted in Fig.\ref{fig:networks}.

We validate our theory in two stages. In the first stage, we consider single-component bubbles composed of pure H$_2$. The goal is to demonstrate that our theory not only generalizes the formulation of \cite{bueno2024AWR}, but also removes key limitations that enhance predictive accuracy. In the second stage, we consider two-component bubbles composed of H$_2$ and CO$_2$. Table \ref{tab:sims} lists the cases considered in each stage, which we detail next. We use $S^b_{tot}$ to denote the global bubble saturation in a network, which stands in contrast to local saturation $S^b$ in individual pores. Recall that $y^{mix}_1$ is the global (or bulk) mole fraction of H$_2$ of all bubbles combined.

\begin{table}[h!]
	\centering
	\caption{Validation set used to compare the theory to PNM. Pore networks consist of homogeneous, heterogeneous uncorrelated, and heterogeneous correlated pore sizes. Cases are divided into one- and two-component scenarios, each given a description and a code name for easy reference. The $l_c$, $S^b_{tot}$, and $y^{mix}_1$ denote correlation length, total bubble saturation in the network, and bulk composition of the bubble population, respectively. Pop 1--3, Random $y_1$, and Small-pore $y_1$ are defined in the text.}
	\label{tab:sims}
	\renewcommand{\arraystretch}{1.2}
	\hspace{-13pt}
	\begin{tabular}{@{}|c@{\hspace{6pt}}|l wc{1.2cm} wc{1.2cm} wc{1.4cm} l@{}}
		\cline{2-6}
		\multicolumn{1}{@{}c@{\hspace{6pt}}|}{} & \hspace{2pt}Description & $l_c$ & $S^b_{tot}$ & $y_1^{mix}$ & Code name \\[-0.0ex] 
		\hline
		\multirow{4}{*}{\rotatebox[origin=c]{90}{One component}} 
		& \hspace{2pt}Homogeneous, Fully occupied & $\infty$ & 30\% & 1 & Hom-One-Full \\ 
		& \hspace{2pt}Uncorrelated, Fully occupied & 0 & 30\% & 1 & Unc-One-Full \\ 
		& \hspace{2pt}Correlated, Fully occupied & $3\,l_t$ & 60\% & 1 & Cor-One-Full\\ 
		& \hspace{2pt}Correlated, Partially occupied & $3\,l_t$ & 12\% & 1 & Cor-One-Part\\ 
		\hline
		\multirow{15}{*}{\rotatebox[origin=c]{90}{Two components}} 
		& \hspace{2pt}Homogeneous, Pop 1, Random $y_1$ & $\infty$ & 60\% & 0.50 & Hom-Pop1-yRand \\ 
		& \hspace{2pt}Homogeneous, Pop 2, Random $y_1$ & $\infty$ & 60\% & 0.75 & Hom-Pop2-yRand \\ 
		& \hspace{2pt}Homogeneous, Pop 3, Random $y_1$ & $\infty$ & 40\% & 0.49 & Hom-Pop3-yRand \\ 
		\cline{2-6}
		& \hspace{2pt}Uncorrelated, Pop 1, Random $y_1$ & 0 & 60\% & 0.47 & Unc-Pop1-yRand \\ 
		& \hspace{2pt}Uncorrelated, Pop 2, Random $y_1$ & 0 & 60\% & 0.73 & Unc-Pop2-yRand \\ 
		& \hspace{2pt}Uncorrelated, Pop 3, Random $y_1$ & 0 & 39\% & 0.47 & Unc-Pop3-yRand \\ 
		\cline{2-6}
		& \hspace{2pt}Correlated, Pop 1, Random $y_1$ & $3\,l_t$ & 60\% & 0.49 & Cor-Pop1-yRand \\ 
		& \hspace{2pt}Correlated, Pop 2, Random $y_1$ & $3\,l_t$ & 60\% & 0.74 & Cor-Pop2-yRand \\ 
		& \hspace{2pt}Correlated, Pop 3, Random $y_1$ & $3\,l_t$ & 39\% & 0.46 & Cor-Pop3-yRand \\ 
		\cline{2-6}
		& \hspace{2pt}Uncorrelated, Pop 1, Small-pore $y_1$ & 0 & 60\% & 0.49 & Unc-Pop1-ySmall\\ 
		& \hspace{2pt}Uncorrelated, Pop 2, Small-pore $y_1$ & 0 & 60\% & 0.74 & Unc-Pop2-ySmall\\ 
		& \hspace{2pt}Uncorrelated, Pop 3, Small-pore $y_1$ & 0 & 40\% & 0.48 & Unc-Pop3-ySmall\\ 
		\cline{2-6}
		& \hspace{2pt}Correlated, Pop 1, Small-pore $y_1$ & $3\,l_t$ & 60\% & 0.49 & Cor-Pop1-ySmall\\ 
		& \hspace{2pt}Correlated, Pop 2, Small-pore $y_1$ & $3\,l_t$ & 60\% & 0.74 & Cor-Pop2-ySmall\\ 
		& \hspace{2pt}Correlated, Pop 3, Small-pore $y_1$ & $3\,l_t$ & 40\% & 0.48 & Cor-Pop3-ySmall\\ 
		\hline
	\end{tabular}
\end{table}

In stage 1 of our validation, we consider cases in the top segment of Table \ref{tab:sims}, all taken from \cite{bueno2024AWR}. The first three correspond to pure H$_2$ bubbles ($y_1\!=\!1$) occupying \textit{all} pores in the network at fixed local saturation $S^b$. We label these \textit{Hom-One-Full}, \textit{Unc-One-Full}, and \textit{Cor-One-Full}, depending on whether the network is \textit{Hom}, \textit{Unc}, or \textit{Cor}. In \textit{Unc-One-Full} and \textit{Cor-One-Full}, local saturation $S^b$ is uniform across all pores (thus $S^b_{tot}\!=\!S^b$), whereas in \textit{Hom-One-Full}, $S^b$ is randomly drawn from $[\,1,5\,]\%$. These three cases correspond to the more difficult problems in \cite{bueno2024AWR}, which we use to demonstrate that our theory reproduces those results. The fourth case, \textit{Cor-One-Part}, corresponds to a scenario presented as a challenge in \cite{bueno2024AWR} (Section 7.2), whose ripening dynamics that theory could not capture due to large initial bubble spacing. We show our theory overcomes this limitation, thereby generalizing \cite{bueno2024AWR}. The setup places sub-critical bubbles in the largest 20\% of pore sizes, with local saturation sampled uniformly from $S^b\!\in\![\,5,35\,]\%$, and super-critical bubbles in the smallest 20\%, with $S^b$ sampled uniformly from $[\,65,95\,]\%$. Here too, all bubbles are composed of pure H$_2$.

In stage 2 of the validation, we consider cases in segments 2--6 of Table \ref{tab:sims}. These comprise different combinations of network type (\textit{Hom}, \textit{Unc}, \textit{Cor}), bubble population (\textit{Pop1}, \textit{Pop2}, \textit{Pop3}), and spatial distribution of $y_1$ (\textit{yRand}, \textit{ySmall}). Since network types were described above, we focus here on the three bubble populations: In \textit{Pop1}, all pores contain bubbles at local saturation $S^b\!=\!60\%$, with roughly half by volume being pure H$_2$ and the other half pure CO$_2$. \textit{Pop2} is similar, but the pure CO$_2$ bubbles are replaced by bubbles that are an equimolar H$_2$--CO$_2$ mixture. In \textit{Pop3}, roughly half of the bubbles by volume are pure H$_2$ at $S^b\!=\!60\%$, and the other half are pure CO$_2$ at $S^b\!=\!30\%$. For each population, the labels \textit{yRand} and \textit{ySmall} specify how bubbles are assigned to pores. In \textit{yRand}, pores are chosen at random, implying no correlation between pore size $R_p$ and bubble composition $y_1$. In \textit{ySmall}, the smallest pores are filled with H$_2$ first, then the remaining pores are filled with other bubbles (pure CO$_2$ in \textit{Pop1} and \textit{Pop3}, equimolar mixture in \textit{Pop2}). In the \textit{Cor} network, \textit{ySmall} produces spatial correlation in $y_1$, not just $R_p$. Hereafter, we refer to each of the resulting 15 cases by the code names listed in Table \ref{tab:sims} (e.g., \textit{Cor-Pop2-yRand}).

The above initial conditions for two-component ripening are motivated by underground hydrogen storage. Cyclic injections of H$_2$ are often preceded by a cushion gas like CO$_2$, which can mix with subsequent H$_2$ injections. In addition, cyclic injections can cause capillary trapping of bubbles within preferential flow paths (larger pores), diverting subsequent H$_2$ into smaller pores. This motivates the equimolar mixture in \textit{Pop2} and the \textit{ySmall} placement. We note that placing H$_2$ bubbles in larger pores would be a less challenging test of our theory, since early-stage mass transfer is exclusively from CO$_2$ to H$_2$ due to differences in solubility and diffusion coefficient, causing more bubble deformation when H$_2$ resides in smaller pores.

In all cases, the temperature $T$ is set to 333 K, wetting-phase pressure $P_w$ to 10$^8$ dynes/cm$^2$, surface tension $\sigma$ to 32 dynes/cm, molecular diffusion coefficients to $D_1\!=\!5.13\times10^{-5}$ cm$^2$/s for H$_2$ and $D_2\!=\!1.6\times10^{-5}$ cm$^2$/s for CO$_2$, and Henry's constants to $H_1\!=\!7.1\times 10^{10}$ dynes/cm$^2$ for H$_2$ and $H_2\!=\!0.16\times 10^{10}$ dynes/cm$^2$ for CO$_2$. For each case in Table \ref{tab:sims}, the temporal evolution of the bubble population due to Ostwald ripening is predicted by the theory and compared to PNM simulation. The population balance Eq.\ref{eq:populationBalance} is discretized over the phase space using the finite-volume method with a first-order upwind scheme. Although Eq.\ref{eq:populationBalance} must be solved numerically, it is far more efficient than PNM because there is no limit to the number of bubbles that can be simulated. Moreover, networks with large correlation lengths in pore size require prohibitively large representative elementary volumes (REVs) in the PNM, which is not a limitation for the theory.

\section{Results} \label{sec:results}

\subsection{One-component ripening} \label{sec:results_one}
\begin{figure}[h]
	\centering
	\includegraphics[width=\textwidth]{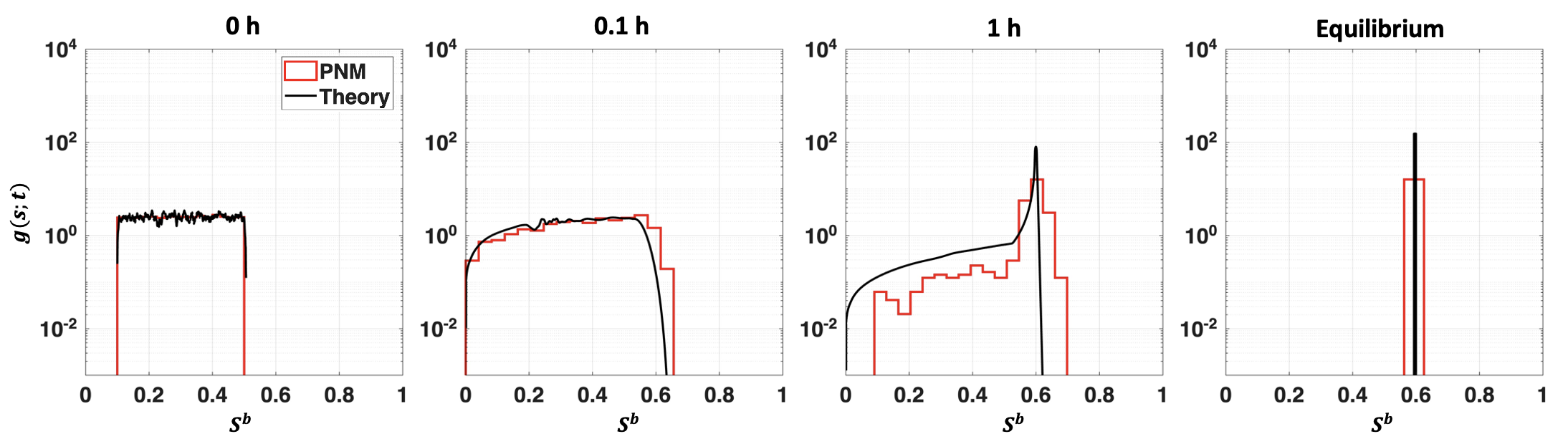}
	\caption{Comparison of the bubble number density $g(s;t)$ predicted by theory (white contours) and PNM (heat map). Since bubbles are single-component ($y_1\!=\!1$) and network is homogeneous, the phase space is one-dimensional with $S^b$ axis. Snapshots are shown at $t\!=\!0$, 0.1, 10 h, and equilibrium for the \textit{Hom-One-Full} case in Table \ref{tab:sims}. The theory reproduces results from \cite{bueno2024AWR}.}
	\label{fig:one-hom-full}
\end{figure}

\begin{figure}[h]
	\centering
	\includegraphics[width=\textwidth]{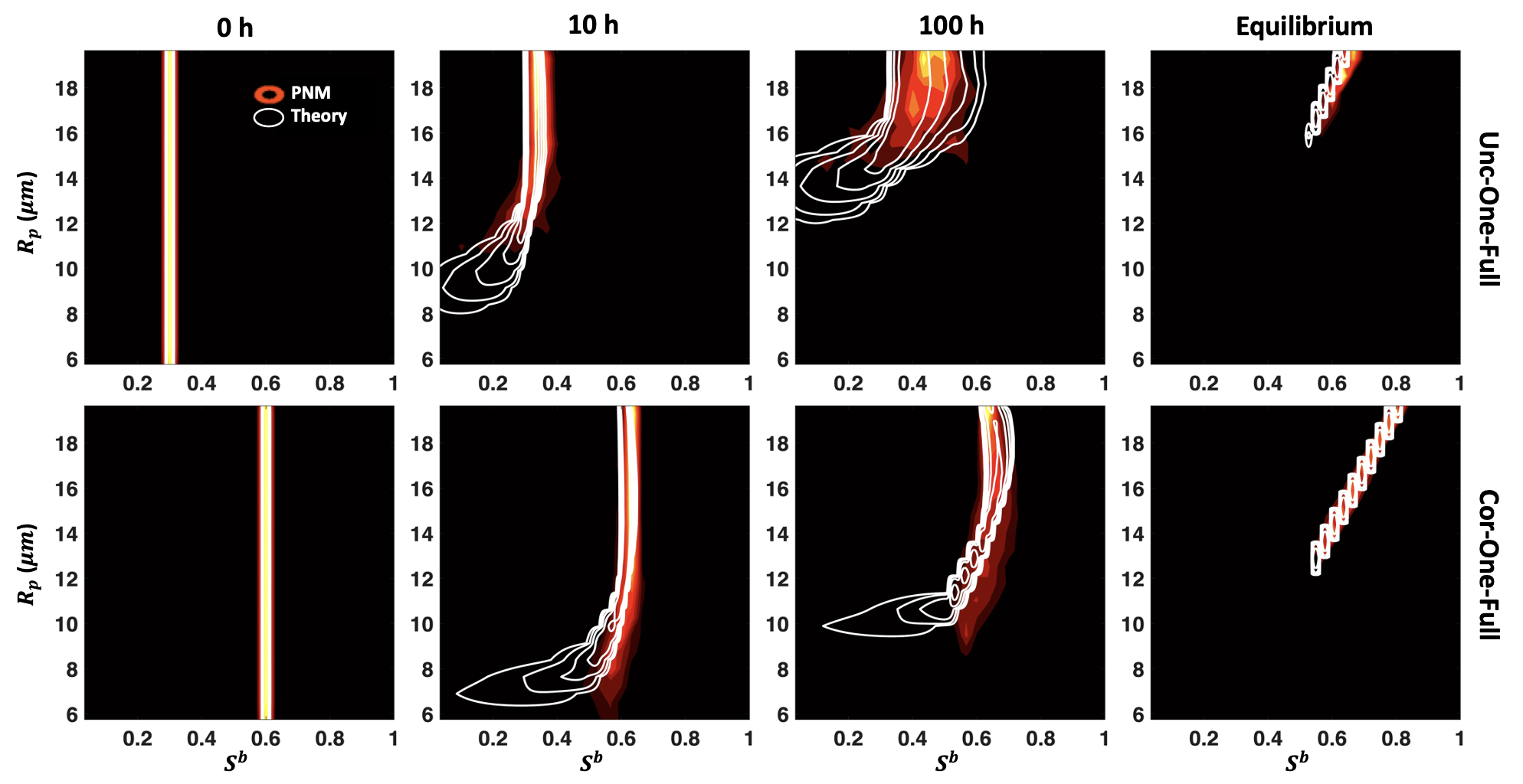}
	\caption{Comparison of the bubble number density $g(s;t)$ predicted by theory (white contours) and PNM (heat map). Since bubbles are single-component ($y_1\!=\!1$), the phase space is two-dimensional with $R_p$ and $S^b$ axes. Snapshots are shown at $t\!=\!0$, 10, 100 h, and equilibrium for the \textit{Unc-One-Full} and \textit{Cor-One-Full} cases in Table \ref{tab:sims}. The theory reproduces results from \cite{bueno2024AWR}.}
	\label{fig:one-het-full}
\end{figure}

We begin by comparing the theory to PNM for the one-component ripening cases listed in Table \ref{tab:sims}. Focusing on \textit{Hom-One-Full}, Fig.\ref{fig:one-hom-full} shows the theory successfully predicts the temporal evolution of $g(s;t)$. Since the network is homogeneous with uniform $R_p$, and bubbles are composed of one component with $y_1\!=\!1$, the phase space is one-dimensional, consisting of only the $S^b$ axis. Focusing next on \textit{Unc-One-Full} and \textit{Cor-One-Full}, Fig.\ref{fig:one-het-full} compares the evolution of $g(s;t)$ from the theory to the PNM, where good agreement is again observed. The PNM is depited by a heat map, and the theory by white contour lines. Here, the phase space is two-dimensional with axes $S^b$ and $R_p$, because $y_1\!=\!1$.
At $t\!=\!0$, $g(s;t)$ is a vertical stripe since all bubbles have the same saturation. Over time, the higher curvature in smaller pores drives mass flux toward larger pores, causing $g(s;t)$ to shift left (toward $S^b\!=\!0$) at small $R_p$ and right at large $R_p$. At equilibrium, curvature is uniform across all bubbles, producing the oblique line in Fig.\ref{fig:one-het-full}. The results in Figs.\ref{fig:one-hom-full}--\ref{fig:one-het-full} for \textit{Hom-One-Full}, \textit{Unc-One-Full}, and \textit{Cor-One-Full} were also successfully predicted by the theory of \cite{bueno2024AWR}.

\begin{figure}[t]
	\centering
	\includegraphics[width=\textwidth]{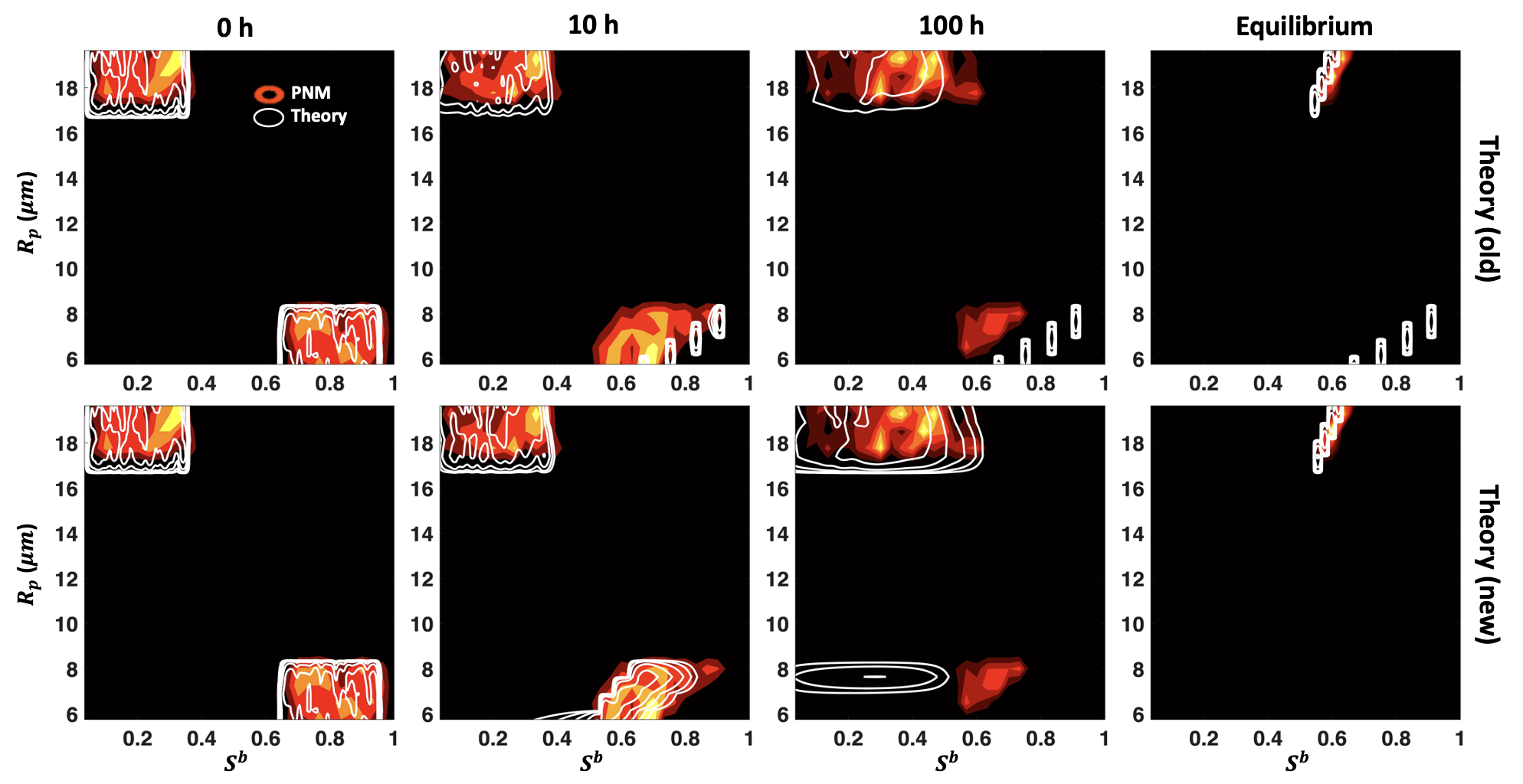}
	\caption{Comparison of the bubble number density $g(s;t)$ predicted by theory (white contours) and PNM (heat map) for the \textit{Cor-One-Part} case. Snapshots are at $t\!=\!0$, 10, 100 h, and equilibrium. Top row: theory of \cite{bueno2024AWR}, which predicts two groups evolving independently (unphysical). Bottom row: our theory, which captures interactions between distant bubbles accurately.}
	\label{fig:one-het-part}
\end{figure}

\begin{figure}[h!]
	\centering
	\includegraphics[width=0.9\textwidth]{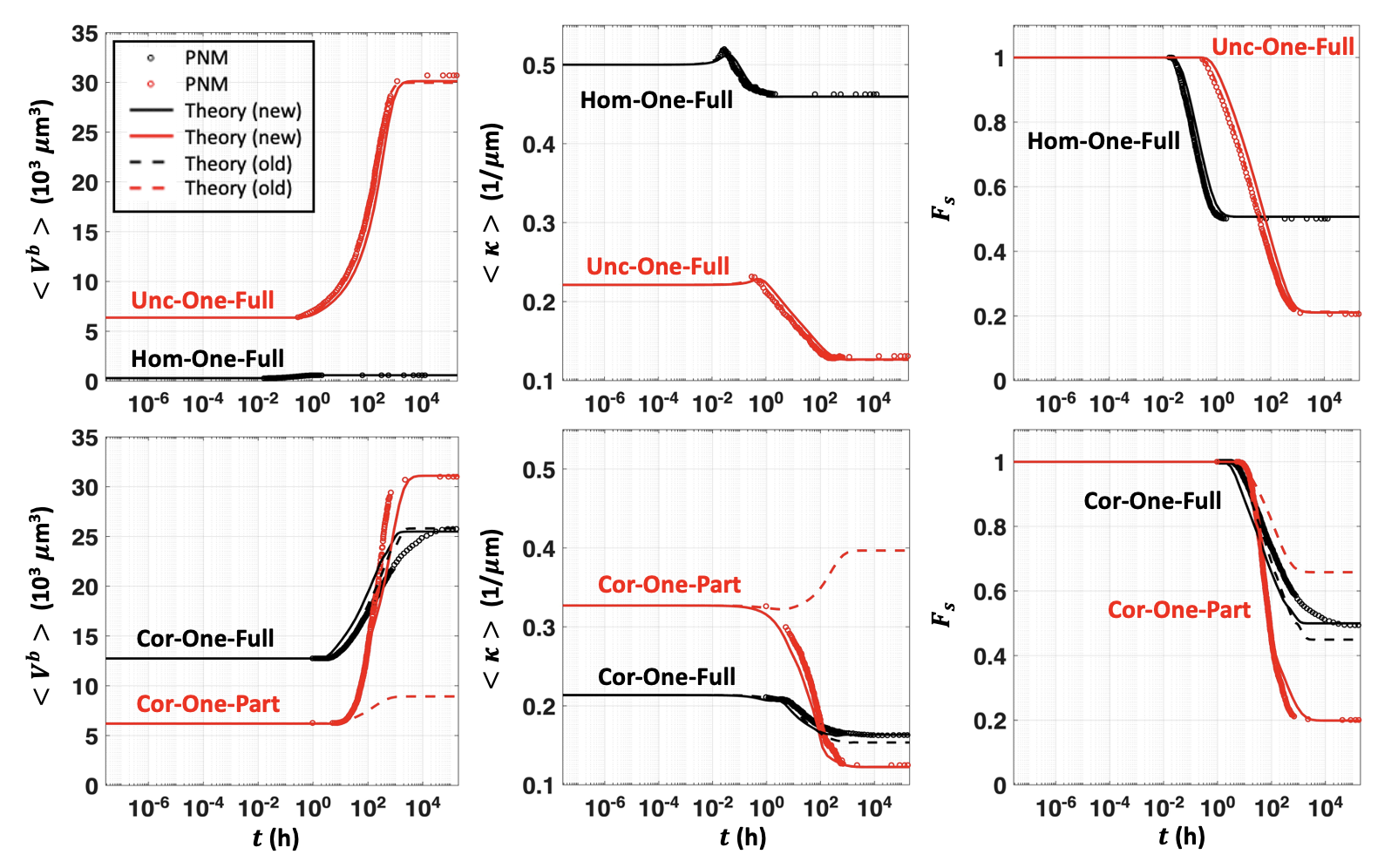}
	\caption{Average bubble volume $\langle V^b\rangle$, mean curvature $\langle\kappa\rangle$, and survived bubble fraction $F_s$ versus time for the one-component ripening cases in Table \ref{tab:sims}. The proposed theory (solid lines) shows good agreement with PNM (symbols). Predictions from \cite{bueno2024AWR} (dashed lines) are slightly worse for \textit{Hom-One-Full}, \textit{Unc-One-Full}, and \textit{Cor-One-Full}, but completely off for \textit{Cor-One-Part}.}
		\label{fig:one-mean}
\end{figure}

We now turn to \textit{Cor-One-Part}, which was presented as a challenge in \cite{bueno2024AWR}. Fig.\ref{fig:one-het-part} shows the complex initial condition of $g(s;t)$ and its subsequent evolution in time. The top row corresponds to the theory of \cite{bueno2024AWR}, which fails to predict this evolution, resulting in an isolated cluster of bubbles in the bottom-right of the phase space that remains disconnected or ``unaware'' of other bubbles in the population. This failure is due to the fundamental inability of that theory to capture interactions between bubbles that are multiple throats apart. Our theory in the bottom row of Fig.\ref{fig:one-het-part} removes this limitation, and is therefore superior to that of \cite{bueno2024AWR}.

Fig.\ref{fig:one-mean} shows the corresponding bulk properties of the above populations versus time. They consist of the average bubble volume $\langle V^b\rangle$, mean curvature $\langle\kappa\rangle$, and survived fraction of bubbles $F_s$ (i.e., current number of bubbles divided by the number of bubbles at $t\!=\!0$). These properties are readily computed from $g(s;t)$:
\begin{equation} \label{eq:mean_props}
	\langle V^b \rangle = \frac{1}{n_b(t)}
	       \int_\Omega V^b_s\, g(s;t)\,ds
	\,,\quad
	\langle \kappa \rangle = \frac{1}{n_b(t)}
	       \int_\Omega \kappa_s\, g(s;t)\,ds
	\,,\quad
	F_s = \frac{n_b(t)}{n_b(0)}
	\,,\quad
	n_b(t) = \int_\Omega g(s;t)\,ds
\end{equation}
Good agreement between our theory and PNM is observed in all cases. Fig.\ref{fig:one-mean} also includes predictions from \cite{bueno2024AWR}, which are slightly worse for the first three cases (\textit{Hom-One-Full}, \textit{Unc-One-Full}, \textit{Cor-One-Full}) but completely off for \textit{Cor-One-Part}.
In summary, Figs.\ref{fig:one-hom-full}--\ref{fig:one-mean} demonstrate that our theory not only successfully reproduces results from the theory of \cite{bueno2024AWR} for single-component ripening but also generalizes it by removing key limitations. The next section takes this further by capturing two-component ripening dynamics.


\subsection{Two-component ripening}

\subsubsection{Homogeneous network, uncorrelated composition}
\label{sec:two-hom-yRand}

\begin{figure}[t!]
	\centering
	\includegraphics[width=\textwidth]{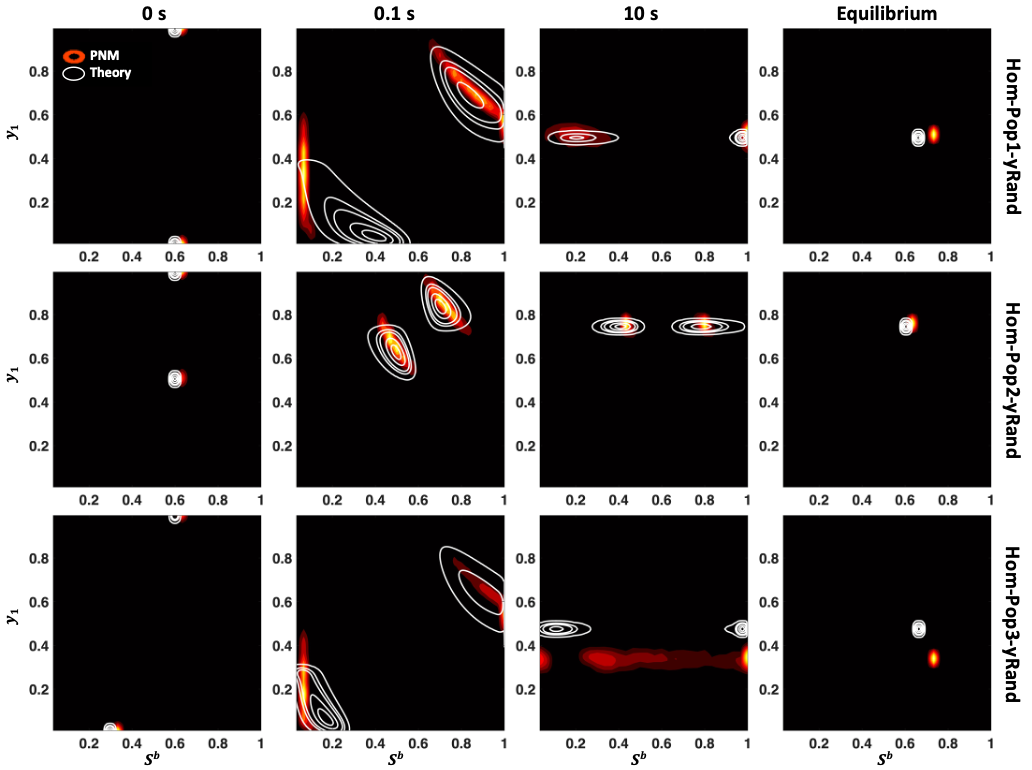}
	\caption{Comparison of the bubble number density $g(s;t)$ predicted by the theory (white contours) and PNM (heat map) for the two-component ripening cases \textit{Hom-Pop1-yRand}, \textit{Hom-Pop2-yRand}, and \textit{Hom-Pop3-yRand} in Table \ref{tab:sims}. Since the network is homogeneous with uniform $R_p$, the phase space is 2D with $y_1$ and $S^b$ axes. Snapshots are at $t\!=\!0$, 0.1, 10 s, and equilibrium.}
	\label{fig:two-hom-ys}
\end{figure}

\begin{figure}[t!]
	\centering
	\includegraphics[width=\textwidth]{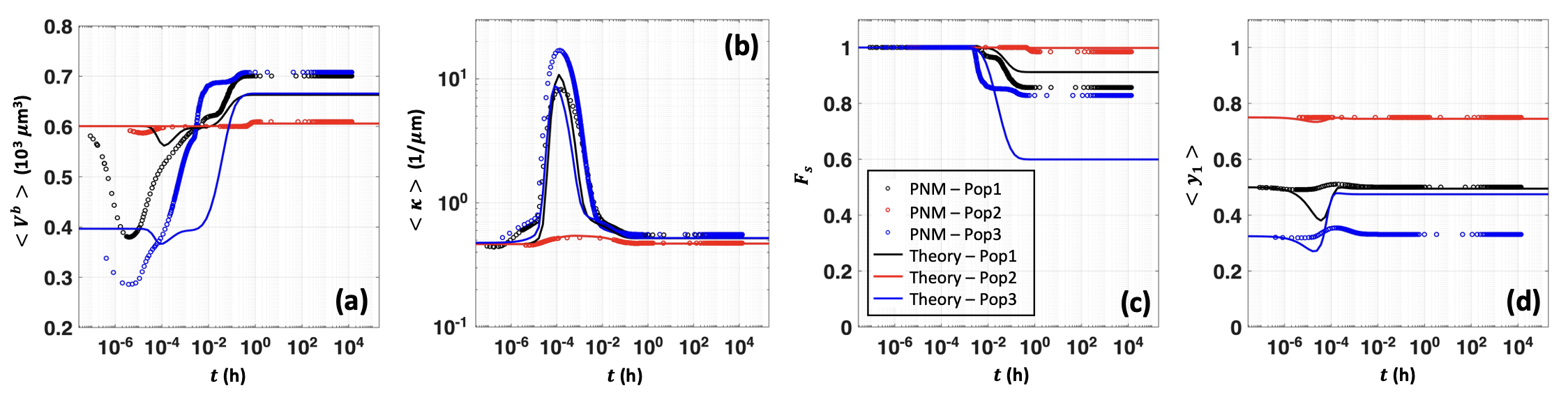}
	\caption{Average bubble volume $\langle V^b\rangle$, mean curvature $\langle\kappa\rangle$, survived fraction $F_s$, and mean composition $\langle y_1 \rangle$ versus time for two-component ripening cases \textit{Hom-Pop1-yRand}, \textit{Hom-Pop2-yRand}, and \textit{Hom-Pop3-yRand} in Table \ref{tab:sims}. Theory (solid lines) agrees with PNM (symbols) except for $\langle V^b\rangle$ at early times, and $F_s$ and $\langle y_1\rangle$ in \textit{Pop3}. Deviations in $\langle V^b\rangle$ and $\langle y_1\rangle$ arise because PNM accounts for solute storage in water, while the theory does not. Deviation in $F_s$ reflects instability of sub-critical ripening.}
	\label{fig:two-hom-mean}
\end{figure}

We begin our comparison of the theory against PNM for two-component ripening by considering cases in the second segment of Table \ref{tab:sims}, namely, \textit{Hom-Pop1-yRand}, \textit{Hom-Pop2-yRand}, and \textit{Hom-Pop3-yRand}. These correspond to the homogeneous network with random placement of bubbles at $t\!=\!0$, i.e., no correlation between $y_1$ and $R_p$. Fig.\ref{fig:two-hom-ys} shows the evolution of bubble number density $g(s;t)$ versus time, where we see good agreement between theory and PNM. Since $R_p$ is uniform, the phase space is two-dimensional with $y_1$ and $S^b$ axes.
At $t\!=\!0$, H$_2$ and CO$_2$ bubbles in \textit{Pop1} have local saturation $S^b\!=\!60\%$. \textit{Pop2} is the same, but instead of pure CO$_2$ bubbles, we have equimolar H$_2$--CO$_2$ bubbles. In \textit{Pop3}, H$_2$ bubbles have $S^b\!=\!60\%$, while CO$_2$ bubbles have $S^b\!=\!30\%$. In all three cases, as bubbles evolve, $g(s;t)$ approaches a horizontal line of constant $y_1$, corresponding to the equilibrium bubble composition. Due to differences in solubility and diffusion coefficient, mass transfer is initially directed from CO$_2$ bubbles to H$_2$ bubbles, causing the $S^b$ of H$_2$ bubbles to grow.
This can lead to highly deformed bubbles during the initial stages of ripening.

Once bubble compositions have nearly homogenized---a stage called \textit{partitioning} \cite{bueno2023AWR}---one group (initially pure H$_2$) has high $S^b$ while the other (initially CO$_2$ or equimolar mixture) has low $S^b$. Thereafter, ripening is driven by curvature differences in a stage called \textit{coevolution} \cite{bueno2023AWR}, causing the two groups to approach each other along the constant-$y_1$ line. At equilibrium, only one cluster remains, with all bubbles having the same composition and curvature (thus saturation, since $R_p$ is uniform). In Fig.\ref{fig:two-hom-ys}, the difference in $S^b$ at equilibrium between the theory and PNM is 4--5\%, and the difference in $y_1$ is 0.5\% for \textit{Pop1} and \textit{Pop2}, but 14\% for \textit{Pop3}.
The larger $y_1$ discrepancy in \textit{Pop3} arises because its CO$_2$ bubbles are sub-critical ($S^b\!=\!30\%$) and reside in large pores, leaving substantial water volume that absorbs H$_2$ during partitioning. The PNM captures this, lowering the equilibrium $y_1$ predicted, whereas the theory does not as it assigns no storage capacity to water.

Fig.\ref{fig:two-hom-mean} shows bulk properties $\langle V^b\rangle$, $\langle\kappa\rangle$, $F_s$, and mean composition $\langle y_1 \rangle$ versus time, where $\langle y_1 \rangle$ is computed analogously to Eq.\ref{eq:mean_props}. Note this is identical to $y^{mf}_1$ in Eq.\ref{eq:ymf}. The agreement is decent except for early stages of $\langle V^b\rangle$, and late stages of $F_s$ and $\langle y_1 \rangle$ for \textit{Pop3}. Focusing on $\langle V^b\rangle$, we see a temporary sag in PNM that is present but much less pronounced in the theory. This sag in PNM has two causes: (1) The wetting phase can store dissolved solute, which introduces a lag in mass transfer between H$_2$ and CO$_2$/mixture bubbles on account of differences in solubility and diffusion coefficient. This effect is \textit{not} captured by the theory, as mass exchange between bubbles is instantaneous; (2) The rapid initial influx of mass into H$_2$ bubbles causes them to grow and become deformed, resulting in the observed peak in $\langle\kappa\rangle$. Since bubbles are gaseous, their compressibility causes a temporary decrease in mean volume, before deformed bubbles relax back to lower curvatures. This effect \textit{is} captured by our theory, which is why the spike in $\langle\kappa\rangle$ is predicted accurately.

Focusing next on $F_s$ and $\langle y_1 \rangle$, we see good agreement for \textit{Pop1} and \textit{Pop2}, but significant error for \textit{Pop3}. The $\langle y_1 \rangle$ deviation in \textit{Pop3} occurs early ($\sim\!10^{-4}$ s), consistent with the water storage effect discussed above. The $F_s$ deviation occurs later ($\sim\!10^{0}$ s), after coevolution is complete, and reflects a different mechanism: in \textit{Pop3}, CO$_2$ bubbles are sub-critical (spherical) from the start and remain so as they shrink during partitioning. The subsequent curvature-driven ripening of such spherical bubbles is unstable \cite{bueno2023AWR}, with survival depending on local neighborhood configuration. In the theory, sub-critical bubbles interact only with the mean field, through which they tend to lose mass to super-critical (deformed) bubbles. In PNM, sub-critical bubbles also exchange mass with neighboring sub-critical bubbles, some of which gain mass at the expense of others. This allows more bubbles to survive in PNM than in the theory, explaining why the theory underpredicts $F_s$.


\subsubsection{Heterogeneous uncorrelated network, uncorrelated composition}
\label{sec:two-unc-yRand}

\begin{figure}[t!]
	\centering
	\includegraphics[width=\textwidth]{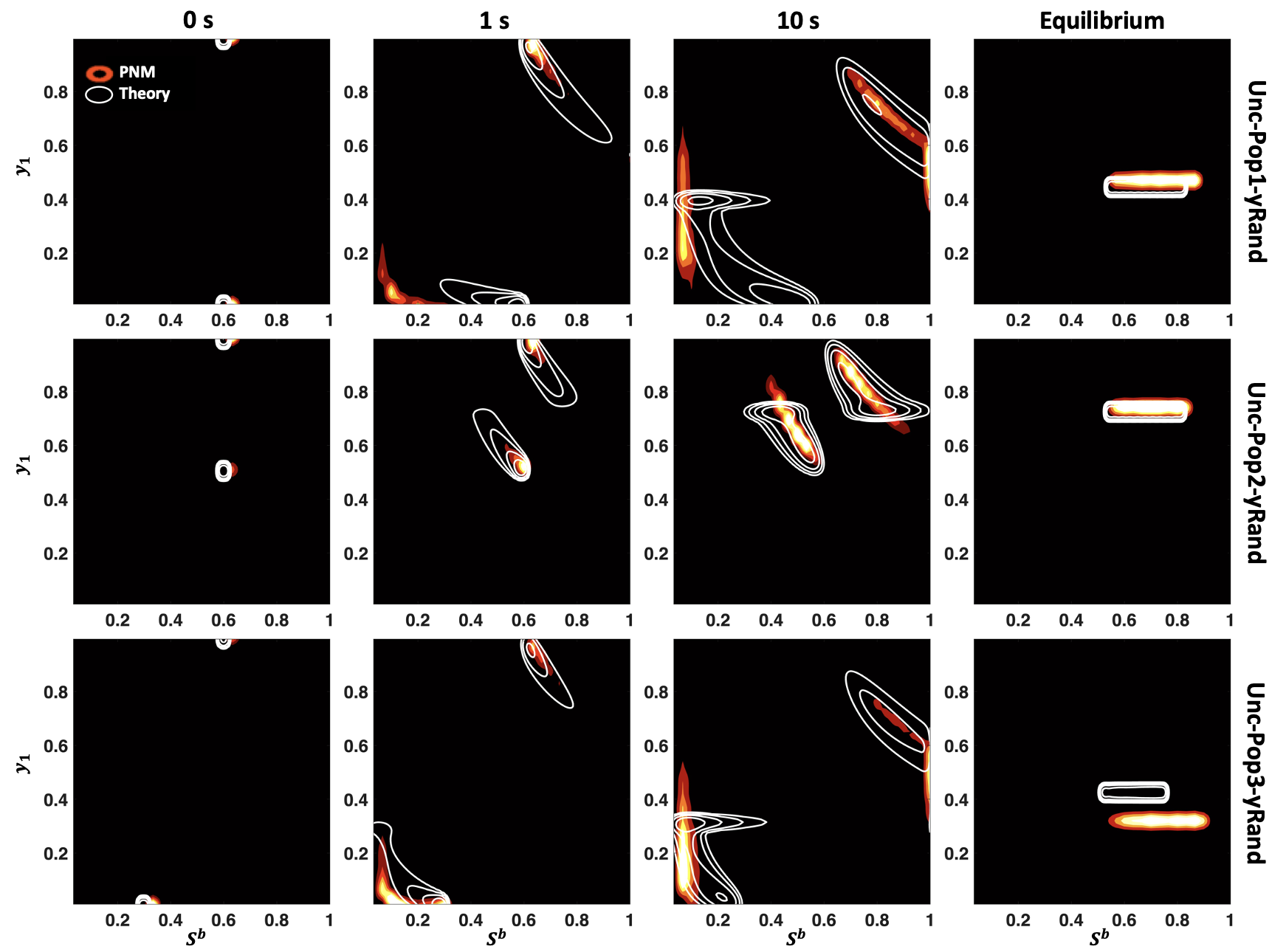}
	\caption{Projection of the bubble number density $g(s;t)$ onto the $S^b$--$y_1$ plane, predicted by the theory (white contours) and PNM (heat map) for the two-component ripening cases \textit{Unc-Pop1-yRand}, \textit{Unc-Pop2-yRand}, and \textit{Unc-Pop3-yRand} in Table \ref{tab:sims}. Network is heterogeneous and uncorrelated with random bubble placement. Snapshots are at $t\!=\!0$, 1, 10 s, and equilibrium.}
	\label{fig:two-unc-ys}
\end{figure}

\begin{figure}[t!]
	\centering
	\includegraphics[width=\textwidth]{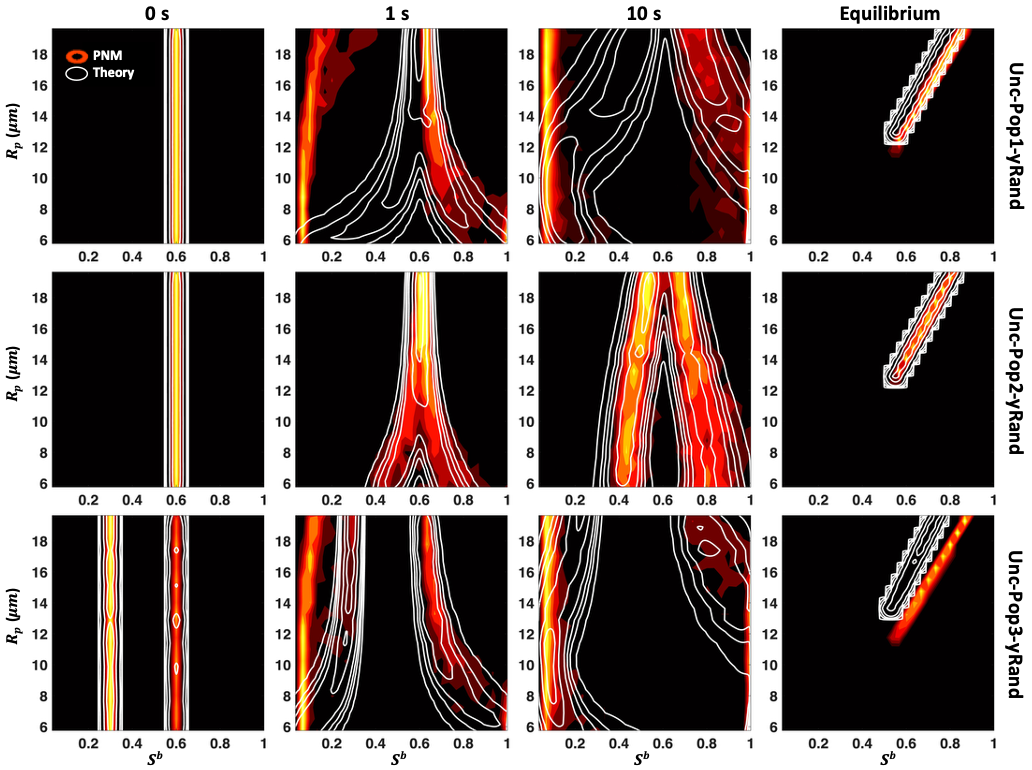}
	\caption{Projection of the bubble number density $g(s;t)$ onto the $R_p$--$S^b$ plane, predicted by the theory (white contours) and PNM (heat map) for the two-component ripening cases \textit{Unc-Pop1-yRand}, \textit{Unc-Pop2-yRand}, and \textit{Unc-Pop3-yRand} in Table \ref{tab:sims}. Network is heterogeneous and uncorrelated with random bubble placement. Snapshots are at $t\!=\!0$, 1, 10 s, and equilibrium.}
	\label{fig:two-unc-sr}
\end{figure}

\begin{figure}[t!]
	\centering
	\includegraphics[width=\textwidth]{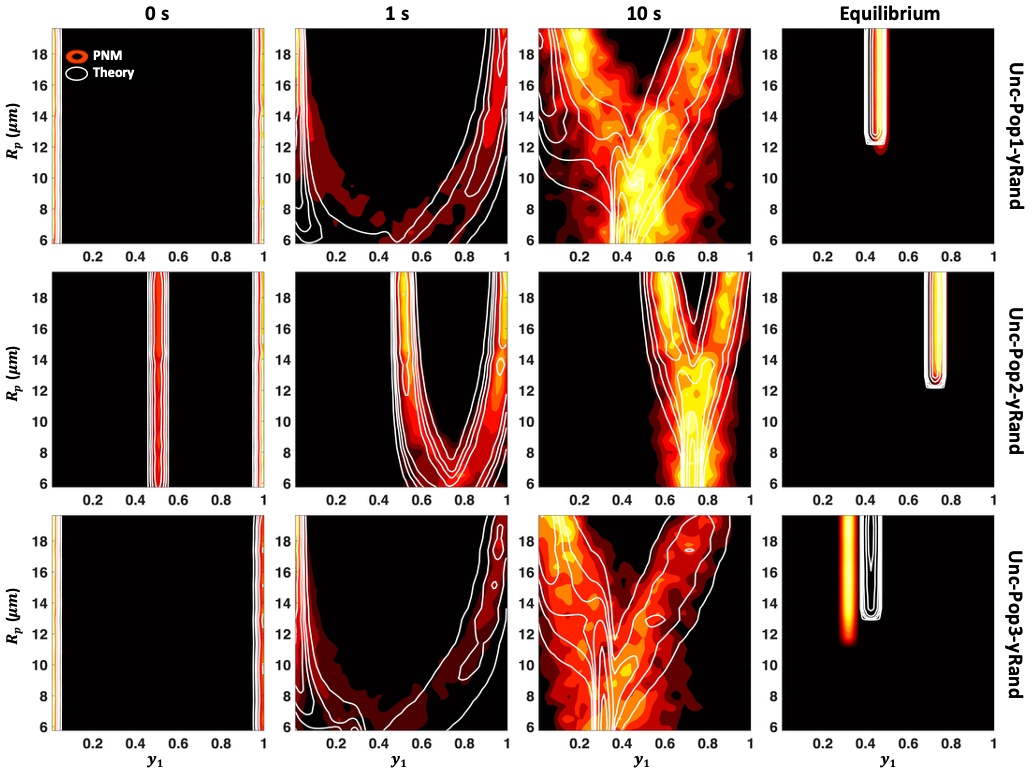}
	\caption{Projection of the bubble number density $g(s;t)$ onto the $R_p$--$y_1$ plane, predicted by the theory (white contours) and PNM (heat map) for the two-component ripening cases \textit{Unc-Pop1-yRand}, \textit{Unc-Pop2-yRand}, and \textit{Unc-Pop3-yRand} in Table \ref{tab:sims}. Network is heterogeneous and uncorrelated with random bubble placement. Snapshots are at $t\!=\!0$, 1, 10 s, and equilibrium.}
	\label{fig:two-unc-yr}
\end{figure}

\begin{figure}[t!]
	\centering
	\includegraphics[width=\textwidth]{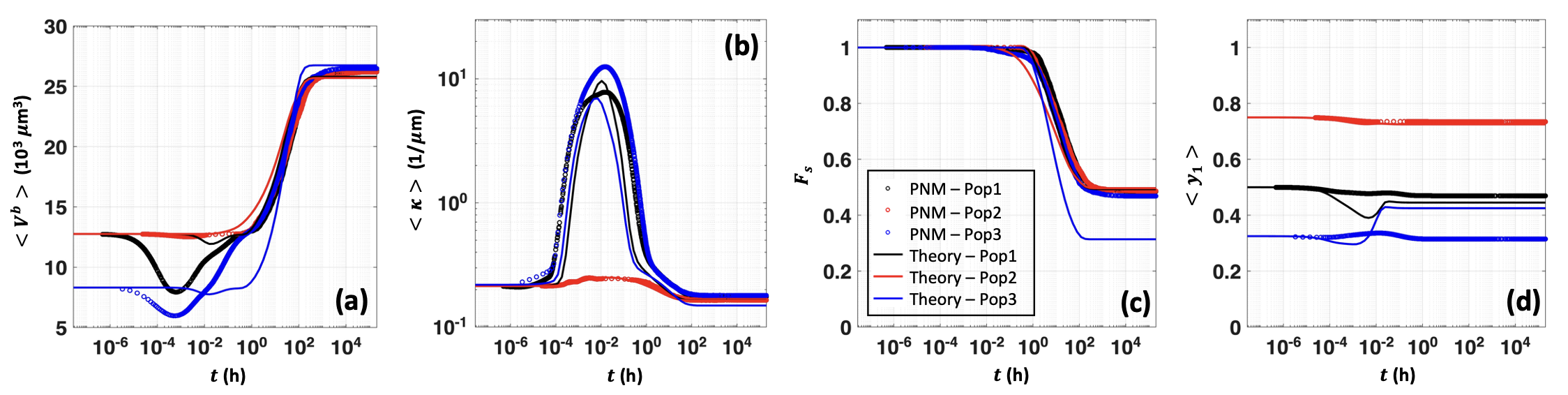}
	\caption{Bulk properties $\langle V^b\rangle$, $\langle\kappa\rangle$, $F_s$, and $\langle y_1 \rangle$ versus time for the two-component ripening cases \textit{Unc-Pop1-yRand}, \textit{Unc-Pop2-yRand}, and \textit{Unc-Pop3-yRand} in Table \ref{tab:sims}. The proposed theory (solid lines) shows good agreement with PNM (symbols).}
	\label{fig:two-unc-mean}
\end{figure}

We now focus on cases in the third segment of Table \ref{tab:sims}, namely, \textit{Unc-Pop1-yRand}, \textit{Unc-Pop2-yRand}, and \textit{Unc-Pop3-yRand}. These correspond to the heterogeneous and uncorrelated network with random placement of bubbles at $t\!=\!0$, i.e., no correlation between $y_1$ and $R_p$. Now, $g(s;t)$ lives in a three-dimensional phase space. To visualize it, we consider three projections (or marginals) of $g(s;t)$ onto the $S^b$--$y_1$, $R_p$--$S^b$, and $R_p$--$y_1$ planes. The $S^b$--$y_1$ projection is shown in Fig.\ref{fig:two-unc-ys} at $t\!=\!0$, 1 s, 10 s, and equilibrium. The evolution of $g(s;t)$ is very similar to Fig.\ref{fig:two-hom-ys}, where H$_2$ and CO$_2$/mixture bubbles first approach a line of constant $y_1$, then approach each other whilst staying close to that line. One difference compared to the homogeneous network in the previous section is that the evolution here is slower. Another is that at equilibrium, bubbles attain different $S^b$ to ensure uniform curvature due to $R_p$ variability. Overall, the theory is in good agreement with PNM. The largest discrepancy corresponds to \textit{Pop3} for reasons already discussed in Section \ref{sec:two-hom-yRand}.

The corresponding $R_p$--$S^b$ projection is shown in Fig.\ref{fig:two-unc-sr}. At $t\!=\!0$, $g(s;t)$ appears as a vertical stripe in \textit{Pop1} and \textit{Pop2} because $S^b\!=\!60\%$ is uniform across all bubbles, but as two stripes in \textit{Pop3} because $S^b\!=\!60\%$ holds for H$_2$ bubbles and $S^b\!=\!30\%$ for CO$_2$ bubbles. As time progresses, $g(s;t)$ splits into two stripes corresponding to the CO$_2$/mixture (left) and H$_2$ (right) bubbles, since each pore size is occupied by both bubble types at $t\!=\!0$. Once bubble compositions have homogenized, the two stripes approach each other and merge into an oblique line similar to that in Fig.\ref{fig:one-het-full}. This is because for the same curvature, smaller pores require less bubble deformation, thus lower $S^b$. Here too, the agreement between theory and PNM is good. The observations and takeaways remain similar for the $R_p$--$y_1$ projection shown in Fig.\ref{fig:two-unc-yr}.

Finally, Fig.\ref{fig:two-unc-mean} shows how bulk properties $\langle V^b \rangle$, $\langle \kappa \rangle$, $F_s$, and $\langle y_1 \rangle$ evolve with time. Overall, theory and PNM are in good agreement, with larger deviations observed in \textit{Pop3}. Similar to Fig.\ref{fig:two-hom-mean}, the $\langle y_1 \rangle$ deviation in \textit{Pop3} arises because PNM accounts for solute storage in water, whereas the theory does not. The $F_s$ deviation reflects instability of sub-critical bubble ripening. Namely, in \textit{Pop3}, CO$_2$ bubbles are spherical from the start and remain so during partitioning, after which their survival depends on local neighborhood configuration rather than interaction with the mean field.
That said, $F_s$ and $\langle y_1\rangle$ are in slightly better agreement here than in the homogeneous network, since the variability in $R_p$ likely provides some stabilization.

\subsubsection{Heterogeneous correlated network, uncorrelated composition}
\label{sec:two-cor-yRand}
\begin{figure}[t!]
	\centering
	\includegraphics[width=\textwidth]{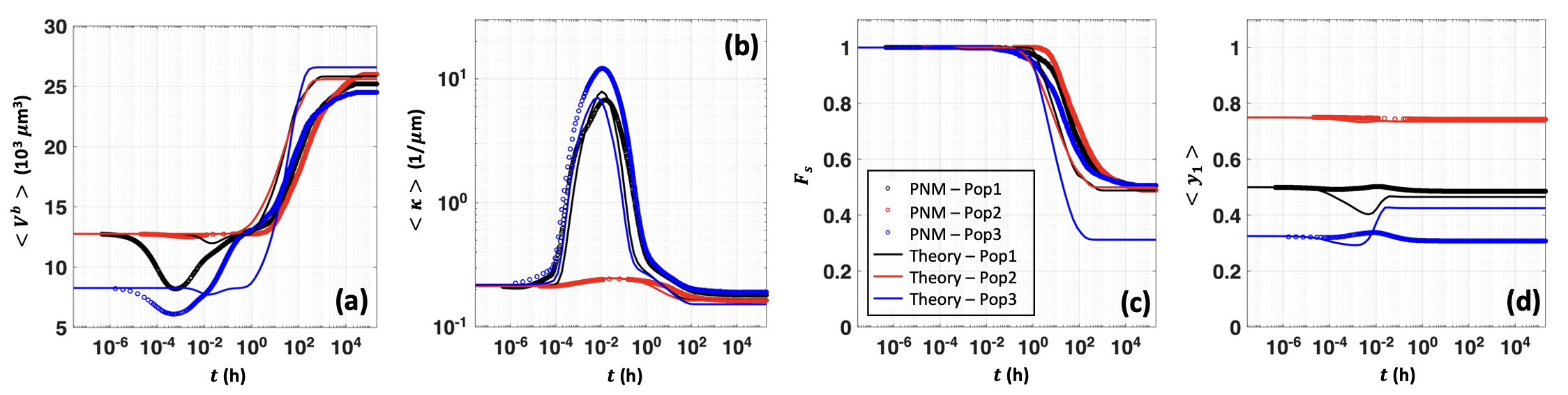}
	\caption{Bulk properties $\langle V^b\rangle$, $\langle\kappa\rangle$, $F_s$, and $\langle y_1 \rangle$ versus time for the two-component ripening cases \textit{Cor-Pop1-yRand}, \textit{Cor-Pop2-yRand}, and \textit{Cor-Pop3-yRand} in Table \ref{tab:sims}. The network is correlated and bubble placement is random. The proposed theory (solid lines) shows good agreement with PNM (symbols). Results are nearly indistinguishable from Fig.\ref{fig:two-unc-mean} for uncorrelated network.}
	\label{fig:two-cor-mean}
\end{figure}

The results for the cases in the fourth segment of Table \ref{tab:sims} are very similar to those from the previous section. These cases are \textit{Cor-Pop1-yRand}, \textit{Cor-Pop2-yRand}, and \textit{Cor-Pop3-yRand}, corresponding to the heterogeneous and spatially correlated network with random bubble placement, i.e., no correlation between $y_1$ and $R_p$. Due to this similarity with the uncorrelated network, we only present the bulk properties in Fig.\ref{fig:two-cor-mean}, which is nearly indistinguishable from Fig.\ref{fig:two-unc-mean}. The only difference is that late-time ripening dynamics are slightly slower in the correlated network (see $F_s$). The likely explanation is that spatial correlations in $R_p$ play little role in early ripening, which is driven primarily by gradients in bubble composition. What matters is which bubbles (H$_2$ or CO$_2$/mixture) are initially neighbors—or equivalently, how $y_1$ is arranged spatially. This arrangement is random both here and in the previous section, hence the small difference in results. At late times, however, when ripening is driven mainly by gradients in curvature, larger correlation entails smaller gradients in $\kappa$ (since $R_p$ controls $\kappa$ through Eq.\ref{eq:kappa_vb}), thus slower mass transfer between bubbles.

\subsubsection{Heterogeneous uncorrelated network, correlated composition}
\label{sec:two-unc-ySmall}
\begin{figure}[t!]
	\centering
	\includegraphics[width=\textwidth]{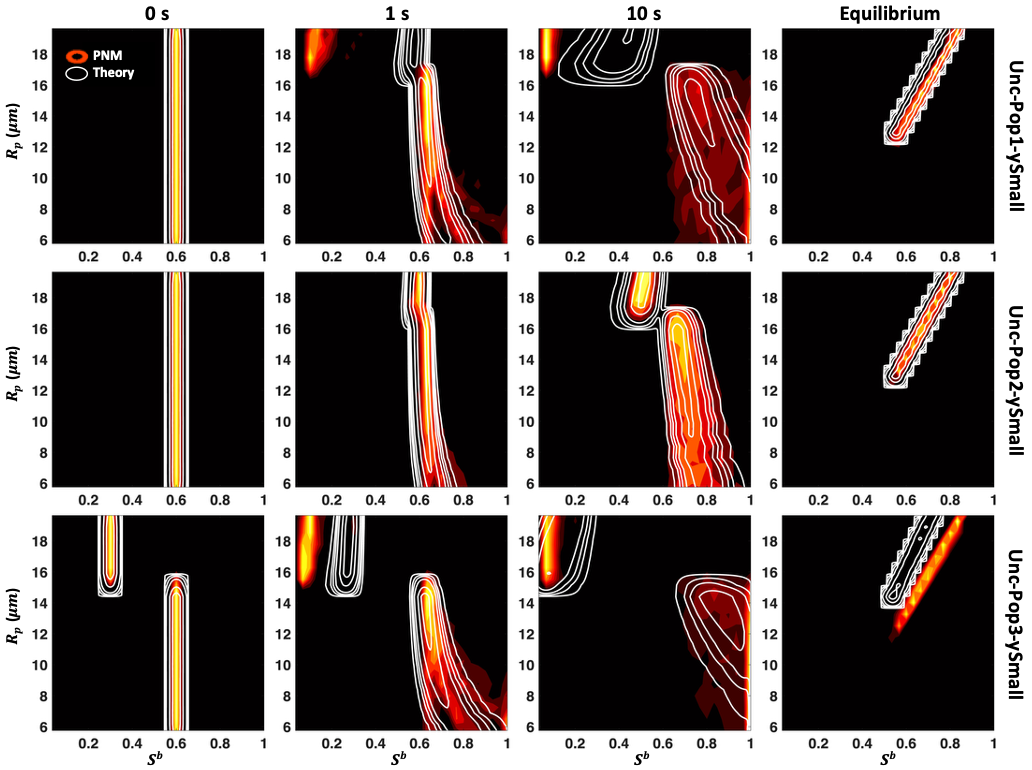}
	\caption{Projection of the bubble number density $g(s;t)$ onto the $R_p$--$S^b$ plane, predicted by theory (white contours) and PNM (heat map) for two-component ripening cases \textit{Unc-Pop1-ySmall}, \textit{Unc-Pop2-ySmall}, and \textit{Unc-Pop3-ySmall} in Table \ref{tab:sims}. Network is heterogeneous and uncorrelated with H$_2$ bubbles placed in small pores. Snapshots are at $t\!=\!0$, 1, 10 s, and equilibrium.}
	\label{fig:two-unc-ysm-sr}
\end{figure}

\begin{figure}[t!]
	\centering
	\includegraphics[width=\textwidth]{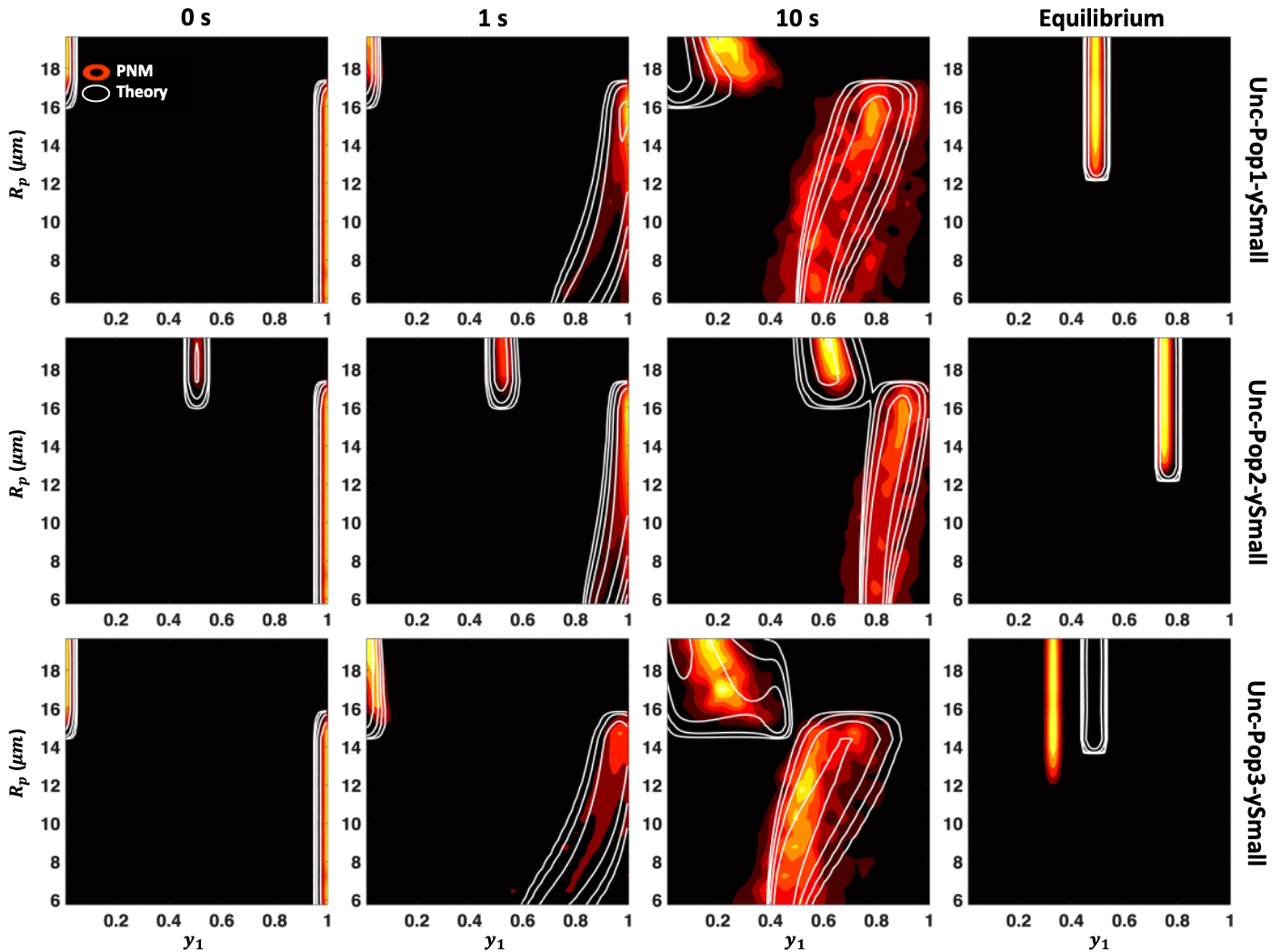}
	\caption{Projection of the bubble number density $g(s;t)$ onto the $R_p$--$y_1$ plane, predicted by theory (white contours) and PNM (heat map) for two-component ripening cases \textit{Unc-Pop1-ySmall}, \textit{Unc-Pop2-ySmall}, and \textit{Unc-Pop3-ySmall} in Table \ref{tab:sims}. Network is heterogeneous and uncorrelated with H$_2$ bubbles placed in small pores. Snapshots are at $t\!=\!0$, 1, 10 s, and equilibrium.}
	\label{fig:two-unc-ysm-yr}
\end{figure}

We now turn to cases in the fifth segment of Table \ref{tab:sims}, namely, \textit{Unc-Pop1-ySmall}, \textit{Unc-Pop2-ySmall}, and \textit{Unc-Pop3-ySmall}. These correspond to the heterogeneous uncorrelated network with H$_2$ bubbles placed in small pores and CO$_2$/mixture bubbles in large pores. Since the two bubble groups have equal bulk volume, H$_2$ bubbles are more numerous. Similar to Section \ref{sec:two-unc-yRand}, we consider projections of $g(s;t)$ onto the $S^b$--$y_1$, $R_p$--$S^b$, and $R_p$--$y_1$ planes. The $S^b$--$y_1$ projection is omitted because it is very similar to Fig.\ref{fig:two-unc-ys}. Figs.\ref{fig:two-unc-ysm-sr}--\ref{fig:two-unc-ysm-yr} illustrate the remaining two projections, where we see qualitatively different evolution patterns compared to prior sections when the initial arrangement of $y_1$ was spatially random. The agreement between theory and PNM is good, with errors larger in \textit{Pop3} due to reasons already discussed in Section \ref{sec:two-hom-yRand}.

\begin{figure}[t!]
	\centering
	\includegraphics[width=\textwidth]{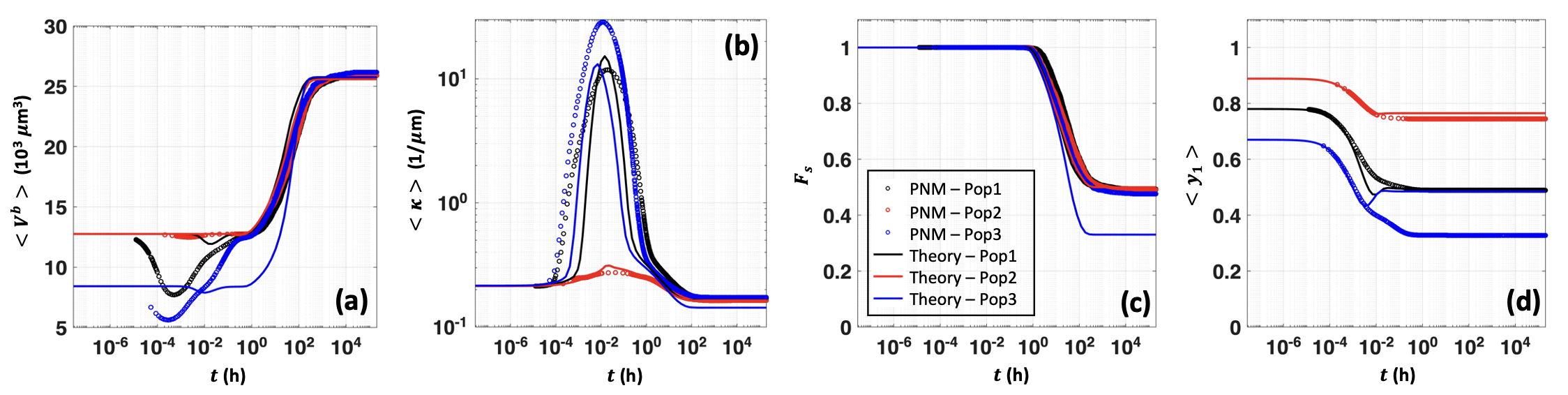}
	\caption{Bulk properties $\langle V^b\rangle$, $\langle\kappa\rangle$, $F_s$, and $\langle y_1 \rangle$ versus time for two-component ripening cases \textit{Unc-Pop1-ySmall}, \textit{Unc-Pop2-ySmall}, and \textit{Unc-Pop3-ySmall} in Table \ref{tab:sims}. The proposed theory (solid lines) shows good agreement with PNM (symbols).}
	\label{fig:two-unc-ysm-mean}
\end{figure}

Bulk properties are plotted in Fig.\ref{fig:two-unc-ysm-mean}. The main difference compared to Fig.\ref{fig:two-unc-mean} concerns $\langle y_1 \rangle$: while roughly constant in Fig.\ref{fig:two-unc-mean}, it now decreases with time. Note that $\langle y_1 \rangle$, which equals $y^{mf}_1$ by Eq.\ref{eq:ymf}, is \textit{not} the same as $y^{mix}_1$. By Corollary \ref{cor:ymix}, $y^{mix}_1$ is constant over time in the theory, and only as $t\!\rightarrow\!\infty$ does $y^{mf}_1\!\rightarrow\!y^{mix}_1$.\footnote{From \ref{ap:proof_cor}, $\gamma\!\rightarrow\!1$ as $t\!\rightarrow\!\infty$. Therefore, both $y^{mf,*}_1$ and $y^{mf}_1$ converge to $y^{mix}_1$.} The values of $y^{mix}_1$ for each case are listed in Table \ref{tab:sims}. One can verify that the equilibrium values of $\langle y_1 \rangle$ in Fig.\ref{fig:two-unc-ysm-mean} match the $y^{mix}_1$ values in Table \ref{tab:sims}.
The discrepancy in equilibrium $\langle y_1 \rangle$ for \textit{Pop3} arises because some H$_2$ is absorbed by water in PNM---an effect not captured by the theory---resulting in lower $\langle y_1 \rangle$ predicted by PNM. The deviation in $F_s$ reflects the instability of sub-critical ripening, as discussed in Section \ref{sec:two-hom-yRand}.


\subsubsection{Heterogeneous correlated network, correlated composition}
\label{sec:two-cor-ySmall}
\begin{figure}[t!]
	\centering
	\includegraphics[width=\textwidth]{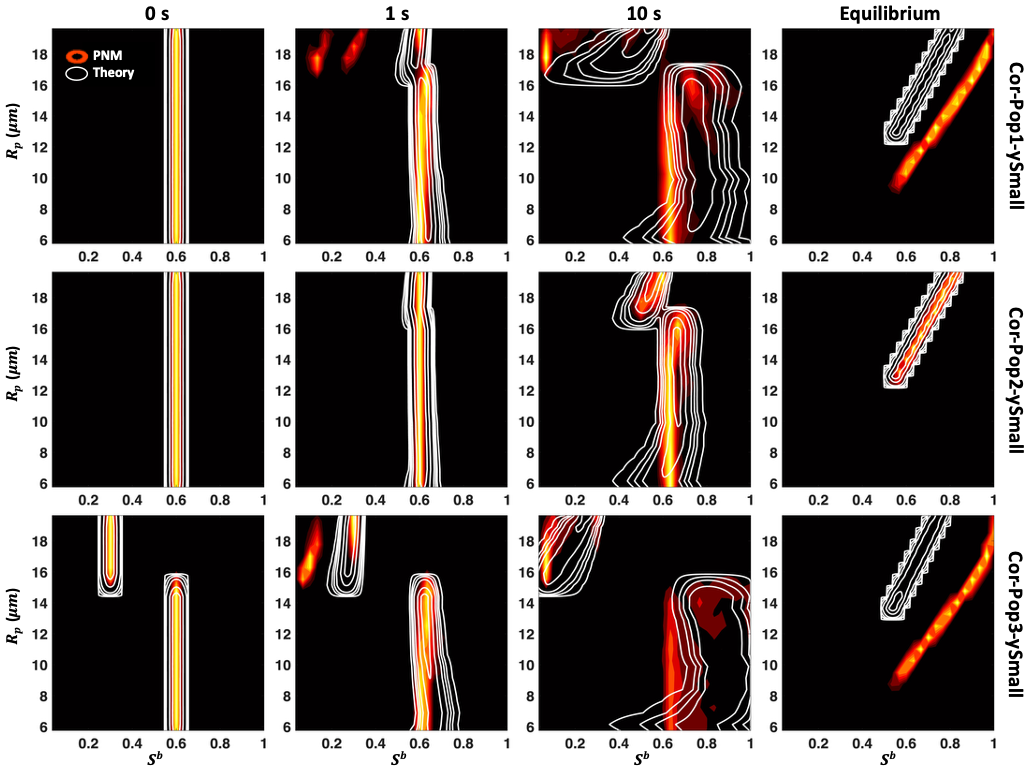}
	\caption{Projection of the bubble number density $g(s;t)$ onto the $R_p$--$S^b$ plane, predicted by theory (white contours) and PNM (heat map) for two-component ripening cases \textit{Cor-Pop1-ySmall}, \textit{Cor-Pop2-ySmall}, and \textit{Cor-Pop3-ySmall} in Table \ref{tab:sims}. Network is heterogeneous and correlated with H$_2$ bubbles placed in small pores. Snapshots are at $t\!=\!0$, 1, 10 s, and equilibrium.}
	\label{fig:two-cor-ysm-sr}
\end{figure}

\begin{figure}[t!]
	\centering
	\includegraphics[width=\textwidth]{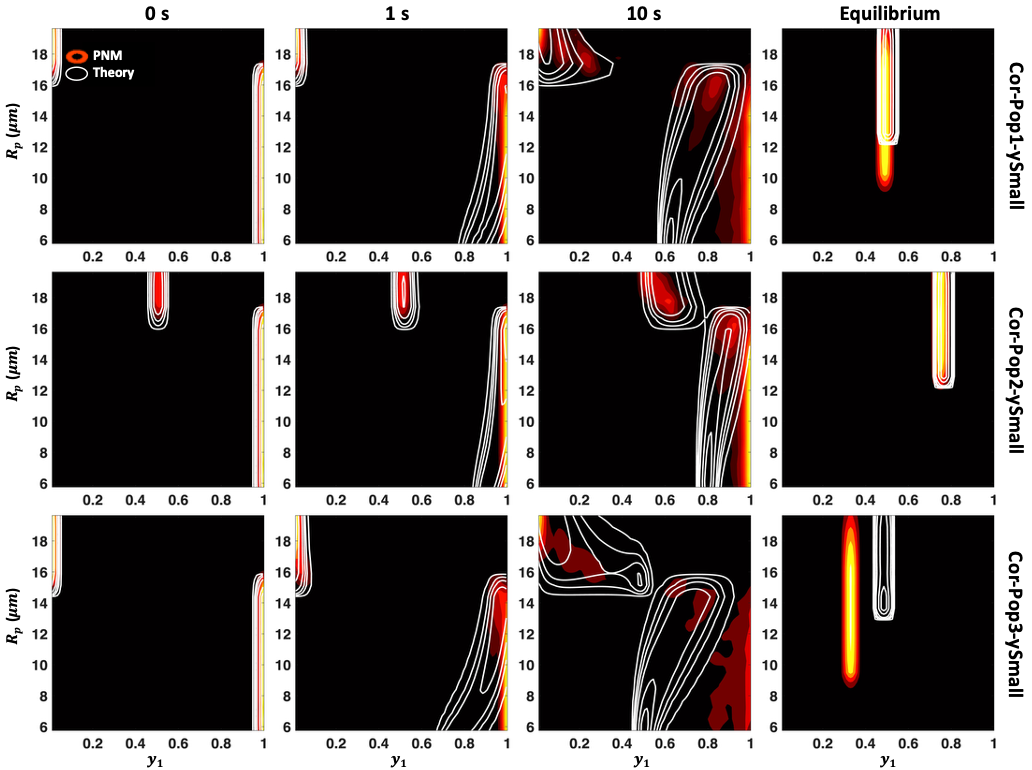}
	\caption{Projection of the bubble number density $g(s;t)$ onto the $R_p$--$y_1$ plane, predicted by theory (white contours) and PNM (heat map) for two-component ripening cases \textit{Cor-Pop1-ySmall}, \textit{Cor-Pop2-ySmall}, and \textit{Cor-Pop3-ySmall} in Table \ref{tab:sims}. Network is heterogeneous and correlated with H$_2$ bubbles placed in small pores. Snapshots are at $t\!=\!0$, 1, 10 s, and equilibrium.}
	\label{fig:two-cor-ysm-yr}
\end{figure}

\begin{figure}[t!]
	\centering
	\includegraphics[width=\textwidth]{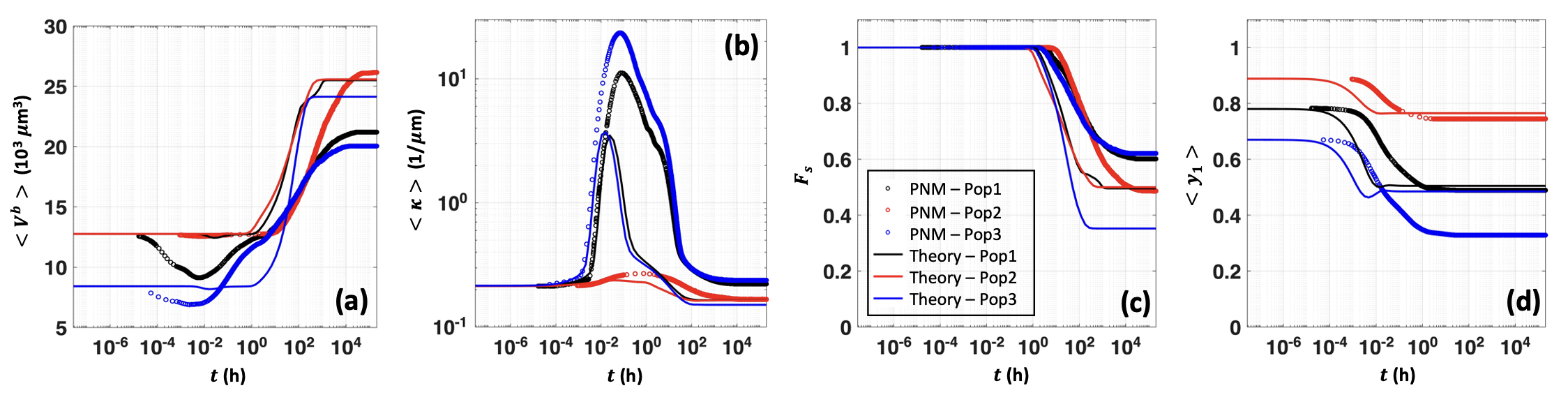}
	\caption{Bulk properties $\langle V^b\rangle$, $\langle\kappa\rangle$, $F_s$, and $\langle y_1 \rangle$ versus time for two-component ripening cases \textit{Cor-Pop1-ySmall}, \textit{Cor-Pop2-ySmall}, and \textit{Cor-Pop3-ySmall} in Table \ref{tab:sims}. The proposed theory (solid lines) shows satisfactory agreement with PNM (symbols).}
	\label{fig:two-cor-ysm-mean}
\end{figure}

We now examine the final segment of Table \ref{tab:sims}: \textit{Cor-Pop1-ySmall}, \textit{Cor-Pop2-ySmall}, and \textit{Cor-Pop3-ySmall}, which correspond to the heterogeneous correlated network with H$_2$ bubbles placed in small pores and CO$_2$/mixture bubbles in large pores. These cases are the most challenging because both $R_p$ and $y_1$ are spatially correlated. As in the previous section, we omit the $S^b$--$y_1$ projection of $g(s;t)$ because it is very similar to Fig.\ref{fig:two-unc-ys}. The remaining two projections, $R_p$--$S^b$ and $R_p$--$y_1$, are shown in Figs.\ref{fig:two-cor-ysm-sr}--\ref{fig:two-cor-ysm-yr}.
We see satisfactory agreement between theory and PNM, but errors are significantly larger than in previous sections. This is especially apparent in the equilibrium configurations of \textit{Pop1} and \textit{Pop3}, and in the theory overestimating the speed of ripening. Fig.\ref{fig:two-cor-ysm-mean} shows bulk properties versus time, with consistent observations.

The most likely explanation relates to two fundamental assumptions in the theory. First, bubbles at state $s$ interact with a mean-field having composition $y^{mf}_1$ and curvature $\kappa^{mf}_s$---equivalently $(y_1 P^b)^{mf}_s$. Only $\kappa^{mf}_s$ accounts for spatial correlations through Eq.\ref{eq:mf_approx}, hence the subscript $s$. The mean-field composition $y^{mf}_1$, computed via Eq.\ref{eq:ymf}, is $s$-independent and informed by \textit{all} bubbles in the population, whether proximate or distant relative to the $s$-bubble. This causes overestimation of the phase velocities through Eq.\ref{eq:closure_final_2}, directly impacting $dy_{1,s}/dt$ and indirectly impacting $dS^b_s/dt$ through coupling terms, thus accelerating ripening. Second, spatial correlations accounted for in $\kappa^{mf}_s$ are of pore size $R_p$, not composition $y_1$. This assumption was necessary to render the computation of mean-field properties tractable via approximations made between Eqs.\ref{eq:mf_exact} and \ref{eq:mf_approx}. The errors in Figs.\ref{fig:two-cor-ysm-sr}--\ref{fig:two-cor-ysm-yr} are the price we pay for these simplifications. The reason \textit{Pop2} predictions are more accurate is that chemical-potential gradients between H$_2$ and mixture bubbles are less than gradients between H$_2$ and CO$_2$ bubbles in \textit{Pop1} and \textit{Pop3}, resulting in weaker disequilibrium.


\section{Discussion}\label{sec:discussion}
\subsection{Limitations of the theory} \label{sec:disc_lims}
This work presents the first theory for multicomponent ripening of bubbles in porous media, generalizing that of \cite{bueno2024AWR} for single-component ripening. It also simplifies the formulation and removes key limitations of that theory, as demonstrated in Section \ref{sec:results_one}. While we focused on H$_2$ and CO$_2$ bubbles, other gases relevant to underground hydrogen storage (e.g., N$_2$, CH$_4$) or otherwise could be used without loss of generality. Below, we discuss pending limitations of the theory to guide future research.
They are divided into \textit{methodological}, which exist within the assumptions of the theory and may be addressed by reformulating certain parts, and \textit{foundational}, which stem from core assumptions that may require a completely new formulation.

\subsubsection{Methodological limitations} \label{sec:disc_lims_meth}
Even under the assumptions of the theory, Section \ref{sec:two-cor-ySmall} showed that predicting ripening of bubble populations faces three remaining challenges: (1) capturing both kinetics and equilibrium state is difficult when composition is spatially correlated (\textit{Cor-ySmall}); (2) late-time predictions of bubble composition $\langle y_1 \rangle$ are overestimated when the wetting phase is preferentially undersaturated in one component at $t\!=\!0$ (\textit{Pop3}); (3) early-time predictions of bubble volume $\langle V^b \rangle$ exhibit a temporary sag whose magnitude is underestimated.

Addressing the first limitation requires alternative approximations to those in Eqs.\ref{eq:mf_exact}--\ref{eq:mf_approx} and Eq.\ref{eq:ymf}. The former postulates dominant control of pore-size correlations over other spatial correlations, while the latter removes state-dependence from $y^{mf}_1$ and replaces the closure Eq.\ref{eq:closure_raw} with Eq.\ref{eq:closure_final} to avoid numerical difficulties and ensure convergence to equilibrium. The analysis in \ref{app:closure_reform} shows this reformulation is necessary for Eulerian (grid-based) approaches used to discretize and solve the population balance Eq.\ref{eq:populationBalance} (e.g., finite volumes herein).
A Lagrangian (particle-based) method could obviate such reformulation and use the more accurate closure Eq.\ref{eq:closure_raw} directly. This in turn would enable a state-dependent $y^{mf}_{1,s}$ that accounts for spatial correlations in composition.
The second and third limitations arise because the wetting phase has no storage capacity for dissolved solute, rendering mass transfer between bubbles effectively instantaneous. This causes overprediction of $\langle y_1 \rangle$ and the inability to capture the sag in $\langle V^b \rangle$. One remedy is to augment Eq.\ref{eq:closure_final} with an auxiliary balance equation for dissolved solute in the wetting phase. These remedies are left as future work.

While the theory was tested for two-component bubbles, extension to multiple components is straightforward: each new component adds another dimension to the phase space and an extra phase velocity to Eq.\ref{eq:closure_final}. Finally, contact angle and pore shape can be generalized beyond those assumed here (zero contact angle, semi-cubic pores), as they enter the theory solely through the curvature-volume relation in Eq.\ref{eq:kappa_vb}.

\subsubsection{Foundational limitations} \label{sec:disc_lims_found}
Foundational limitations question the very choice of phase-space axes we have made for the theory, namely, $R_p$, $S^b$, and $y_1$. The most significant is the assumption that each bubble occupies no more than one pore, which renders the use of pore size $R_p$ as a state coordinate inappropriate. However, a clear alternative is not self-evident because even if a bubble occupies only two pores, it begs the question: which pores?---since their sizes matter. Let the bubble occupy three or more pores and the question grows with combinatorial complexity. Another limitation of the same type is when bubbles, even if confined to one pore, occupy pores with different shape or wettability (i.e., contact angle). Put differently, there is heterogeneity in the curvature-volume relation Eq.\ref{eq:kappa_vb}, which the theory assumes to be uniform across all pores. Introducing a separate axis for each shape or contact angle would make the theory intractable. A viable option might be to parameterize the curvature-volume relation with contact angle and shape-controlling constants and add those as extra phase-space axes. Either way, an increase in dimensionality of the phase space seems unavoidable.

\subsection{Computational efficiency and upscaling}
\label{sec:disc_comp}
Since the population balance Eq.\ref{eq:populationBalance} must be discretized and solved numerically (here using the finite-volume method), readers might wonder if there are true gains in computational cost over solving the problem spatially using PNM. The answer is an emphatic \textit{yes} for two reasons: (1) there is no limit to the number of bubbles that can be simulated in the theory, whereas the cost of PNM scales with the number of bubbles; (2) when the correlation length $l_c$ of the porous medium is large, truly enormous pore networks are needed to ensure ripening predictions are representative of an REV, which is computationally prohibitive. The theory's cost is independent of $l_c$, enabling predictions out of reach with PNM or any other pore-scale model. For the results in Section \ref{sec:results}, the theory was faster by a factor of 192 for homogeneous networks and 24 for heterogeneous networks. These speedups were achieved despite using a very fine mesh for the phase space, which from experience we have found unnecessary, meaning even higher acceleration is possible.

\subsection{Implications for underground hydrogen storage}
\label{sec:disc_uhs}
The proposed theory has practical applications in underground hydrogen storage, where major concerns include: (1) inability to recover injected H$_2$ due to capillary trapping; (2) loss of purity of recovered H$_2$; and (3) maintaining high storage capacity and injectivity (i.e., low flow resistance) across multiple injection-withdrawal cycles.
The theory predicts which pore sizes are occupied by bubbles, their volumes, and compositions at all times. When coupled to a permeability model like Carman-Kozeny \cite{kozeny1927bundle, carman1937bundle}, where the porous medium is represented by a bundle of capillary tubes of radii $R_p$, predictions of $g(s;t)$ can be translated into estimates of relative permeability evolution. This would enable prediction of storage capacity and injectivity versus time, and could be used to maximize H$_2$ retrieval and minimize purity loss in cyclic injections.

The \textit{ySmall} cases in Section \ref{sec:valid} were motivated by the scenario where cushion gas (e.g., CO$_2$, CH$_4$, N$_2$), possibly mixed with H$_2$, gets trapped in preferential flow paths (larger pores) during the first injection cycle as it is chased by H$_2$. The H$_2$ injected in future cycles is diverted into secondary flow paths (smaller pores) not yet blocked by trapped bubbles. If this conceptualization is correct, maximizing H$_2$ recovery would suggest choosing a cushion gas that feeds into H$_2$ bubbles, causing them to grow and coalesce into mobile ganglia. CO$_2$ would be such a candidate: Section \ref{sec:valid} shows mass transfer is directed from CO$_2$ to H$_2$ bubbles within minutes. While this would promote mobilization, it would also entail H$_2$-purity loss. Alternatively, a cushion gas like CH$_4$ would draw mass out of H$_2$ bubbles, moving H$_2$ from smaller pores into larger pores where CH$_4$ resides. Recovery through these preferential flow paths would then encounter minimal flow resistance—though here too, purity loss is inevitable to relocate the H$_2$.
Future work should determine initial conditions that arise from cyclic injections in a given microstructure to enable optimizing storage operations.


\section{Conclusion}
\label{sec:conclusion} 
We have presented the first kinetic theory for multicomponent Ostwald ripening of bubbles in porous media. The formulation describes a trapped bubble population with a number-density function $g(s;t)$ in a three-dimensional statistical space of bubble states $s$. The coordinates of this space consist of the size of the occupied pore $R_p$, occupied fraction of the pore volume $S^b$ (bubble saturation), and bubble composition $y_1$. The theory predicts the evolution of $g(s;t)$ in time through a population balance equation. Closure is achieved through mean-field approximations that embed the physics of ripening, account for spatial correlations in pore size, and ensure convergence to a physically meaningful equilibrium with provable mass conservation.

The theory generalizes that of \cite{bueno2024AWR} for single-component ripening, removing key limitations such as the inability to capture interactions between distant bubbles. Systematic validation against pore-network simulations across 19 cases---spanning homogeneous, heterogeneous, correlated, and uncorrelated networks with various bubble populations---shows good agreement with PNM in both the evolution of $g(s;t)$ and bulk properties of the population, without any adjustable parameters. The most challenging cases involve spatially correlated bubble composition, where errors arise due to methodological limitations of the theory discussed in Section \ref{sec:disc_lims_meth}. Unlike PNM, the theory allows for arbitrarily large bubble populations and arbitrary correlation lengths in pore size, enabling predictions beyond PNM's reach. This makes the theory a practical tool for predicting Ostwald ripening in a variety of applications, including underground hydrogen storage.


\section*{Acknowledgments}
This material is based upon work supported by the National Science Foundation, United States under Grant No. CBET-2348723. N.B. and L.A. gratefully acknowledge funding support from the William A. Fustos Family Professorship in Energy and Mineral Engineering at the Pennsylvania State University. We acknowledge the Institute for Computational and Data Sciences (ICDS) at Penn State University for computational resources.

\section*{Declaration of interests}
The authors declare that they have no known competing financial interests or personal relationships that could have appeared to influence the work reported in this paper.

\appendix
\setcounter{figure}{0} 
\setcounter{table}{0} 
\section{Derivation of areal porosity} \label{app:app_porosity}
We derive an expression for areal porosity $\phi_a$ in Eq.\ref{eq:fluxarea} of a network with average coordination number $z$, tortuosity $\tau$, throat cross-sectional area $A_t$, and throat length $l_t$. Pick one pore and draw a cylinder (in 2D) or sphere (in 3D) of radius $l$ centered at it. How many throats does the surface of this cylinder/sphere intersect? Denoting this number by $z_n$, areal porosity $\phi_a$ can be estimated as the ratio of the total cross-sectional area of intersected throats $A_T\!=\!z_n A_t$ to the surface area of the cylinder/sphere itself $A_S\!=\!2\pi l h$ in 2D ($h$ is network's out-of-plane thickness) or $A_S\!=\!4\pi l^2$ in 3D. Thus, all we need is an expression for $z_n$. We proceed by estimating the number of pores within an annulus from radial distance $l$ to $l+l_t$ as $n_p\!=\!A_Sl_t/(l_t^2 h)$ in 2D and $n_p\!=\!A_Sl_t/l_t^3$ in 3D, where the denominator represents the bulk area/volume associated with each pore. From each of these $n_p$ pores, roughly $z/2$ throats emanate out of the cylinder/sphere, and $z/2$ point inward. The latter are the ones intersected by the cylinder/sphere, resulting in $z_n\!=\!n_p z/2$. Substituting for $n_p$ and inserting $z_n$ into $\phi_a\!=\!z_nA_t/A_S$ from above, we get $\phi_a\!=\!C A_t/l_t$ with $C\!=\!z/(2h)$ in 2D and $C\!=\!z/(2l_t)$ in 3D.

\section{Rationale for reformulating Eq.\ref{eq:closure_raw} into Eq.\ref{eq:closure_final}} \label{app:closure_reform}
The phase-velocity expressions for $dS^b_s/dt$ and $dy_{1,s}/dt$ in Eq.\ref{eq:closure_raw} are numerically problematic when an Eulerian (grid-based) method (like finite volume herein) is used to solve the population balance Eq.\ref{eq:populationBalance}. Our goal is to demonstrate this difficulty with an example, and show why the reformulated expressions in Eq.\ref{eq:closure_final} do not suffer from it. To begin, consider a bubble population residing in a \textit{homogeneous} network (uniform $R_p$) for simplicity. Furthermore, assume the majority of bubbles in the network ($1-\varepsilon$ fraction of them, where $\varepsilon\!\ll\!1$) have mutually equilibrated except a small minority ($\varepsilon$ fraction) that are at state $s^\ast$. For such a system, mean-field properties are close (arbitrarily as $\varepsilon\!\rightarrow\!0$) to the equilibrated majority's state---we therefore do not distinguish between the majority and mean field and use the superscript $(\cdot)^{mf}$ for both. Given this setup, let us now analyze how the $s^\ast$-bubbles would evolve under Eq.\ref{eq:closure_raw} to equilibrate with the mean field.

\begin{figure}[t!]
	\centering
	\includegraphics[width=0.75\textwidth]{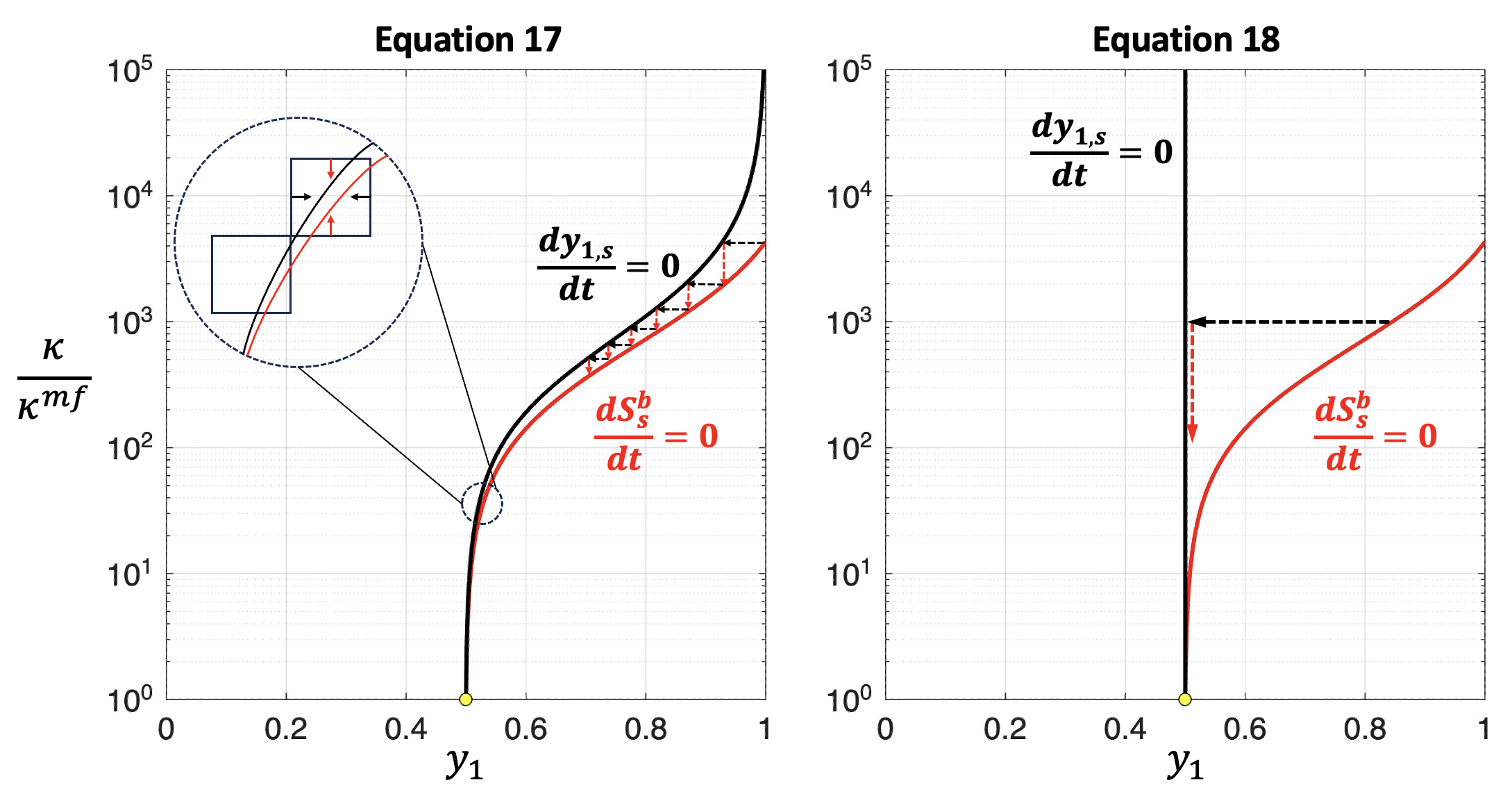}
	\caption{(Left) Schematic of the curves defined by $dS^b_s/dt\!=\!0$ and $dy_{1,s}/dt\!=\!0$ for Eq.\ref{eq:closure_raw} in phase space for a homogeneous network. The phase space is 2D because pore size ($R_p$) is uniform, with equivalent coordinates $(\kappa/\kappa^{mf},y_1)$ instead of $(S^b,y_1)$. The two curves form an ever narrowing corridor that bubbles must traverse to reach equilibrium (yellow dot). At some point, the corridor's width is smaller than the grid used to discretize the population balance Eq.\ref{eq:populationBalance}, halting further progress towards the yellow dot. (Right) The reformulation in Eq.\ref{eq:closure_final} removes this drawback by repositioning the black line to widen the corridor.}
	\label{fig:ReformClosure}
\end{figure}

At equilibrium, $s^\ast$-bubbles will satisfy $dS^b_{s^\ast}/dt\!=\!0$ and $dy_{1,{s^\ast}}/dt\!=\!0$, with $s^\ast$ corresponding to the mean-field state $s^{mf}$. This point, $s^{mf}$, is the intersection of two curves: one defined by $dS^b_s/dt\!=\!0$ and the other by $dy_{1,s}/dt\!=\!0$---$s$ here is a free index. Fig.\ref{fig:ReformClosure} visualizes this in the slightly altered phase space with coordinates $(\kappa/\kappa_{mf},y_1)$, instead of $(S^b,y_1)$. We pause to note that the majority of bubbles must be super-critical (non-spherical) because otherwise they would not be in equilibrium (contradiction); ripening of spherical bubbles is unstable to perturbations \cite{mehmani2022JCP}. This makes the coordinate mapping $S^b\!\mapsto\!\kappa/\kappa^{mf}$ well-posed, otherwise the map would not be injective (i.e., spherical and deformed bubbles with different $S^b$ could map onto the same $\kappa$). The left panel of Fig.\ref{fig:ReformClosure} depicts the $dS^b_s/dt\!=\!0$ curve in red and the $dy_{1,s}/dt\!=\!0$ curve in black.

To see why $dS^b_s/dt\!=\!0$ and $dy_{1,s}/dt\!=\!0$ correspond to curves, use them to set the LHS of Eq.\ref{eq:closure_raw} to zero:
\begin{subequations} \label{eq:ReformClosure_1}
\begin{align}
	0 &= \frac{D_1}{H_1} \left((y_1P^b)^{mf} - (y_{1}P^b)_s\right) + 
	    \frac{D_2}{H_2} \left((y_2P^b)^{mf} - (y_{2}P^b)_s\right)
	    \label{eq:ReformClosure_1a}
	\\
	0 &= y_{2,s}\frac{D_1}{H_1} \left((y_1P^b)^{mf} - (y_{1}P^b)_s\right) - 
	    y_{1,s}\frac{D_2}{H_2} \left((y_2P^b)^{mf} - (y_{2}P^b)_s\right)
	    \label{eq:ReformClosure_1b}
\end{align}
\end{subequations}
Substituting $\smash{y_{2,s}\!=\! 1 - y_{1,s}}$ and $\smash{P^b_s\!=\!P_w + \sigma\kappa_s}$, and recognizing $\smash{(y_1P^b)^{mf}\!\approx\!y^{mf}_{1}(P^b)^{mf}}$ due to the population being in near equilibrium, the conclusion follows.
Notice we have dropped the subscript $s$ from $(\cdot)^{mf}_s$ because the mean field here does not depend on $s$.
Now, we claim that regardless of where the $s^\ast$-bubbles are located in the $(\kappa/\kappa_{mf},y_1)$ plane, they will move towards $s^{mf}$ depicted by the yellow dot in Fig.\ref{fig:ReformClosure}. Moreover, if at any point during that movement $s^\ast$ falls inside the wedge defined by the space between the black and red curves in Fig.\ref{fig:ReformClosure} (left panel), then it will stay inside that wedge as it approaches $s^{mf}$. But since this movement forces $s^\ast$ through an ever narrowing corridor, at some point the gap between the two curves becomes smaller than the size of a single (finite-volume) grid used to discretize Eq.\ref{eq:populationBalance}. This is shown by the zoom-in inset in Fig.\ref{fig:ReformClosure}. Because phase velocities on the faces of this grid point \textit{into} the grid, as their centroids fall outside the wedge, the $s^\ast$-bubbles become trapped within that grid and their progress towards $s^{mf}$ gets halted.

To prove this claim, we examine the signs of $dS^b_s/dt$ and $dy_{1,s}/dt$ everywhere on the $(\kappa/\kappa_{mf},y_1)$ plane. Clearly, if $dS^b_s/dt\!=\!0$ defines a curve (red), splitting the plane into two regions, either $dS^b_s/dt\!>\!0$ or $dS^b_s/dt\!<\!0$ must hold in each region. To determine the sign of each region, it suffices to evaluate the sign of $dS^b_{s'}/dt$ at an arbitrary point $s'$ away from the red curve in Fig.\ref{fig:ReformClosure} (left panel).
We pick this point to lie on the vertical line $y_{1,s'}\!=\!y^{mf}_1$ above the yellow dot ($s^{mf}$). Substituting this into the RHS of Eq.\ref{eq:closure_raw_1} yields:
\begin{equation} \label{eq:ReformClosure_2}
	\frac{dS^b_{s'}}{dt} 
	     =  C_1 \left(P^{b,mf} - P^b_{s'}\right)
		 =  C_1  \left(\kappa^{mf} - \kappa_{s'}\right)
\end{equation}
where $C_1\!>\!0$.\footnote{Because $d\kappa_{s'}/dS^b_{s'}\!>\!0$ in Eq.\ref{eq:closure_raw_1} on account of bubbles being super-critical.} The second equality follows from $P^b\!=\!P_w + \sigma\kappa$. Since $\kappa_{s'}\!>\!\kappa^{mf}$ at the point we have chosen, $dS^b_{s'}/dt\!<\!0$. Therefore, $dS^b_s/dt\!<\!0$ must hold for all $s$ in the region left of the red curve in Fig.\ref{fig:ReformClosure} (left panel), and $dS^b_s/dt\!>\!0$ in the region to its right.
An identical argument proves that $dy_{1,s}/dt\!>\!0$ must hold left of the black curve, and $dy_{1,s}/dt\!<\!0$ to its right. However, this time the arbitrary point is chosen on the horizontal line $\kappa_{s'}\!=\!\kappa^{mf}$, implying $P^b_{s'}\!=\!P^{b,mf}$. Substituting this into Eq.\ref{eq:closure_raw_2} yields:
\begin{equation} \label{eq:ReformClosure_3}
	\frac{dy_{1,s'}}{dt} 
	     =  C_2 \left(y_1^{mf} - y_{1,s'}\right)
\end{equation}
where $C_2\!>\!0$, and the sign of each region is determined.

What we have shown above is that the vertical component of the phase velocity $dS^b_s/dt$, which is proportional to $d\kappa_s/dt$ by Eq.\ref{eq:kappa_vb} and the chain rule, points down in the region left of the red curve and points up to the right of it. Similarly, the horizontal component of the phase velocity $dy_{1,s}/dt$ points right in the region left of the black curve and points left to the right of it. This means once $s^\ast$-bubbles fall into the wedge between the red and black curves, they cannot escape it and the numerical problem above is inevitable.

The reformulation of Eq.\ref{eq:closure_raw} into Eq.\ref{eq:closure_final} solves this problem by repositioning the black curve to the vertical line shown in the right panel of Fig.\ref{fig:ReformClosure}. This broadens the wedge and prevents the gap between the red and black curves from becoming smaller than the grid size unless $s^\ast$ is very close to the yellow dot (equilibrium).
Empirically, we observed the same problem in heterogeneous networks, which are far less tractable because the phase space is 3D.
As discussed in Section \ref{sec:disc_lims_meth}, the reformulation of Eq.\ref{eq:closure_raw}, which comes with approximation errors, may be avoided if a Lagrangian (particle-based) method is used to solve the population balance Eq.\ref{eq:populationBalance}.

\section{Validating Eq.\ref{eq:bubbleSpacing} for bubble spacing} \label{app:bubbleSpacing}
\begin{figure}
	\centering
	\includegraphics[width=0.7\textwidth]{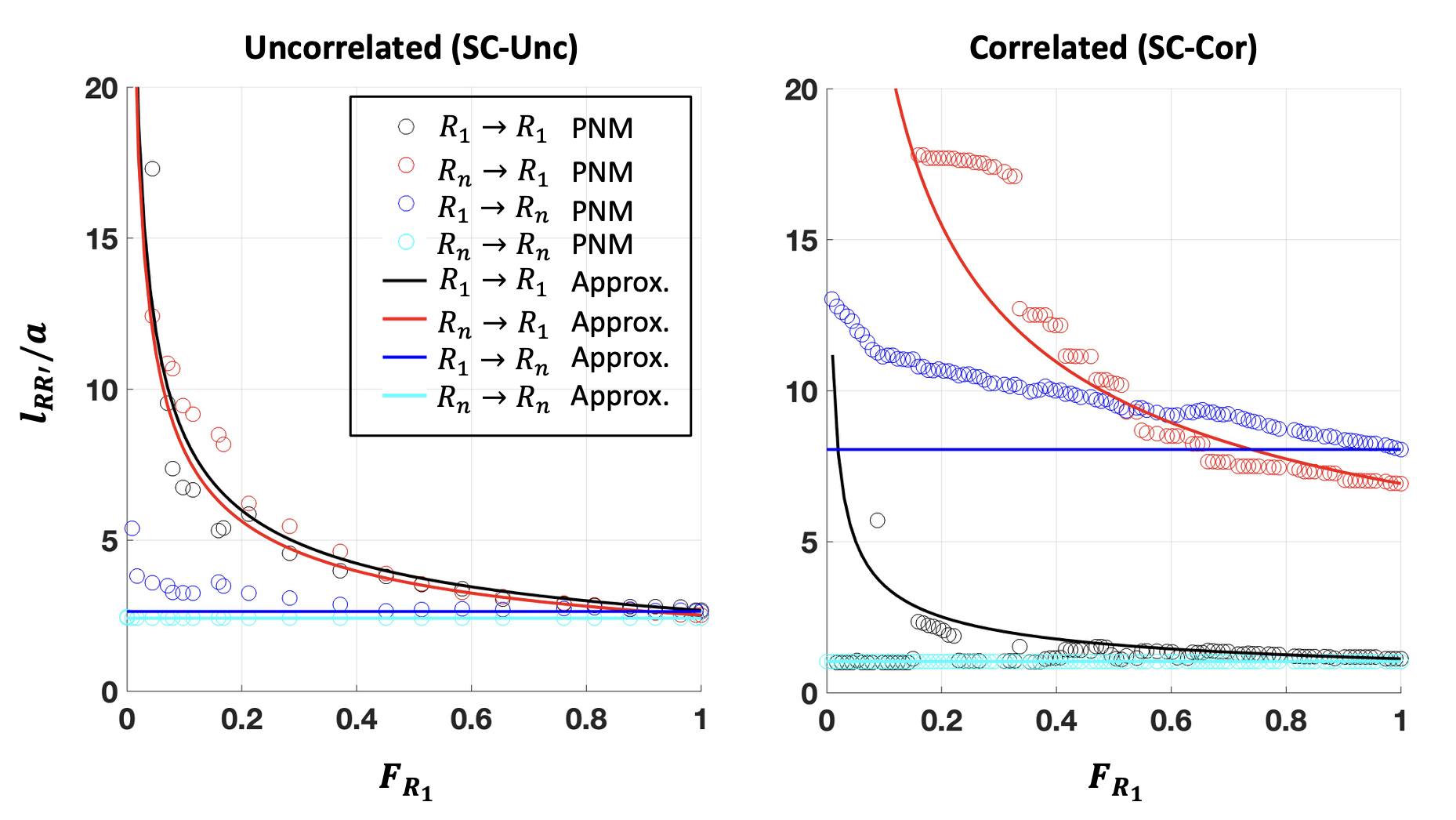}
	\caption{Inter-bubble spacing $l_{RR'}$ predicted from Eq.\ref{eq:bubbleSpacing} in the theory (solid lines) versus PNM using Eq.\ref{eq:lrr_PNM} (circles). For clarity, we use $R_x\!\rightarrow\!R_y$ to represent $l_{R_xR_y}$. Pore sizes are divided into $n\!=\!20$ bins, where $R_1$ is the smallest and $R_n$ the largest.}
	\label{fig:bubspace}
\end{figure}

We validate the $l_{RR'}$ computed from Eq.\ref{eq:bubbleSpacing} against direct computations from the PNM in Section \ref{sec:PNM}. We consider heterogeneous correlated and uncorrelated networks corresponding to the SC-Cor and SC-Unc cases in Section \ref{sec:valid}, respectively. Since the exact position of all bubbles are known in the PNM, we can compute $l_{RR'}$ directly using the following formula corresponding to the variable's definition:
\begin{equation} \label{eq:lrr_PNM}
	l_{RR'} = \langle\, \min_{R'} |x_R - x_{R'}| \,\rangle_{R}
\end{equation}
Namely, $l_{RR'}$ is the average nearest distance from bubbles residing in pores of size $R_p$ to bubbles in pores of size $R'_p$. In Eq.\ref{eq:lrr_PNM}, $x_R$ and $x_{R'}$ denote the positions of these bubbles. The minimum is taken over all bubble-occupied $R'_p$ and the average $\langle\cdot\rangle$ over all bubble-occupied $R_p$. Incidentally, Eq.\ref{eq:lrr_PNM} is also used to compute $d_{RR'}$ in Eq.\ref{eq:bubbleSpacing} with the crucial difference that the minimum and average are taken over all pores in the network regardless of bubble occupancy.
Fig.\ref{fig:bubspace} plots the $l_{RR'}$ obtained from Eq.\ref{eq:bubbleSpacing} against that from Eq.\ref{eq:lrr_PNM}. The comparison is made by dividing pore sizes in the networks into $n\!=\!20$ bins and focusing on the first $R_1$ (smallest pores) and last $R_n$ (largest pores) bins. This pick is motivated by the fact that $R_1$ and $R_n$ have the largest contrast among other pore sizes, thus are farthest apart spatially in the correlated network. This places the most stringent test on Eq.\ref{eq:bubbleSpacing}. In Fig.\ref{fig:bubspace}, we plot $l_{RR'}$ for subscripts $R$ and $R'$ taken from the set $\{R_1,R_n\}$. For clarity's sake, we use the notation $R_x\!\rightarrow\!R_y$ in the legend of Fig.\ref{fig:bubspace} to denote $l_{R_xR_y}$. 

For better visualization, instead of plotting $l_{RR'}$ versus time, the $x$-axis of Fig.\ref{fig:bubspace} denotes the fraction of the smallest bubble-occupied pores in the network, i.e., $F_{R_1}$. Since bubbles in $R_1$ vanish the fastest during ripening, $F_{R_1}$ serves as a proxy for time. For $l_{R_1R_1}$ and $l_{R_nR_1}$, Eq.\ref{eq:bubbleSpacing} is an explicit function of $F_{R_1}$. For $l_{R_1R_n}$ and $l_{R_nR_n}$, which depend explicitly on $F_{R_n}$, we convert to $F_{R_1}$ using the corresponding value from the PNM simulation. To improve visualization, we normalize the $y$-axis by dividing $l_{RR'}$ by the lattice spacing of the network. Fig.\ref{fig:bubspace} shows acceptable agreement between Eq.\ref{eq:bubbleSpacing} and PNM, validating the former's accuracy.

\section{Proof of Theorem \ref{thm:massConsrv}} \label{ap:proof_thm}
Beginning from the leftmost expression in Theorem \ref{thm:massConsrv}, we can write:
\begin{equation} \label{eq:thm_1}
	\begin{split}
	\frac{dn_\alpha^{tot}}{dt} = 
	\frac{d}{dt} \int_\Omega n_{\alpha,s}g_s\, ds = 
	\int_\Omega n_{\alpha,s} \frac{\partial g_s}{\partial t}\, ds = 
   -\int_\Omega n_{\alpha,s} \nabla\cdot(g_s \boldsymbol{u})\, ds = 
    \qquad\qquad\qquad\qquad\qquad\qquad\\
   -\int_\Omega \nabla \cdot (n_{\alpha,s} g_s \boldsymbol{u})\, ds + 
      \int_\Omega g_s\boldsymbol{u} \cdot \nabla n_{\alpha,s}\, ds = 
   -\int_{\partial\Omega} n_{\alpha,s} g_s \boldsymbol{u}\cdot\nu\, d\Gamma + 
      \int_\Omega g_s\boldsymbol{u} \cdot \nabla n_{\alpha,s}\, ds
    \end{split}
\end{equation}
where we have used the abbreviated notation $g_s\!=\!g(s;t)$. The second equality follows from the Leibniz integral rule and the fact that neither $\Omega$ (the entire phase space) nor $n_{\alpha,s}$ depend on time $t$. The third equality followed from using the population balance equation Eq.\ref{eq:populationBalance} to substitute for $\partial g_s/\partial t$. The fourth is the consequence of applying the product rule, and the fifth follows from the divergence theorem where $\nu$ is the outward unit normal on $\partial\Omega$. Noticing $\boldsymbol{u}\cdot\nabla n_{\alpha,s}\!=\!dn_{\alpha,s}/dt$ holds due to Eq.\ref{eq:chain}, we arrive at:
\begin{equation} \label{eq:thm_2}
	\frac{dn_\alpha^{tot}}{dt} = 
   -\int_{\partial\Omega} n_{\alpha,s} g_s \boldsymbol{u}\cdot\nu\, d\Gamma + 
      \int_\Omega g_s \frac{dn_{\alpha,s}}{dt} \, ds
\end{equation}

We must now show that the first integral on the RHS is zero, because then the theorem is proven. Recall from Section \ref{sec:theory_overview} that $\partial\Omega$ consists of six boundaries: $R_p\!=\!0$, $R_p\!=\!R_p^{max}$, $y_1\!=\!0$, $y_1\!=\!1$, $S^b\!=\!0$, and $S^b\!=\!1$. The BCs at all these boundaries is $g_s\!=\!0$, except $S^b\!=\!0$. Therefore, the following holds:
\begin{equation} \label{eq:thm_3}
   \int_{\partial\Omega} n_{\alpha,s} g_s \boldsymbol{u}\cdot\nu\, d\Gamma =
   \lim_{\epsilon \rightarrow 0} 
        \int_{S^b=\epsilon} n_{\alpha,s} g_s \frac{dS^b_s}{dt}\, d\Gamma
\end{equation}
where we have used the fact that $\boldsymbol{u}\cdot\nu\!=\!dS^b_s/dt$ holds on $S^b\!=\!\epsilon$ for fixed $\epsilon\!>\!0$. We now show the integrand on the RHS goes to zero as $\epsilon\!\rightarrow\!0$. More precisely, the integral can be bounded as:
\begin{equation} \label{eq:thm_4}
   \left|\,\int_{S^b=\epsilon} n_{\alpha,s} g_s \frac{dS^b_s}{dt}\, 
                                                   d\Gamma\,\right| \,\leq\,
   \int_{S^b=\epsilon} \left|\,n_{\alpha,s} g_s \frac{dS^b_s}{dt}\,
                                       \right|\, d\Gamma \,\leq\,
   \left\|\,n_{\alpha,s} \frac{dS^b_s}{dt}\right\|_\infty 
      \int_{S^b=\epsilon} \left|\,g_s\,\right|\, d\Gamma
\end{equation}
through a straightforward application of H\"older's inequality. For brevity, we denote the portion of $\Omega$ where $S^b\!=\!\epsilon$ (i.e., integration domain) by $\Gamma_\epsilon$. The rightmost integral of $|g_s|$ on $\Gamma_\epsilon$ ($L_1$ norm) is bounded since $g_s\!\geq\!0$ and the integral on $\Omega\supset\Gamma_\epsilon$ is bounded, equaling the number of bubbles in the population $n_b$. 

We thus focus on the $L_\infty$ norm of $n_{\alpha,s}dS^b_s/dt$ over $\Gamma_\epsilon$ in Eq.\ref{eq:thm_4}. As $\epsilon\!\rightarrow\!0$, it is evident $S^b_s\!\rightarrow\!0$. In this limit, bubbles are circular/spherical in shape whose curvature scales like $\kappa_s\sim (S^b_s)^{-1/d}$. We use the symbol $a\sim b$ to mean: there exist constants $C_1$ and $C_2$ such that $C_1 b\leq a \leq C_2 b$. From $P^b_s\!=\!P_w + \sigma\kappa_s$, we have $P^b_s\sim (S^b_s)^{-1/d}$. Since $V^b_s\!=\!V_{p,s}S^b_s$, we also have $V^b_s\sim S^b_s$. From the ideal gas law, $n_{\alpha,s}\!=\!y_{\alpha,s}P^b_sV^b_s/RT$ holds, and because $y_{\alpha,s}\sim 1$ we can write $n_{\alpha,s}\sim (S^b_s)^{1-1/d}$. Recalling Eq.\ref{eq:closure_final_1} for $dS^b_s/dt$ in Section \ref{sec:theory_closure}:
\begin{equation} \label{eq:thm_5}
\frac{d S^b_{s}}{dt} = 
	\frac{1}{V_{p,s}} \left(\rho^b_s + \frac{S^b_{s}}{RT} \sigma\frac{d\kappa_{s}}{dS^b_s} \right)^{-1} 
	\rho_w G_s \sum^2_{\alpha=1} \frac{D_{\alpha}}{H_\alpha} 
	                  \left((y_{\alpha} P^b)^{mf}_{s} - y_{\alpha,s}  P^b_s\right)
\end{equation}
we can now prove that $dS^b_s/dt\lesssim 1$ holds, where the symbol $a\lesssim b$ means: there exists a constant $C$ such that $a\leq Cb$. To demonstrate, note that $S^b_s d\kappa_s/dS^b_s\sim\kappa_s\sim (S^b_s)^{-1/d}$, which leads to the scaling:
\begin{equation} \label{eq:thm_6}
	\frac{1}{V_{p,s}} \left(\rho^b_s + \frac{S^b_{s}}{RT} \sigma\frac{d\kappa_{s}}{dS^b_s} \right)^{-1} 
	\rho_w G_s \sim (S^b_s)^{1/d}
\end{equation}
for the term before the summation in Eq.\ref{eq:thm_5}. For the summation itself, we can write:
\begin{equation} \label{eq:thm_7}
	\left|\sum^2_{\alpha=1} \frac{D_{\alpha}}{H_\alpha} 
	                  \left((y_{\alpha} P^b)^{mf}_{s} - 
	                         y_{\alpha,s}  P^b_s\right)\right|
	\leq
	\left(\sum^2_{\alpha=1} \left(\frac{D_{\alpha}}
	                              {H_\alpha}\right)^2\right)^{1/2}
	\left(\sum^2_{\alpha=1} \left((y_{\alpha} P^b)^{mf}_{s} - 
	                         y_{\alpha,s}  P^b_s\right)^2\right)^{1/2}
	\lesssim
	P^b_s \sim (S^b_s)^{-1/d}
\end{equation}
The first inequality follows from Cauchy-Schwarz. The second is the result of applying $|a - b|\leq |a + b| \leq |a| + |b|$ for $a,b\!\geq\!0$ to $|(y_\alpha P^b)^{mf}_s - y_{\alpha,s} P^b_s|$ followed by using $y_{\alpha,s}\sim y^{mf}_\alpha \sim 1$ and the fact that $P^{b,mf}_s$ is bounded by $P^b_s$. The latter is because $P^b_s$ dominates $P^{b,mf}_s$ in the limit $S^b_s\!\rightarrow\!0$. Hence, the summation in Eq.\ref{eq:thm_5} can be bounded from above by a term that scales like $(S^b_s)^{-1/d}$. Together with Eq.\ref{eq:thm_6}, this proves $dS^b_s/dt\lesssim 1$.

Given $dS^b_s/dt\lesssim 1$ and $n_{\alpha,s}\sim (S^b_s)^{1-1/d}$ from the above analysis, we can write:
\begin{equation} \label{eq:thm_8}
	\left\|\,n_{\alpha,s} \frac{dS^b_s}{dt}\right\|_\infty 
	\lesssim (S^b_s)^{1-1/d} = \epsilon^{1-1/d} \rightarrow 0
\end{equation}
The RHS follows from the fact that the norm is taken over $\Gamma_\epsilon$ ($S^b\!=\!\epsilon$) and $\epsilon\!\rightarrow\!0$. We have thus proven:
\begin{equation} \label{eq:thm_9}
   \int_{\partial\Omega} n_{\alpha,s} g_s \boldsymbol{u}\cdot\nu\, d\Gamma =
   0
\end{equation}
through Eq.\ref{eq:thm_3}, which in turn proves Theorem \ref{thm:massConsrv} due to Eq.\ref{eq:thm_2}.

\section{Proof of Corollary \ref{cor:ymix}} \label{ap:proof_cor}
A consequence of Theorem \ref{thm:massConsrv} is:
\begin{equation} \label{eq:cor_1}
	\int_\Omega n_{\alpha,s}g(s;t)\, ds = 
	\int_\Omega n_{\alpha,s}g(s;0)\, ds = C_\alpha
\end{equation}
where $C_\alpha$ is a constant that depends on $\alpha$ but not time $t$. Summing Eq.\ref{eq:cor_1} over $\alpha\!\in\{1,2\}$ yields:
\begin{equation} \label{eq:cor_2}
	\int_\Omega n_{s}g(s;t)\, ds = \int_\Omega n_{s}g(s;0)\, ds = C
\end{equation}
where
\begin{equation} \label{eq:cor_3}
	n_s = \sum^2_{\alpha=1} n_{\alpha,s}
\end{equation}
and $C$ is a constant independent of $\alpha$ and $t$.

Substituting Eqs.\ref{eq:cor_1}-\ref{eq:cor_2} into the definition of $y^{mix}_\alpha$ in Eq.\ref{eq:ymix} and noting $n_{\alpha,s}\!=\!y_{\alpha,s}n_s$, we get:
\begin{equation} \label{eq:cor_4}
	y^{mix}_\alpha = 
	   {\int_s y_{\alpha,s}\, n_s\, g(s;t)\, ds}\;/
	   {\int_s n_s\, g(s;t)\, ds} =
	   C_\alpha/C
\end{equation}
implying $y^{mix}_\alpha$ remains constant for all $t$. This prove the first assertion in Corollary \ref{cor:ymix}.

To prove the second, notice at $t\!\rightarrow\!\infty$ when the bubble population has reached equilibrium, all bubble compositions and curvatures must be identical. Namely, $y_{1,s}\!=\!y^{mf}_1$ and $y_{1,s}P^b_s\!=\!(y_1P^b)^{mf}_s$ (thus $\kappa_s\!=\!\kappa^{mf}_s$) must hold for all $s$. Otherwise, phase velocities $dS^b_s/dt$ and/or $dy_{1,s}/dt$ computed from Eq.\ref{eq:closure_final} would be non-zero at some $s$, contradicting the fact that the population has reached steady-state equilibrium. Denoting these equilibrium properties by $y_1^\infty$ and $(y_1P^b)^{\infty}$, it then follows:
\begin{equation} \label{eq:cor_5}
	\lim_{t\rightarrow\infty} y_{1,s} = 
	    \lim_{t\rightarrow\infty} y^{mf}_1 = y^{\infty}_1
	\,,\quad\quad
	\lim_{t\rightarrow\infty} y_{1,s}P^b_s = 
	    \lim_{t\rightarrow\infty} (y_1 P^b)^{mf}_s = (y_1P^b)^{\infty}
\end{equation}
Using these relations, we now take the limit of Eq.\ref{eq:gamma} at $t\!\rightarrow\!\infty$, which yields $\gamma\!\rightarrow\!1$. Similarly, taking the limit of $y^{mix}_1$ in Eq.\ref{eq:ymix}, we obtain $y^{mix}_1\!\rightarrow\!y^{\infty}_1$. However, since we established that $y^{mix}_1$ remains constant for all $t$, then $y^{mix}_1\!=\!y^{\infty}_1$ must hold. Using $\gamma\!\rightarrow\!1$ and $y^{mf}_1\!\rightarrow\!y^{mix}_1$ in taking the limit of Eq.\ref{eq:correct} proves $y^{mf,\ast}_1\!\rightarrow\!y^{mix}_1$.


\bibliography{References}

@article{bueno2024AWR,
	author = {Nicolas Bueno and Luis Ayala and Yashar Mehmani},
	doi = {10.1016/j.advwatres.2023.104826},
	issn = {0309-1708},
	journal = {Advances in Water Resources},
	publisher = {Elsevier Ltd},
	title = {A generalized kinetic theory of Ostwald ripening in porous media},
	volume = {193},
	year = {2024},
}

@article{bueno2023AWR,
	author = {Nicolas Bueno and Luis Ayala and Yashar Mehmani},
	doi = {10.1016/j.advwatres.2023.104581},
	issn = {0309-1708},
	journal = {Advances in Water Resources},
	publisher = {Elsevier Ltd},
	title = {Ostwald ripening of multi-component bubbles in porous media: A theory and a pore-scale model of how bubble populations equilibrate},
	volume = {182},
	year = {2023},
}

@article{mehmani2022JCP,
   author = {Yashar Mehmani and Ke Xu},
   doi = {10.1016/j.jcp.2022.111041},
   issn = {10902716},
   journal = {Journal of Computational Physics},
   month = {5},
   publisher = {Academic Press Inc.},
   title = {Pore-network modeling of Ostwald ripening in porous media: How do trapped bubbles equilibrate?},
   volume = {457},
   year = {2022},
}

@article{mehmani2022AWR,
   author = {Yashar Mehmani and Ke Xu},
   doi = {10.1016/j.advwatres.2022.104223},
   issn = {03091708},
   journal = {Advances in Water Resources},
   month = {8},
   publisher = {Elsevier Ltd},
   title = {Capillary equilibration of trapped ganglia in porous media: A pore-network modeling approach},
   volume = {166},
   year = {2022},
}

@article{xu2019GravGRL,
  title={Gravity-induced bubble ripening in porous media and its impact on capillary trapping stability},
  author={Xu, Ke and Mehmani, Yashar and Shang, Luoran and Xiong, Qingrong},
  journal={Geophysical Research Letters},
  volume={46},
  number={23},
  pages={13804--13813},
  year={2019},
  publisher={Wiley Online Library}
}

@article{yu2023GRLkinetics,
  title={Bubble Coarsening Kinetics in Porous Media},
  author={Yu, Yuehongjiang and Wang, Chuanxi and Liu, Junning and Mao, Sheng and Mehmani, Yashar and Xu, Ke},
  journal={Geophysical Research Letters},
  volume={50},
  number={1},
  pages={e2022GL100757},
  year={2023},
  publisher={Wiley Online Library}
}

@article{mehmani2024deplete,
  title={A continuum theory of diffusive bubble depletion in porous media},
  author={Mehmani, Yashar and Xu, Ke},
  journal={Advances in Water Resources},
  volume={185},
  pages={104625},
  year={2024},
  publisher={Elsevier}
}

@article{salehpour2025micro,
  title={Ostwald Ripening in Underground Gas Storage},
  author={Salehpour, Mohammad and Lan, Tian and Bueno, Nicolas and Laku, Md Zahidul Islam and Mehmani, Yashar and Zhao, Benzhong},
  journal={arXiv preprint arXiv:2509.00044},
  year={2025}
}

@article{xu2017PRL,
   title={Egalitarianism among bubbles in porous media: an ostwald ripening derived anticoarsening phenomenon},
  author={Xu, Ke and Bonnecaze, Roger and Balhoff, Matthew},
  journal={Physical review letters},
  volume={119},
  number={26},
  pages={264502},
  year={2017},
  publisher={APS}
}

@article{feng2022KeGravTIPM,
  title={Comprehensive Darcy-Scale Analysis of Ripening in Porous Media},
  author={Feng, Yitian and Wang, Chuanxi and Jin, Xu and Xu, Ke},
  journal={Transport in Porous Media},
  volume={144},
  number={1},
  pages={301--316},
  year={2022},
  publisher={Springer}
}

@article{deChalendar2018pnm,
  title={Pore-scale modelling of Ostwald ripening},
  author={de Chalendar, Jacques A and Garing, Charlotte and Benson, Sally M},
  journal={Journal of Fluid Mechanics},
  volume={835},
  pages={363--392},
  year={2018},
  publisher={Cambridge University Press}
}

@article{joewondo2022pnm,
  title={Nonuniform collective dissolution of bubbles in regular pore networks},
  author={Joewondo, Nerine and Garbin, Valeria and Pini, Ronny},
  journal={Transport in Porous Media},
  volume={141},
  number={3},
  pages={649--666},
  year={2022},
  publisher={Springer}
}

@article{joewondo2023exp,
  title={Experimental evidence of the effect of solute concentration on the collective evolution of bubbles in a regular pore-network},
  author={Joewondo, Nerine and Garbin, Valeria and Pini, Ronny},
  journal={Chemical Engineering Research and Design},
  volume={192},
  pages={82--90},
  year={2023},
  publisher={Elsevier}
}

@article{singh2023pore,
  title={Pore-scale Ostwald ripening of gas bubbles in the presence of oil and water in porous media},
  author={Singh, Deepak and Friis, Helmer Andr{\'e} and Jettestuen, Espen and Helland, Johan Olav},
  journal={Journal of Colloid and Interface Science},
  volume={647},
  pages={331--343},
  year={2023},
  publisher={Elsevier}
}

@article{singh2024ostwald,
  title={Ostwald ripening of gas bubbles in porous media: Impact of pore geometry and spatial bubble distribution},
  author={Singh, Deepak and Friis, Helmer Andr{\'e} and Jettestuen, Espen and Helland, Johan Olav},
  journal={Advances in Water Resources},
  volume={187},
  pages={104688},
  year={2024},
  publisher={Elsevier}
}

@article{garing2017PcExp,
  title={Pore-scale capillary pressure analysis using multi-scale X-ray micromotography},
  author={Garing, Charlotte and de Chalendar, Jacques A and Voltolini, Marco and Ajo-Franklin, Jonathan B and Benson, Sally M},
  journal={Advances in Water Resources},
  volume={104},
  pages={223--241},
  year={2017},
  publisher={Elsevier}
}

@article{yaxin2020JFM,
  title={A continuum-scale representation of Ostwald ripening in heterogeneous porous media},
  author={Li, Yaxin and Garing, Charlotte and Benson, Sally M},
  journal={Journal of Fluid Mechanics},
  volume={889},
  pages={A14},
  year={2020},
  publisher={Cambridge University Press}
}

@article{yaxin2022TIPM,
  title={Long-term redistribution of residual gas due to non-convective transport in the aqueous phase},
  author={Li, Yaxin and Orr Jr, Franklin M and Benson, Sally M},
  journal={Transport in Porous Media},
  volume={141},
  number={1},
  pages={231--253},
  year={2022},
  publisher={Springer}
}

@article{blunt2022ostwald,
  title={Ostwald ripening and gravitational equilibrium: Implications for long-term subsurface gas storage},
  author={Blunt, Martin J},
  journal={Physical Review E},
  volume={106},
  number={4},
  pages={045103},
  year={2022},
  publisher={APS}
}

@article{zhang2023trap,
  title={Pore-scale observations of hydrogen trapping and migration in porous rock: Demonstrating the effect of Ostwald ripening},
  author={Zhang, Yihuai and Bijeljic, Branko and Gao, Ying and Goodarzi, Sepideh and Foroughi, Sajjad and Blunt, Martin J},
  journal={Geophysical Research Letters},
  volume={50},
  number={7},
  pages={e2022GL102383},
  year={2023},
  publisher={Wiley Online Library}
}

@article{song2025XrayH2,
  title={Visualized experiments on the hydrogen transports and bubble ripening mechanism in porous reservoir of underground hydrogen storage},
  author={Song, Rui and Feng, Daiying and Hui, Gang and Liu, Jianjun and Yang, Chunhe},
  journal={International Journal of Hydrogen Energy},
  volume={105},
  pages={326--344},
  year={2025},
  publisher={Elsevier}
}

@article{boon2024XrayH2,
  title={Multiscale experimental study of H/brine multiphase flow in porous rock characterizing relative permeability hysteresis, hydrogen dissolution, and Ostwald ripening},
  author={Boon, Maartje and Rademaker, Tim and Winardhi, Chandra Widyananda and Hajibeygi, Hadi},
  journal={Scientific Reports},
  volume={14},
  number={1},
  pages={30170},
  year={2024},
  publisher={Nature Publishing Group UK London}
}

@Article{ostwald1897,
  title={Studien {\"u}ber die Bildung und Umwandlung fester K{\"o}rper},
  author={Ostwald, Wilhelm},
  journal={Zeitschrift f{\"u}r physikalische Chemie},
  volume={22},
  number={1},
  pages={289--330},
  year={1897},
  publisher={Oldenbourg Wissenschaftsverlag}
}

@article{lifshitz1961orig,
  title={The kinetics of precipitation from supersaturated solid solutions},
  author={Lifshitz, Ilya M and Slyozov, Vitaly V},
  journal={Journal of physics and chemistry of solids},
  volume={19},
  number={1-2},
  pages={35--50},
  year={1961},
  publisher={Elsevier}
}

@article{wanger1961orig,
  title={Theorie der Alterung von Niederschlagen durch Umlosen},
  author={Wanger, C},
  journal={Z. Elektrochem.},
  volume={65},
  pages={581--591},
  year={1961}
}

@article{voorhees1985,
  title={The theory of Ostwald ripening},
  author={Voorhees, Peter W},
  journal={Journal of Statistical Physics},
  volume={38},
  pages={231--252},
  year={1985},
  publisher={Springer}
}

@article{voorhees1992,
   author = {P W Voorhees},
   journal = {Annual Reviews Material Science},
   pages = {197-215},
   title = {Ostwald ripening of two-phase mixtures},
   volume = {22},
   url = {www.annualreviews.org},
   year = {1992},
}

@article{bray2002theory,
  title={Theory of phase-ordering kinetics},
  author={Bray, Alan J},
  journal={Advances in Physics},
  volume={51},
  number={2},
  pages={481--587},
  year={2002},
  publisher={Taylor \& Francis}
}

@article{kim2018alloy,
  title={Ostwald ripening of spheroidal particles in multicomponent alloys},
  author={Kim, Kyoungdoc and Voorhees, Peter W},
  journal={Acta Materialia},
  volume={152},
  pages={327--337},
  year={2018},
  publisher={Elsevier}
}

@article{philippe2013alloy,
  title={Ostwald ripening in multicomponent alloys},
  author={Philippe, T and Voorhees, Peter W},
  journal={Acta Materialia},
  volume={61},
  number={11},
  pages={4237--4244},
  year={2013},
  publisher={Elsevier}
}

@article{morral1994alloy,
  title={Particle coarsening in binary and multicomponent alloys},
  author={Morral, JE and Purdy, GR},
  journal={Scripta Metallurgica et Materialia;(United States)},
  volume={30},
  number={7},
  year={1994}
}

@article{kukushkin1996alloy,
  title={Theory of the Ostwald ripening in multicomponent systems under non-isothermal conditions},
  author={Kukushkin, SA and Slyozov, VV},
  journal={Journal of Physics and Chemistry of Solids},
  volume={57},
  number={2},
  pages={195--204},
  year={1996},
  publisher={Elsevier}
}

@article{carman1937bundle,
  title={Fluid flow through granular beds},
  author={Carman, Philip Crosbie},
  journal={Trans. Inst. Chem. Eng. London},
  volume={15},
  pages={150--156},
  year={1937}
}

@article{kozeny1927bundle,
  title={Ueber kapillare leitung des wassers im boden},
  author={Kozeny, Josef},
  journal={Sitzungsberichte der Akademie der Wissenschaften in Wien},
  volume={136},
  pages={271},
  year={1927}
}

@article{hanson2022DOEreport,
title={Subsurface hydrogen and natural gas storage: State of knowledge and research recommendations report},
  author={Hanson, Angela Goodman and Kutchko, Barbara and Lackey, Greg and Gulliver, Djuna and Strazisar, Brian R and Tinker, Kara A and Haeri, Foad and Wright, Ruishu and Huerta, Nicolas and Baek, Seunghwan and others},
  year={2022},
  publisher={National Energy Technology Laboratory (NETL), USA}
}

@article{bachu2008co2,
  title={CO2 storage in geological media: Role, means, status and barriers to deployment},
  author={Bachu, Stefan},
  journal={Progress in energy and combustion science},
  volume={34},
  number={2},
  pages={254--273},
  year={2008},
  publisher={Elsevier}
}

@article{imhoff1996NAPL,
  title={Dissolution fingering during the solubilization of nonaqueous phase liquids in saturated porous media: 2. Experimental observations},
  author={Imhoff, Paul T and Thyrum, Geoffrey P and Miller, Case T},
  journal={Water Resources Research},
  volume={32},
  number={7},
  pages={1929--1942},
  year={1996},
  publisher={Wiley Online Library}
}

@article{lee2020FCbazyl,
  title={Accelerating bubble detachment in porous transport layers with patterned through-pores},
  author={Lee, Jason K and Lee, ChungHyuk and Fahy, Kieran F and Kim, Pascal J and Krause, Kevin and LaManna, Jacob M and Baltic, Elias and Jacobson, David L and Hussey, Daniel S and Bazylak, Aimy},
  journal={ACS Applied Energy Materials},
  volume={3},
  number={10},
  pages={9676--9684},
  year={2020},
  publisher={ACS Publications}
}

@article{lu2010FC,
  title={Water management studies in PEM fuel cells, part III: Dynamic breakthrough and intermittent drainage characteristics from GDLs with and without MPLs},
  author={Lu, Zijie and Daino, Michael M and Rath, Cody and Kandlikar, Satish G},
  journal={International Journal of Hydrogen Energy},
  volume={35},
  number={9},
  pages={4222--4233},
  year={2010},
  publisher={Elsevier}
}

@article{holocher2003dissol,
  title={Kinetic model of gas bubble dissolution in groundwater and its implications for the dissolved gas composition},
  author={Holocher, Johannes and Peeters, Frank and Aeschbach-Hertig, Werner and Kinzelbach, Wolfgang and Kipfer, Rolf},
  journal={Environmental Science \& Technology},
  volume={37},
  number={7},
  pages={1337--1343},
  year={2003},
  publisher={ACS Publications}
}

\end{document}